\documentclass[aps,prd,twocolumn,showpacs,superscriptaddress,groupedaddress,nofootinbib]{revtex4}  
\usepackage{graphicx}  
\usepackage{dcolumn}   
\usepackage{bm}        
\usepackage{mathrsfs,mathtools}
\usepackage{amsmath,amssymb}   
\usepackage{multirow}
\usepackage{aas_macros}
\usepackage{ulem}

\usepackage{color}

\def\kpc{{\rm kpc}}

\def\TeV{{\rm TeV}}

\def\Msun{\ensuremath{M_{\odot}}}

\newcommand{\be}{\begin{equation}}
\newcommand{\ee}{\end{equation}}
\newcommand{\ba}{\begin{eqnarray}}
\newcommand{\ea}{\end{eqnarray}}

\def\gtsima{$\; \buildrel > \over \sim \;$}
\def\ltsima{$\; \buildrel < \over \sim \;$}
\def\gsim{\lower.5ex\hbox{\gtsima}}
\def\lsim{\lower.5ex\hbox{\ltsima}}
\def\simgt{\lower.5ex\hbox{\gtsima}}
\def\simlt{\lower.5ex\hbox{\ltsima}}
\def\simpr{\lower.5ex\hbox{\prosima}}
\def\mean#1{\left< #1 \right>}

\begin{document}

\title{Cosmic ray $e^{+} e^{-}$  spectrum excess and peak feature observed by the DAMPE experiment from dark matter}

\author{Hong-Bo Jin}
\affiliation{Key Laboratory of Computational Astrophysics, National Astronomical Observatories, \\
Chinese Academy of Sciences, 20A Datun Road, Chaoyang District, Beijing, 100012, China}
\affiliation{School of Astronomy and Space Science, University of Chinese Academy of Sciences}
\author{Bin Yue\footnote{Co-first author}}
\affiliation{Key Laboratory of Computational Astrophysics, National Astronomical Observatories, \\
Chinese Academy of Sciences, 20A Datun Road, Chaoyang District, Beijing, 100012, China}
\affiliation{School of Astronomy and Space Science, University of Chinese Academy of Sciences}
\author{Xin Zhang} 
\affiliation{Key Laboratory of Computational Astrophysics, National Astronomical Observatories, \\
Chinese Academy of Sciences, 20A Datun Road, Chaoyang District, Beijing, 100012, China}
\affiliation{School of Astronomy and Space Science, University of Chinese Academy of Sciences}
\author{Xuelei Chen\footnote{Corresponding author}}
\affiliation{Key Laboratory of Computational Astrophysics, National Astronomical Observatories, \\
Chinese Academy of Sciences, 20A Datun Road, Chaoyang District, Beijing, 100012, China}
\affiliation{School of Astronomy and Space Science, University of Chinese Academy of Sciences}
\affiliation{Center of High Energy Physics, Peking University, Beijing 100871, China}
 
\date{\today}

\begin{abstract}
The Chinese satellite Wukong, also known as the DArk Matter Particle Explorer (DAMPE) 
experiment, has released its observation data of the cosmic ray (CR) electrons and positrons. 
The data shows an excess in the energy spectrum up to TeV energy, and possibly also a 
peak-like fine structure at $\sim$1.4 TeV. We investigate the scenario that the source 
of the excess comes from the annihilation or decay of dark matter particles. We consider 
the W$^+$W$^-$ channel and direct $e^+e^-$ channel (model A), and 
the double $\tau^+\tau^-$ channel  and direct $e^+e^-$ channel (model B). 
We find that the annihilation or decay of diffuse dark matter particles in the Galactic halo can give 
excellent (for the W$^+$W$^-$ channel) or at least good (for the double  $\tau^+\tau^-$ channel) fits to the broad excess.
However, the annihilation cross-section is of the order of $10^{-23}$ cm$^3$ s$^{-1}$,
larger than required for obtaining the correct relic abundance when dark matter froze out. 
We then study whether the narrow peak at $\sim$1.4 TeV could be explained by a nearby subhalo, which thanks to the 
smaller distance, could supply $e^+e^-$ within a narrow energy range. We find that in order to produce a peak width
less than the DAMPE energy bin width (0.2 TeV), the source must be located within $r\lsim 0.53~\kpc$.
Our global fit models do not produce the peak-like feature, instead at $\sim$1.4 TeV the spectrum show 
either a slope or a cliff-like feature. However, if less than optimal fit to the data is allowed, the
 peak-like feature could be generated. Furthermore, an excellent fit with peak could be
 obtained with model B if the background is rescaled.  If the dark matter decay and annihilation rates are 
 determined using the broad excess, the required subhalo mass could be $\sim10^{5}~M_\odot$ for decay model with dark matter  
 particle lifetime $7.3\times10^{25}$ s,  or $\sim10^{4.5}~M_\odot$ for annihilation  model with cross-section $10^{-23}$ cm$^3$~s$^{-1}$ and a shallower density profile slope $\alpha=1.2$, or $\sim10^{2.5}~M_\odot$ if the density profile slope is a steep as $\alpha=1.7$.
However, considering the distribution profile of dark matter subhalos in our Milky Way's host halo, the probability for the existence of a such nearby subhalo as massive as given above is   very low.

\end{abstract}

\maketitle

\section{Introduction}
\label{sec:intro}

According to recent astronomical observations, about $26\%$ of the total cosmic density is made up 
by non-baryonic dark matter, much more than the $4.7\%$ baryonic matter we know of  
\cite{Planck2016cos}. The dark matter plays a crucial role in large scale structure growth and 
galaxy formation, but its presence is only inferred from its gravitational effects, while its nature is 
still unknown. A simple and plausible conjecture is that the dark matter are 
made up of unknown particles, which are electromagnetically neutral but participate in weak 
interactions. Such dark matter particles  may annihilate or decay into standard model particles. 
In the indirect search of weakly massive interacting particle (WIMP) dark matter,  
one looks for the signature of dark matter by observing $\gamma$-ray, energetic neutrino 
or charged cosmic ray (CR) particles as annihilation or decay products (for a review see e.g. \cite{2017arXiv170704591B}).   
As these are also produced by other astrophysical processes, which are not always well understood, 
it is necessary to look for distinct signatures from the dark matter annihilation or decay processes. 
In the case of CR, most CR particles follow power-law distributions, which could be nicely explained 
as  the result of diffusive shock acceleration by supernovae remnants (SNR), followed by a complicated 
transportation process in the Milky Way \cite{2007ARNPS..57..285S}.
On the other hand, the dark matter particle has a specific mass, so the energy of particles 
produced in dark matter annihilation or decay would be distributed in a particular energy range, which may 
produce an excess or deviation from the simple power-law in the CR energy spectrum. In particular, if the 
dark matter particle could annihilate or decay directly to a pair of standard model particles, this may 
produce a narrow line feature, which would be a smoking gun signature of dark matter, as there is no other 
known mechanism to produce such narrow line feature in the multi-GeV energy range.

The CR electrons and positrons have been measured by  a number of balloon 
or space-borne experiments. 
The HEAT \cite{2001ApJ...559..296D}, ATIC \cite{2008Natur.456..362C}, 
PAMELA \cite{2011PhRvL.106t1101A}, Fermi \cite{2009PhRvL.102r1101A,2017PhRvD..95h2007A}, 
and AMS-02 \cite{2014PhRvL.113v1102A} have measured the electron and positron spectrum up to 2 TeV. 
An intriguing excess above a few tens GeV were found by these experiments,  
though the uncertainty remains large at the high energy end where the 
CR flux drops. In addition, the ground based H.E.S.S. experiment observes the electron and positron spectrum 
indirectly at higher energies \cite{2011PhRvL.106p1301A}. 
From joint analysis of these experiment,  a possible break in the energy 
spectrum was found around TeV scale, though in such analysis the systematic uncertainty is 
sizable \cite{2008PhRvL.101z1104A,2009A&A...508..561A}. Such an excess could be generated by 
dark matter  annihilation or decay (e.g. \cite{2009NuPhB.813....1C,Dev2014,2017arXiv171103133P}),
 but to obtain a large excess, the dark matter annihilation rate must be 
higher than usually assumed for achieving the correct abundance when it froze out in the early Universe, 
or are from a relatively nearby source such as a subhalo \cite{2009PhRvL.103c1103B, 2010PhRvD..82h3525F,
2010arXiv1012.0588Z,2011PhRvL.106p1301A,2016PhRvD..93h3513Z,2017PhRvD..96b3006L,2017arXiv171100749L,
2017arXiv171104696Z}. 
Alternatively,  the excess could be produced by 
some energetic astrophysical processes such as pulsars or SNRs which could inject energetic electrons or positrons in the CR energy spectrum
\cite{2015PhRvD..91f3508L,2009PhRvD..80f3005M,2009ApJ...700L.170H,2013PhRvD..88b3013C,2014NuPhS.256....9B,2009PhRvD..80f3003F,2016PTEP.2016b1E01K,Blasi2009a,Blasi2009b,Serpico2012,Mertsch2009,Mertsch2011,Donato2014,Mertsch2014, Tomassetti2012,Tomassetti2015}.

The DAMPE satellite has made a new measurement of the electron and positron spectrum (electrons and 
positrons are not distinguishable in its observation). This satellite is designed to have low background contamination from the 
much greater proton component of the CR, and have high energy resolution so that sharp feature in the 
CR electron and positron spectrum would not be erased by energy resolution issue \cite{2017APh....95....6C}. 
Recently, the DAMPE experiment has released their measurement from 25 GeV to 4.6 TeV \cite{DAMPEpaper}.
Their energy spectrum is broadly consistent with the Fermi-LAT \cite{Abdollahi:2017nat}, however higher than those from the AMS-02 and CALET \cite{CALET2017} at $\gtrsim 70$ GeV.
A break in the energy spectrum is indeed found at about 1 TeV, confirming the earlier result from H.E.S.S. experiment, and 
provides a very precise measurement of CR electron positron energy spectrum at the TeV energy range. 
These results raise new interests on the dark matter contribution to the CR electron and positron spectrum.

In addition to the break, there is also a single peak feature at $\sim$1.4 TeV, with a statistical significance of 
$\sim 3\sigma$.  We should take a cautionary note here: the number of actual events 
detected in this energy bin is  only 93, while for the two adjacent bins the numbers are 74 and 33 respectively, 
so it is possible that the large peak is due to a statistical fluctuation. This would become more clear in the future as more
data is accumulated. Nevertheless, the peak is very intriguing, because in sources of 
astrophysical origin, such as SNRs or pulsars, it is not easy to generate particles with a single energy. 
So if the peak is real, it would be of great importance to the indirect dark matter search. 
Indeed, as we shall see below,  even in the dark matter scenario it is not easy to produce a narrow peak in the 
charged particle spectrum, because the electrons and positrons produced in dark matter annihilation or decay will 
diffuse into broad energy distributions during their propagation. So to generate a narrow peak in the spectrum, the 
source must be located at a very small distance where the diffusion does not take long time. 
One such possibility is a dark matter subhalo which is accidentally located near the Solar system. 

In this paper, we shall investigate the scenario that the dark matter annihilation or decay serve
as possible source of the excess CR electrons and positrons in light of the new DAMPE data.
The paper is organized as follows: In Sec. \ref{methods} we describe our method for computing the CR electron and positron 
spectrum. 
In Sec. \ref{broad} we present the results from fitting 
the broad spectral excess with the Galactic halo diffuse dark matter contribution. In Sec. \ref{peak}, we study whether 
the peak feature at $\sim$1.4 TeV could be generated by a nearby subhalo. In Sec. \ref{subhalo} we investigate what is the 
required mass for such a subhalo, and the constraint on the model from inverse Compton scattered  photons
and CR anisotropy.  We conclude in Sec. \ref{summary}.

\section{Methods}\label{methods}

The propagation of charged CR particles in the Milky Way can be described as a diffusion process, with interaction 
with other particles, and energy loss by radiation and collision, as well as re-acceleration in the turbulent magnetic field 
\cite{Ginzburg:1990sk}. To quantitatively model this process, the {\tt GALPROP}\footnote {\url{https://galprop.stanford.edu/}} software package has been developed, 
which solves the propagation equation with a Crank-Nicholson implicit second-order scheme \cite{Strong:1998pw}. 
The primary electrons are mostly produced in SNRs. Their injection spectrum is 
modeled as a simple power-law form, as is expected from the diffuse shock acceleration mechanism \cite{Blandford:1987pw}.
In the propagation models, the source item of the primary electrons is also often described as a broken power-law spectrum 
multiplied by the assumed spatial distribution given in the cylindrical coordinate \cite{Strong:1998pw}. The density of CR 
electron source is modeled as an exponential disc.  
The secondary electrons and positrons are produced during collisions of CR nucleons (protons dominant) 
with the interstellar gas  \cite{Vladimirov2011_GALPROP}. The distribution of the interstellar gas, which 
collide with CR particles and generate secondary particles, is derived from the observational data of H and CO, etc. \cite{Moskalenko:2001ya,1996A&A...308L..21S}. 
The CR propagation time can be constrained observationally by the ratio between primary particles and the 
secondary particles, such as the Be$^{10}$/Be$^{9}$ and B/C ratios in the cosmic ray. Alternatively, with the advent of 
high quality data from the AMS-02 experiment, the CR model parameters may also be determined solely from the 
proton \cite{Aguilar:2015ooa} and B/C \cite{Aguilar:2016vqr} data of AMS-02, as is demonstrated first in  
Ref. \cite{Jin:2014ica}, which yield tighter constraints.  In this paper, we shall also follow this approach. 

If some CR particles are produced by  dark matter annihilation or decay, their propagation processes are the same, only 
the injection source distribution and energy spectrum differ.  For dark matter annihilation, we model the source term as 
\begin{align}\label{eq:ann-source}
q(\boldsymbol{r},p)=\frac{\rho(\boldsymbol{r})^2}{2 m_{\chi}^2}\langle \sigma v \rangle 
\sum_X \eta_X \frac{dN^{(X)}}{dp} ,
\end{align}
where $\langle \sigma v \rangle$ is the velocity-averaged dark matter annihilation cross-section multiplied by dark matter relative velocity
(referred to as cross-section), $\rho(\boldsymbol{r})$ is the dark matter density distribution function, and $dN^{(X)}/dp$ is the 
injection energy spectrum  of  secondary particles from dark matter annihilating into standard model
 final states through all possible  intermediate states $X$ 
with $\eta_X$ the corresponding branching fractions.   Similarly, the source term for decay can be modeled as
\begin{align}\label{eq:decay-source}
q(\boldsymbol{r},p)=\frac{\rho(\boldsymbol{r})}{m_{\chi}} ~\Gamma_\chi~
\sum_X \eta_X \frac{dN^{(X)}}{dp} ,
\end{align}
where $\Gamma_\chi$ is the decay rate. 
The density profile $\rho(\boldsymbol{r})$ of the dark matter halos were derived in N-body simulation, such as NFW \cite{Navarro:1996gj}, 
isothermal \cite{Bergstrom:1997fj}, and Moore \cite{Moore:1999nt} models. In this paper 
we use the Einasto \cite{Einasto:2009zd} profile to model the dark matter distribution. The energy spectrum of electrons and 
positrons of each channel is calculated by PYTHIA v8.175 \cite{Sjostrand:2007gs}. The hadronic model 
interaction part of the code is also modified to improve the treatment of secondary particles \cite{Jin:2017iwg}.

In the dark matter particle annihilation or decay, the electrons and positrons can be produced directly, 
here we shall call this the direct $e^+e^-$ channel. Note that the annihilation or decay may produce an $e^+e^-$ pair, 
or either an $e^+$ or an $e^-$ plus other particles, but on the whole 
produce equal numbers of $e^+$ and $e^-$. For our investigation these differences do not matter, though in the 
former case the energy of the electron or positron would be exactly $m_\chi$(annihilation) or $m_\chi/2$ (decay),
and the primary energy spectrum of electrons and positrons is a very narrow  peak  around  $E_0=m_\chi$ (annihilation) 
or $E_0=m_\chi/2$ (decay) for the pair production. The electrons and positrons may also be produced 
by annihilation or decay into intermediate particles, such as the  W$^+$W$^-$ pairs or $\tau^+\tau^-$ pairs, which 
further decay to produce electrons and positrons. In these cases the energies are generally more broadly
distributed.  In this paper we shall consider 1) model A: electrons and positrons are produced through  the W$^+$W$^-$ channel
and the direct $e^+e^-$ channel; and 2) model B:  through  the double $\tau^+\tau^-$  channel and direct $e^+e^-$ channel. In previous analysis, it has been found that in fitting the broad 
spectrum excess measured by the AMS-02 experiment, the  other channels usually do not yield as good fit
 as the ones listed above  \cite{Jin:2017iwg}. 
The branching ratio of direct $e^+e^-$ channel to W$^+$W$^-$ channel (double $\tau^+\tau^-$ channel), $f$, 
is treated a free parameter to explore the favored final state.

According to their origins, we divide the CR electrons and positrons into three components : 1) an astrophysical background, 
this is the primary electrons from distant astrophysical sources and the secondary electrons and positrons generated when 
these primary electrons propagate in the space between sources and observers; 2) a diffuse dark matter component, including 
the electrons and positrons produced by the annihilation or decay of the diffuse dark matter particles in the Galactic halo, and the secondary electrons and positrons generated during their propagation;
3) a component from a nearby subhalo, which may accidentally to be located at a very close distance so that it 
significantly alter the CR energy spectrum.

\subsection{The broad spectrum}
\label{subsec:broadspectrum}

Precise measurements of CR electrons and positrons were reported by the AMS-02 experiment \cite{Aguilar:2014mma}. They 
found that in the measured flux of electrons plus positrons, the positron fraction reaches the maximum of 15.9\% at 305 GeV 
\cite{Accardo:2014lma}. There is a positron excess against the astrophysical background, which is usually interpreted as 
produced by either pulsars or dark matter annihilation or decay 
\cite{Jin:2014ica, Jin:2013nta,Liu:2013vha,Chen:2015uha,Zhou:2016eul}.
 
If equal amount of electrons and positrons at the same energy are produced by the additional sources such as pulsars, 
dark matter annihilation or decay, then by subtracting the same amount of electrons as positrons from the total spectrum, 
one can obtain the contribution of  the astrophysical primary electrons. In Ref. \cite{Chen:2014nea}, the primary electron 
background is extracted from the AMS-02 experiment, and it indeed follows a power law spectrum as 
expected for astrophysical origin. This knowledge of primary electrons in turn helps us to improve the 
understanding of the additional sources

 In a recent work \cite{Jin:2017iwg}, it is found that the CR electrons and positrons background 
 as measured by AMS-02 agrees well with the prediction by a convection and  re-acceleration diffusion (DCR) model 
 using the CR protons, positrons and B/C data of the same experiment. 
 Based on the two-peaks feature of the CR electrons 
 and positrons background spectrum, they further pointed out that the positron excess in AMS-02 experiment could be 
 interpreted by diffuse dark matter annihilation in Galactic halo, and the most relevant channels are the W$^+$W$^-$( in the bosons and quarks)  
 and double $\tau^+ \tau^-$ (in the leptons)  annihilation channels.

The dark matter profile is chosen as Einasto \cite{Einasto:2009zd}, 
which is described approximately by a power-law density distribution:
\begin{align}
\rho(r)=\rho_\odot \exp
\left[
-\left( \frac{2}{\alpha_E}\right)
\left(\frac{r^{\alpha_E}-r_\odot^{\alpha_E}}{r_s^{\alpha_E}} \right)
\right] ,
\end{align}
with $\alpha_E\approx 0.17$ and $r_s\approx 20$ kpc.  The local dark matter density is fixed at 
$\rho_\odot=0.43 \text{ GeV}\text{ cm}^{-3}$\cite{Salucci:2010qr}. In this model there is no nearby sources such as 
dark matter subhalos, and the CR electrons and positrons coming from the dark matter annihilation is a diffuse distribution
in the Milky Way. The energy spectrum of electrons and positrons of each channel is calculated by PYTHIA v8.175 
\cite{Sjostrand:2007gs}. Then based on the above density profile, the final energy spectrum after propagation is obtained via 
the  GALPROP \cite{Strong:1998fr}.

In this paper, we use the spectrum template generate as \cite{Jin:2017iwg} for the W$^+$W$^-$, double $\tau^+\tau^-$ and direct $e^+e^-$ channels respectively, as the theoretical models for investigating the broad spectrum excess. We leave the normalizations and branching ratio as free parameters.

\subsection{Nearby source}

Even for the dark matter origin scenario, typically the observed CR electron and positron spectrum 
would still be a broad bump rather than a narrow peak.
Charged particles move in curved trajectories in the stochastic interstellar magnetic field, 
so their transportation in the Galactic-scale  distance can be described as a diffusion process. The TeV electrons and positrons also lose 
their energy rapidly by synchrotron radiation and inverse Compton scattering, with a life time of $10^5 \sim 10^6 $ years. 
As a result, even 
if the electrons and positrons are produced at a single energy by dark matter annihilation or decay, the peak would still be 
broadened quickly as they propagate through the space. For the electron and positron spectrum to retain a narrow peak, the 
source must be fairly close by,  so that the peak is not greatly damped by energy loss or momentum space diffusion during the 
propagation. 

The N-body simulations show that in the Galactic dark matter halo there could be a large number of subhalos which were 
produced during the hierarchical structure formation process \cite{1999ApJ...524L..19M,1999ApJ...522...82K}. Many of 
such halos did not form stars so they can not be observed directly (see Ref. \cite{2017ARA&A..55..343B} for a recent review). 
In what follows we compute the 
the contribution to the electron and positron spectrum from a nearby subhalo, to see
if the peak in the energy spectrum could be reproduced. 

Although the CR spectrum from a nearby subhalo could also be simulated with the  {\tt GALPROP} code, in this paper we consider 
to use an analytical solution of the CR diffusion equation from a nearby source. While this is not as accurate and comprehensive as the {\tt GALPROP} code, it allows more intuitive understanding of the physical picture.

We first assume the source produce electrons and positrons with a single energy $E_0$, from 
a point source. The propagation equation of primary CR electrons (ignoring secondary ones produced in particle collisions)  can be written as 
\begin{eqnarray}
\frac{\partial \psi}{\partial t} &=& \nabla \cdot (D_{xx} \nabla \psi -V_c \psi) 
+\frac{\partial}{\partial p}[p^2 D_{pp} \frac{\partial}{\partial p} \frac{\psi}{p^2}] \nonumber\\
&&-\frac{\partial}{\partial p} [\dot{p} \psi -\frac{p}{3}(\nabla \cdot V_c)\psi]
+ \mathbf{S}
\end{eqnarray}
where $\psi(\mathbf{r}, p, t)$ is the number density of electrons per unit energy, $D_{xx}$ and $D_{pp}$ are the 
spatial and momentum diffusion coefficients respectively, $V_c$ the bulk velocity, and $\mathbf{S}$ the source 
distribution. For the present problem, $p \approx E$. 
This is a linear equation, so we may solve the subhalo contribution separately, and just adding 
it to the solution for the whole galaxy.

The diffusion coefficient and other parameters in the cosmic ray propagation model were fitted from other observations
\cite{Trotta2011}, $D_{xx}= D_0 (\rho/\rho_0)^{\delta}$,
where $D_0=6.59\times10^{28}$ cm$^2$ s$^{-1}$, $\rho=pc/Ze$ and $\rho_0=4\times10^3$ MV, 
here we take $\delta=1/3$ which is for Kolmogorov spectrum of interstellar turbulence.
In the TeV energy range, the electrons and positrons lose their energy primarily by synchrotron radiation 
and inverse Compton scattering (ICS).
The energy loss rate for relativistic electrons via synchrotron radiation is
\begin{equation}
-\left(\frac{dE}{dt}\right)_{\rm syn}=6.9\times10^{-25}\gamma^2\left(\frac{B}{\mu\rm G}\right)^2~{\rm [GeV~s^{-1}]},
\end{equation}
where $\gamma=E/m_ec^2$, and $B$ is the magnetic field in units $\mu\rm G$ and we adopt  the typical magnetic field $\sim6 \mu$G in the Solar neighborhood \cite{Haverkorn2015}.

Regarding the ICS, the Klein-Nishina cross-section for the electron-photon scattering is \cite{Lefa2012}
\begin{equation}
\sigma_{\rm KN}(\mathcal{E},\mathcal{E}_0,\gamma)=\frac{3\sigma_{\rm T}}{4\gamma^2\mathcal{E}_0}G(q,\Gamma_e),
\end{equation}
where $\mathcal{E}_0$  and $\mathcal{E}$ are the energy of photons before and after scattering. $\sigma_{\rm T}=6.65\times10^{-25}$ cm$^2$ is the Thompson scattering cross-section.
The function
\begin{equation}
G(q,\Gamma_e)=2q{\rm ln}q+(1+2q)(1-q)+\frac{(\Gamma_e q)^2(1-q)}{2(1+\Gamma_e q)},
\end{equation}
where
\begin{equation}
\Gamma_e=\frac{4\mathcal{E}_0\gamma}{m_e c^2},~~~\qquad~~~q=\frac{\mathcal{E}}{( \gamma m_e c^2-\mathcal{E} )\Gamma_e}.
\end{equation}

The spectrum of the up scattered photons is
\begin{equation}
\frac{dN}{dtd\mathcal{E}}=\int \sigma_{\rm KN} c \frac{dU_{\rm rad}}{\mathcal{E}_0d\mathcal{E}_0}d\mathcal{E}_0,
\end{equation}
and the energy loss rate via ICS
\begin{equation}
-\left(\frac{dE}{dt}\right)_{\rm ICS} \approx \int_{\mathcal{E}_0}^{\mathcal{E}_{\rm max}} \mathcal{E} \frac{dN}{dtd\mathcal{E}} d\mathcal{E},
\end{equation}
the integration is performed to the maximum energy of the up scattered photons,
\begin{equation}
\mathcal{E}_{\rm max}=\frac{4 \mathcal{E}_0   \gamma^2}{1+4\mathcal{E}_0\gamma/(m_ec^2)  }.
\end{equation}

From \cite{Porter2005,Porter2006}, near our Solar system, the interstellar radiation filed has three peaks at the NIR ($\sim 1\mu$m), FIR ($\sim100\mu$m) and CMB($\sim1000\mu$m), with energy density $\sim0.4$ eV cm$^{-3}$, $\sim0.3$ eV cm$^{-3}$ and $\sim0.2$ eV cm$^{-3}$ respectively. Adopting these values we have
\begin{align}
-\left(\frac{dE}{dt}\right)_{\rm ICS}\approx   \sum _i \int_{\mathcal{E}_{0,i}}^{\mathcal{E}_{{\rm max},i}}  d\mathcal{E} \mathcal{E} \frac{3\sigma_T c U_{{\rm rad},i}}{4\gamma^2 \mathcal{E}^2_{0,i}}G(q_i,\Gamma_{e,i})
\label{eq:ICS}
\end{align} 
where $i$ is one of NIR, FIR or CMB respectively.

The final energy loss rate is
\begin{equation}
b= -\frac{dE}{dt}=-\left(\frac{dE}{dt}\right)_{\rm syn}-\left(\frac{dE}{dt}\right)_{\rm ICS}.
\label{eq:b}
\end{equation}
For example, at 1.5 TeV, $b\approx   3.1\times10^{-10} $ GeV s$^{-1}$.
Also, if the source is nearby, the diffusion in momentum space which is due re-acceleration can be neglected, 
as $\frac{D_{pp}}{E}\sim6.2\times10^{-15}~{\rm GeV~s^{-1}} \ll b.$

The dynamical time scale of the Milky Way is of order of $10^8$ year,  so the time scale for the subhalo to pass by
is much longer than the life time of the electrons and positrons. We may then assume a steady state for the CR spectrum is reached, $\partial \psi /\partial t=0$. 
We ignore the bulk velocity $V_c$.
For a point source of monochromatic electron or positron injection, 
$\mathbf{S}= Q \delta(\mathbf{r}) \delta (E-E_0)$,
then the propagation equation could be simplified as
\begin{equation}
D_{xx}\nabla^2\psi+b\frac{\partial}{\partial_E}\psi+Q\delta(r)\delta(E-E_0)=0.
\end{equation} 
This equation can be solved by making a 3D Fourier transform over the spatial dimensions and a 1D Laplace
transform over time. If the source is very nearby, we do not need to consider the global confinement to the Galactic 
disk by the magnetic field. The solution in the spherically symmetric case is given by
\begin{equation}
\psi(E_0, E, r)=\frac{Q}{b}\frac{\exp[-\frac{r^2}{(4\pi D_{xx}/b)(E_0-E) } ] }{[(4\pi D_{xx}/b)(E_0-E)]^{3/2}}. 
\label{psi}
\end{equation}
Then the observed specific intensity of particles number flux is $J\approx\psi c/(4\pi)$.

If the injected electrons and positrons are not of a single energy but has an energy distribution of $dN/dE$ (e.g. for 
W$^+$W$^-$ channel or double $\tau^+\tau^-$ channel), the 
corresponding spectrum could be obtained by 
\begin{equation}
J = \frac{c f}{4\pi(1+f)} \int dE^\prime  \frac{dN}{dE^\prime} G(E^\prime, E, r)
\end{equation}
where the Green function $G(E^\prime, E, r)$ is given by Eq. (\ref{psi}). If the annihilation or decay has both 
direct $e^+e^-$ channel and  other channels, the result is given by the sum of all the channels.

\begin{figure}[htbp]
\centering{
\includegraphics[scale=0.35]{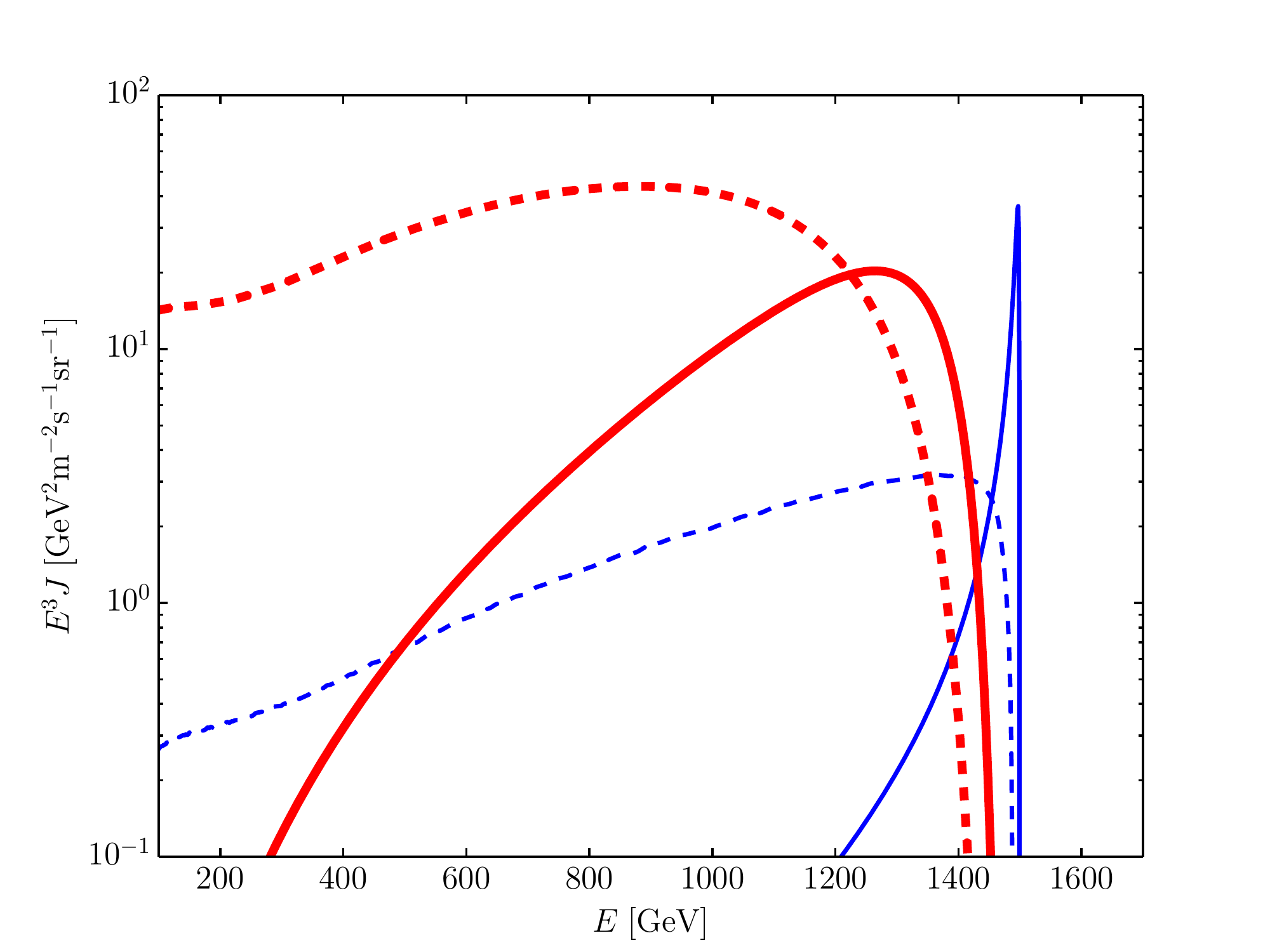}
\includegraphics[scale=0.35]{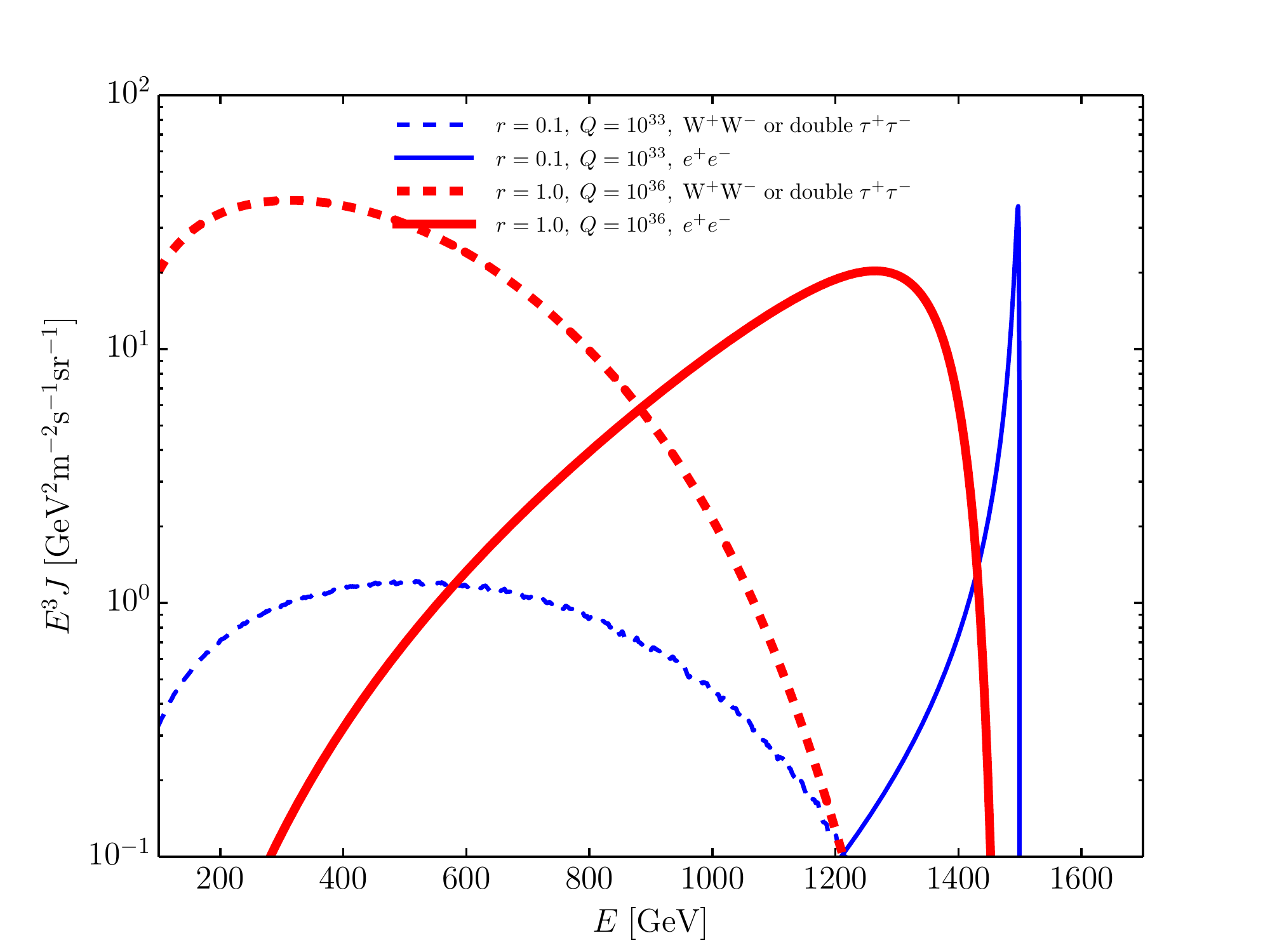}
\caption{The  specific intensity of the electrons and positrons number flux from a nearby subhalo with different distances.  Top panel is for model A while bottom panel is for model B.}
\label{fig:J_e_Q}
}
\end{figure}

We show the electron and positron spectrum produced via the different channels in Fig. \ref{fig:J_e_Q}, 
for model A (top panel) and model B (bottom panel). 
Here we  fix $E_0=1.5$ TeV and $f=0.01$. In each case we show the results for $r=0.1$ and 1 kpc.
The total injection rate $Q$ are adjusted ($Q=10^{33}$ and $10^{36}$ s$^{-1}$ for $r=0.1$ and 1 kpc
respectively) so that the curves can be show on the same plot.
As we expect, the closer the source, the  
narrower the peak.  For $r=0.1~\kpc$ (blue) the peak is very sharp, while for $r=1~\kpc$ (red) the peak is a much 
broader one.  Compared with the direct production case, the electrons and positrons produced via the 
W$^+$W$^-$ or double $\tau^+\tau^-$ channel have broader distributions, as the electrons and positrons produced in this channel are not monochromatic. 
Moreover, the peak is located at higher energy for the $r=0.1~\kpc$ (blue) case than the $r=1~\kpc$ (red) case.
If we want to generate a peak in the spectrum, the direct $e^+ e^-$ channel must be 
used.

We fit the normalization $A$ of the broad spectrum of electrons and positrons due to W$^+$W$^-$ or double $\tau^+\tau^-$ channel and 
the branching ratio $f$ of the direct $e^+e^-$ channel to the W$^+$W$^-$ or double $\tau^+\tau^-$ channel, the distance $r$, and the amount of 
injection $Q$ from the statistics:
\begin{equation}
\chi^2(A,f,r,Q)=\sum \frac{ [\mean{J_i^{\rm nr}(f,r,Q)}+\mean{J_i^{\rm c}(A,f)}-J^i_{\rm obs}]^2 }{\sigma^2_{{\rm stat},i}+\sigma^2_{{\rm sys},i}},
\label{eq:chi2}
\end{equation}
where $\mean{J_i^{\rm nr}(f,r,Q)}$ is the average (i.e. integrating the intensity in the energy bin then divided by the bin width)
mean intensity from the nearby source in the $i$-th energy bin; while
\begin{equation}
 J_i^{\rm c}(A,f)=AJ^{\rm br}_{W/\tau}+fAJ^{\rm br}_e+J_{\rm bg}.
\end{equation}
The astrophysical background  $J_{\rm bg}$,  the broad spectrum template from the annihilation or decay of Galactic diffuse dark matter through the  W$^+$W$^-$ or double $\tau^+\tau^-$ channel, $J^{\rm br}_{W/\tau}$, and through the direct $e^+e^-$ channel, $J^{\rm br}_e$, are all generated using the algorithm presented in \cite{Jin:2017iwg}.
For simplicity we fix $E_0=1.5~\TeV$. Adopting a different value within $\sim 0.1~\TeV$ does not change 
our result. 

\section{Fitting with the Broad Spectrum}\label{broad}

The CR electron and positron spectrum has been measured by a number of  experiments, such as 
the VERITAS \cite{Staszak:2015kza}, H.E.S.S. \cite{Aharonian:2008aa,Aharonian:2009ah} and Fermi-LAT \cite{Abdollahi:2017nat}
before the DAMPE data release. These observations indicate a remarkable break in the spectrum at TeV scale. 
However, the electron and positron data from the ground based atmospheric Cherenkov telescopes such as H.E.S.S. have large uncertainties
from the subtraction of hadronic background and discrimination against gamma rays events \cite{Aharonian:2008aa}. In detail, 
the very-high-energy flux of H.E.S.S. electrons is described by an exponentially cutoff power law with an index of 3.05$\pm$ 0.02 
and a cutoff at 2.1$\pm$ 0.3 TeV in the range of 700 GeV to 5 TeV \cite{Aharonian:2008aa}. 
The low-energy extension of the H.E.S.S. electron measurement  are from 340 GeV to 1.7 TeV with a break energy at about 1 
TeV \cite{Aharonian:2009ah}.

\begin{figure*}[htbp]
\centering{
\includegraphics[scale=0.3]{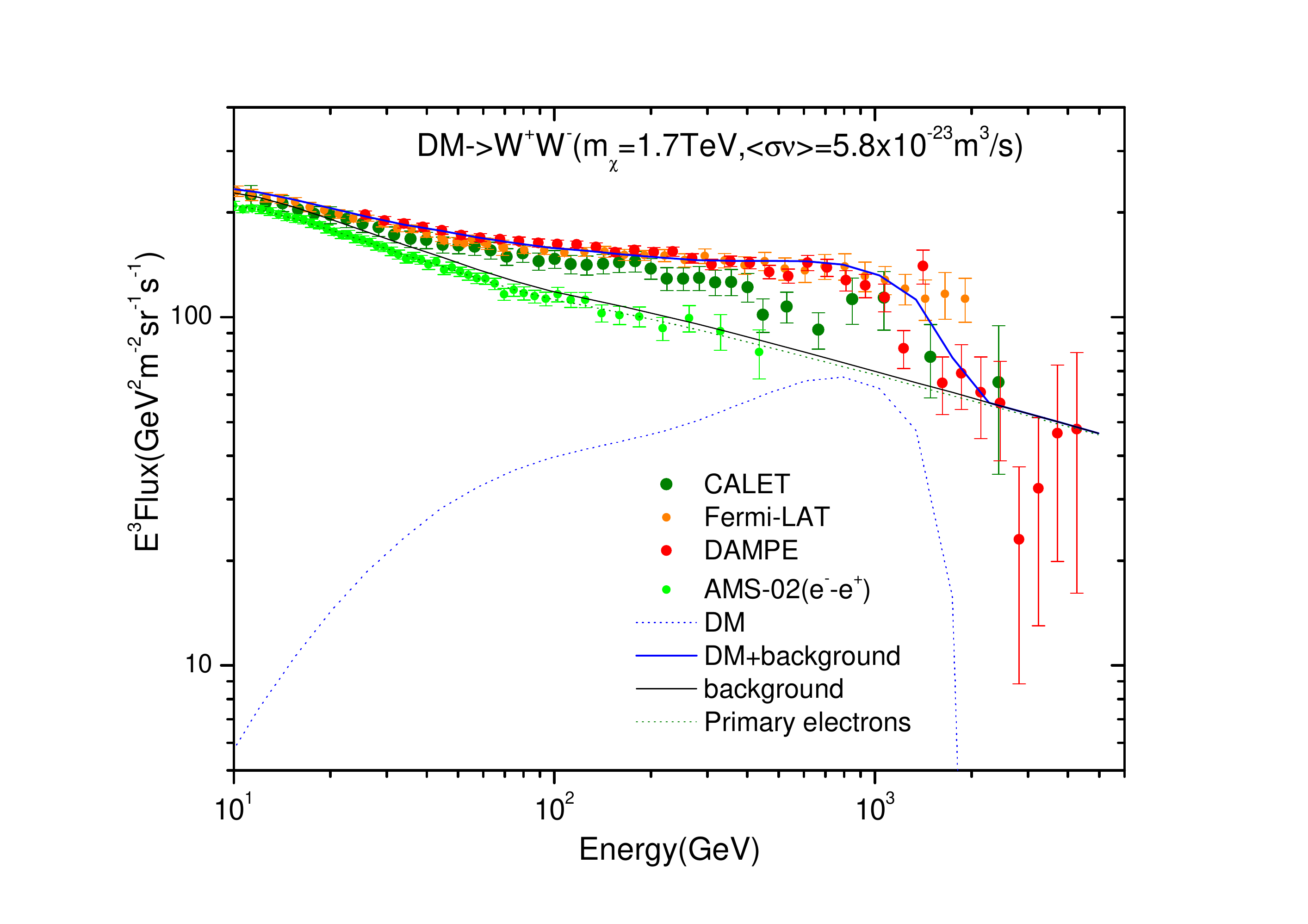}
\includegraphics[scale=0.3]{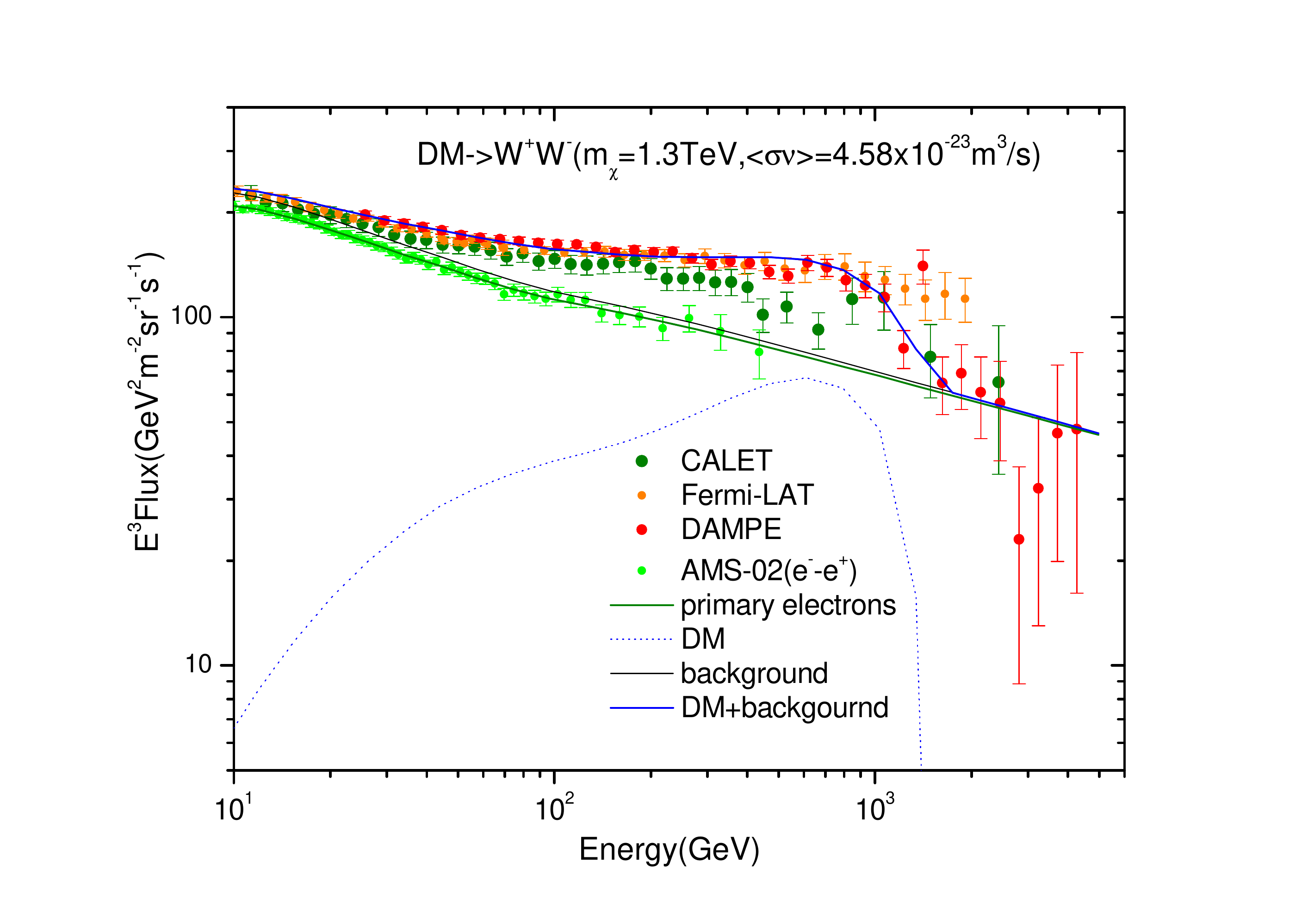}
\includegraphics[scale=0.3]{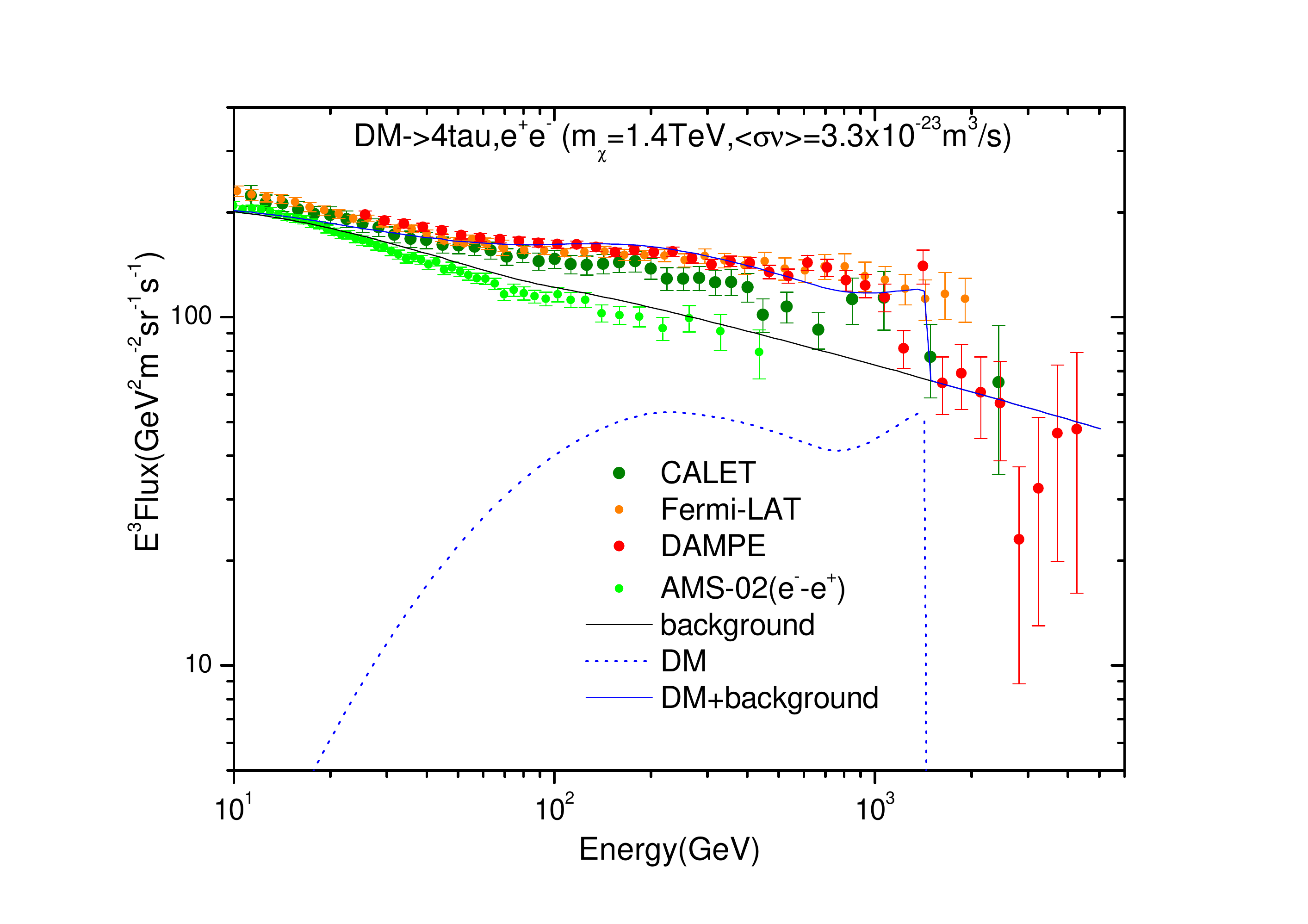}
\includegraphics[scale=0.3]{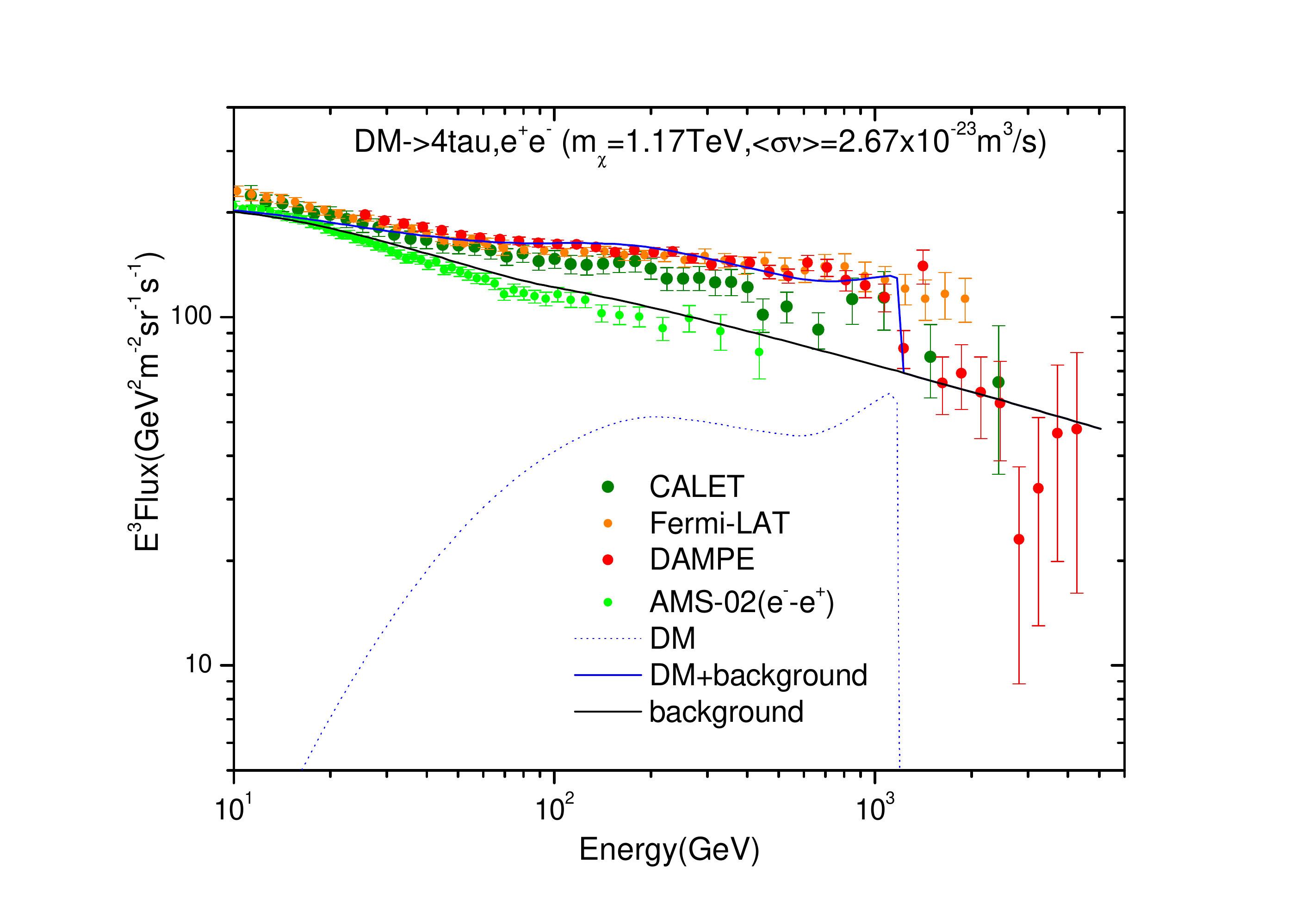}
 \caption{The primary electrons background derived from $(e^- - e^+)$ of the AMS-02 data \cite{Accardo:2014lma} and the total flux of CR electrons and positrons background measured by DAMPE. The curves that fit the data including and excluding the peak point near 1.4 TeV are shown in the left and right panel respectively. We write the best-fit dark matter particle mass $m_\chi$ and the velocity weighted cross-section $\mean{\sigma v}$ in each panel. For top panels, the $\chi^2/{\rm d.o.f}=1.24$ (left) and 1.48 (right) respectively, 
while for the bottom panels  $\chi^2/{\rm d.o.f}=2.33$ and 2.10 respectively.
}
\label{fig:EplusAnni}
}
\end{figure*}

The TeV break of CR electrons and positrons are now confirmed by the highly precise DAMPE observation.
If we assume the primary electron is described by a power-law, while the excess electrons and positrons have the same
spectral distribution, the degeneracy between the background and excess can be broken, and 
the origin of the TeV break is connected to the positron excess \cite{Ackermann:2010ij}. This is 
confirmed in \cite{Chen:2014nea} where the features of the primary electron spectrum is carefully studied, which 
showed a power-law primary electron spectrum without TeV break.

Here, we derive  the excess of electrons under the assumption that the number of excess electrons and positrons are 
the same. We use  the  AMS-02 experiment  $e^--e^+$ data \cite{Aguilar:2014mma} to derive the primary electrons and the
 astrophysical background, and then subtract this background from the DAMPE data to obtain the excess\footnote{
A caveat: there are discrepancies between the CR electrons measured in the various experiments,  
including AMS-02, Fermi-LAT, DAMPE and CALET, and the origin of such discrepancies is not clear, so there is
no fully self-consistent way to use these measurements jointly. In our treatment below, counting on 
the fact that the AMS-02 has the capability of distinguishing electrons and positrons, and the best precision in the CR electron 
flux measurement, we use its data to derive the astrophysical background, but still there is the risk that the excess may be overestimated.
An alternative treatment is discussed  in Sec. \ref{sec:rescaled_bg}.}. 
The primary electrons and the background are shown as green dotted and black solid lines in each panel of Fig. \ref{fig:EplusAnni}. 
At $\gtrsim 100$ GeV they are of similar power-law forms, implying that the amount of secondary electrons and positrons 
generated during the CR propagation is small. Our model background is consistent with Fermi-LAT, CALET and DAMPE 
observational constraints at least at $\lsim 5$ TeV.  At higher energies, the power-law background may break and 
decrease faster due to more efficiency energy loss in both injecting sources and during propagation process in Milky Way. 
At present the higher energy observations still have very large uncertainties, so we only use the background 
value at $\lsim 5$ TeV  to interpret the DAMPE data. 
    
We then compute the excess produced by diffuse dark matter particle annihilations in the Galactic halo. We consider the 
annihilation to electrons and positrons via the W$^+$W$^-$ channel and direct $e^+e^-$ channel (model A) or via the double $\tau^+\tau^-$ channel and   direct $e^+e^-$ channel (model B).

In Fig. \ref{fig:EplusAnni}, we show dark matter annihilation model fit to the DAMPE measured spectrum data. We
plot the primary electrons derived from the AMS-02 $e^- - e^+$ data, the derived background, the dark matter contribution, 
and the dark matter plus background total. The top two panels show the model A results, 
while the bottom two show the model B results. In the left panels of Fig.\ref{fig:EplusAnni}, we try to fit all data points including the peak 
point in the DAMPE data, while in the right panels we exclude the peak point. For model A, by excluding the peak point, the fit to the other 
points is improved. For model B  there is steep drop of the spectrum at $E_0=m_\chi$, where $m_\chi$ is the dark matter particle mass. 
This cliff-like edge fits the peak point better, however the  fitting is worse for the energy just below $E_0$.

We see all these models provide good fit to the data, at least at $\gtrsim100$ GeV.
For model A the $\chi^2_{\rm min}/{\rm d.o.f}=1.24$ and 1.48 when we include or  exclude the 1.4 TeV peak data respectively.
For model B, the corresponding $\chi^2_{\rm min}/{\rm d.o.f}=2.33$ and 2.10.
The model A fit is better than model B,
particularly because at the lower energy part, its spectral shape agrees with the data very 
well.

The best-fit dark matter particle mass are 1.7 TeV and 1.3 TeV respectively for the left and right panel, for model A,
and 1.4 TeV and 1.2 TeV respectively for model B. The required annihilation cross-sections are 
typically $\mean{\sigma v} \sim 10^{-23}$ cm$^3$ s$^{-1}$, within the constraints derived from the AMS-02 
electron data \cite{Jin:2013nta}. The cross-section values are however larger than the one required 
to get correct abundance of WIMPs during the thermal decoupling. To obtain large abundance, the dark matter
must be produced non-thermally in the early Universe, or the cross-section of its annihilation in the current 
time must be somehow enhanced \cite{2009PhRvL.103c1103B,2010PhRvD..82h3525F}. 
Moreover, while the cross-section is derived for electrons and positrons production, we note that these are larger than the 
observational limits on cross-section of annihilation into photons derived from observations of the dwarf spheroidal  
galaxies \cite{Archambault:2017wyh,Ackermann:2015zua}. The required cross-sections here are 
one order of magnitude greater than those of  VERITAS \cite{Archambault:2017wyh}, 
and three order larger than the one from Fermi-LAT \cite{Ackermann:2015zua}. Although the cross-sections of $\gamma$-ray and 
electrons and positrons could be different in principle, we expect them to be of the same order, and 
special mechanism may be needed to achieve the large cross-section to electrons and positrons
while not violating the $\gamma$-ray bound.
  
\begin{figure}[htbp]
\centering{
\includegraphics[scale=0.3]{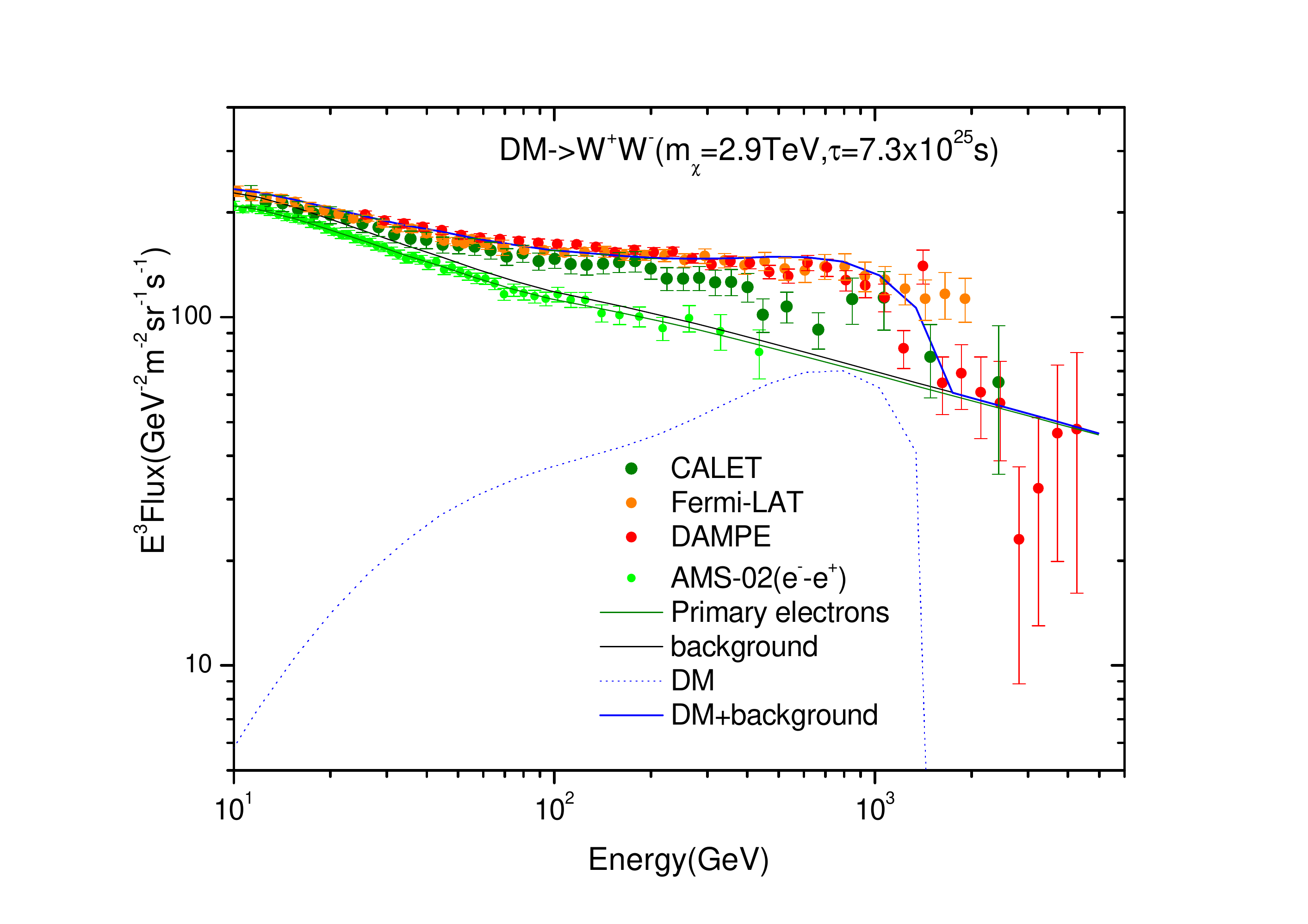}
\caption{Spectrum for dark matter decay via the W$^+$W$^-$ case.}
\label{fig:decay}
}
\end{figure}

In the above we have discussed the case of annihilations. In the case of decay, except for the mass which 
is doubled and the rate is determined  from decay life time instead of the annihilation cross-section, 
others are similar. In Fig. \ref{fig:decay} we show the spectrum of the decay in model A. The spectrum is very close to that of the annihilations, though dark matter particle mass is doubled with $m_\chi= 2.9~\TeV$. The life time of the decaying particle is $\tau = 7.3 \times 10^{25}$ s.

In principle, the dark matter annihilation and decay results may differ, 
because the annihilation rate $ \propto n^2$, while the decay rate $\propto n$, where $n$ is the number density of 
dark matter particles, so the source distribution is in principle different. Nevertheless, in the end we find that
the results are not much different in this case.  As the spectrum for the two cases are quite similar, we shall mainly discuss the annihilation, 
though most of the results are also applicable to decays.

\section{The peak and nearby source}\label{peak}

In the last section we see that the Galactic halo diffuse dark matter annihilation could provide a reasonably good fit to the 
DAMPE $e^+e^-$ spectrum, though it does require a very large annihilation cross-section. However, there is no 
peak produced in the spectrum. Here,  we tentatively treat the peak as real, and 
consider whether such a peak could be produced in dark matter annihilation or decay from a nearby subhalo. 
To do this, we use the {\tt GALPROP} code to compute the primary background and diffuse dark matter contribution 
from the Galactic halo, then adding the contribution from the nearby subhalo. The dark matter parameters and 
subhalo parameters are varied to fit the DAMPE spectrum. 

\subsection{Global Fit Models}
First we consider the two models introduced in Sec. \ref{methods}: model A, in which the  $e^+ e^-$ are produced by the direct channel 
and the W$^+$W$^-$ channel; and model B,  in which the $e^+ e^-$ production are from the direct channel
and the double $\tau^+ \tau^-$ channels. The branching ratio of the direct channel to other channels are taken as free
parameters.

\begin{figure}[htbp]
\centering{
\includegraphics[scale=0.45]{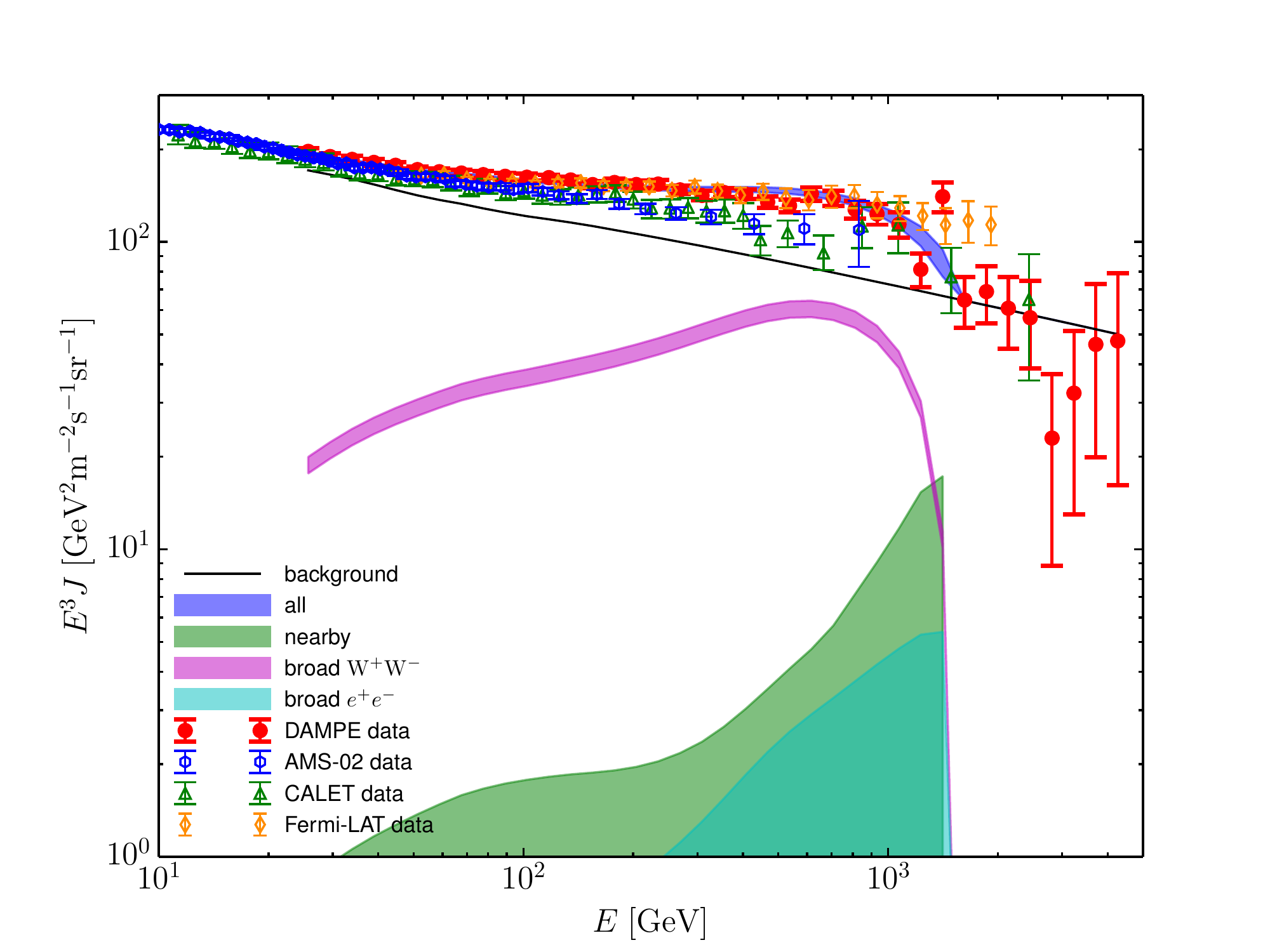}
\caption{The  fit to DAMPE spectrum with model A, with various contributions shown and marked. 
The envelope corresponds to $1\sigma$ in parameter space. As comparisons we also plot the Fermi-LAT, CALET and AMS-02 data.}
\label{fig:J_fit_WW}
}
\end{figure}

\begin{figure}[htbp]
\centering{
\includegraphics[scale=0.45]{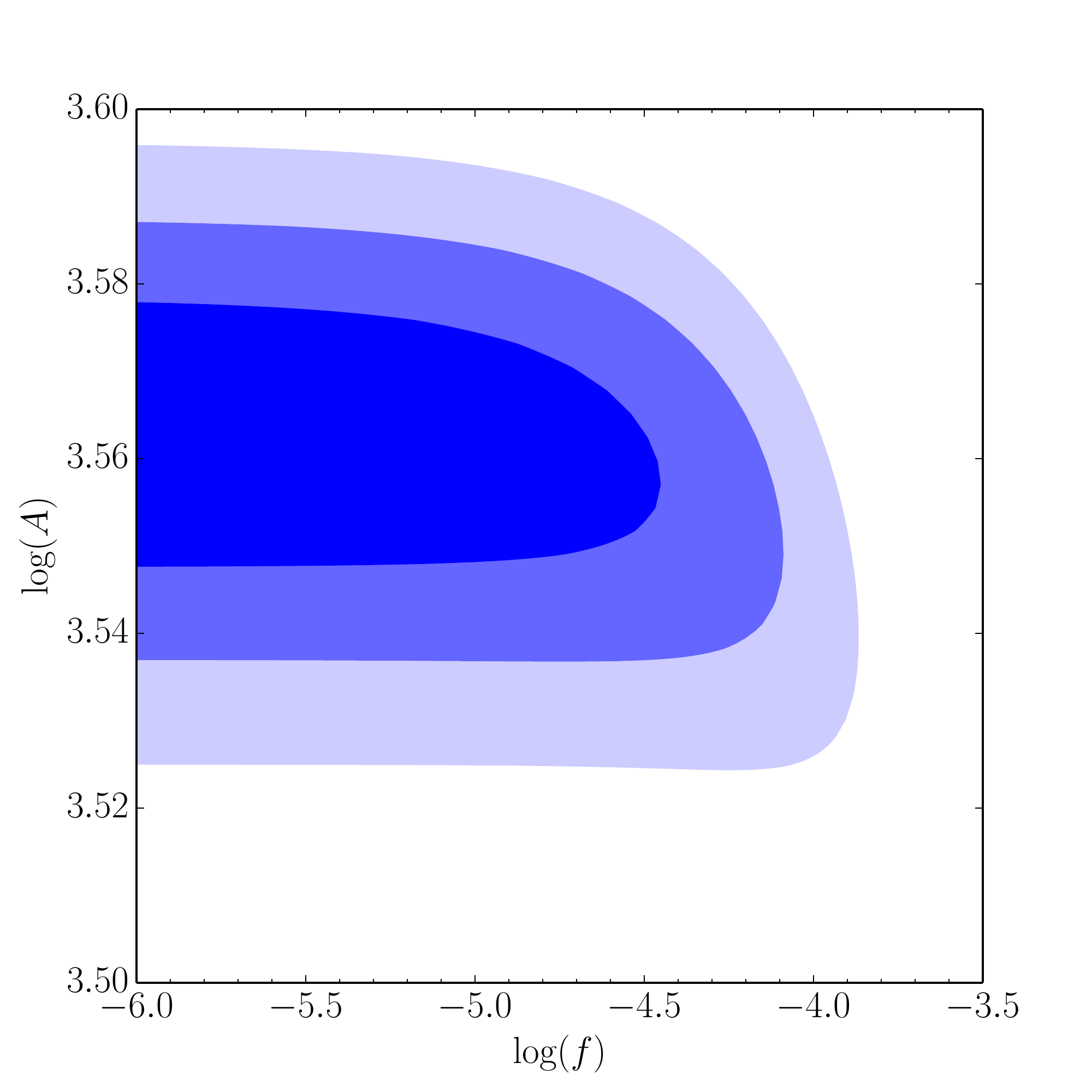}
\caption{The constraints on $f$ and $A$ for model A.}
\label{fig:chi2_2D_b_log_frac2_log_norm1_WW}
}
\end{figure}

{\bf Model A.}   In Fig.\ref{fig:J_fit_WW}, we plot the best
fit to the DAMPE spectrum. The CR background, the various contributions from dark matter annihilations, and the 
total of these are all plotted in the figure.  As we can see from the figure, there is no peak in the total spectrum
curve, which behaves as a gentle break, despite the fact that a subhalo is introduced. Looking into the details, we see
the W$^+$W$^-$ channel broad spectrum (i.e. the one from Galactic halo diffuse dark matter annihilations) gives an excellent fit to 
the broad excess, which are the majority of the data and also have much smaller measurement error bars than the few data 
points near the 1.4 TeV peak.  As a result, the W$^+$W$^-$ channel is the dominant contribution to the DAMPE spectrum. 
The broad $e^+ e^-$ spectrum (i.e. the direct channel contribution from the Galactic halo diffuse dark matter), on the other hand, does not
have the correct shape to fit the broad excess, so its contribution must be suppressed. We plot the ${\rm log}f - \log A$ distribution in Fig. \ref{fig:chi2_2D_b_log_frac2_log_norm1_WW}, with other parameters marginalized.
We find that the branching ratio is limit to be $f \lsim1.1\times10^{-4}$ at $3\sigma$, meaning that the contribution 
from direct $e^+e^-$ channel is quite limited compared with the W$^+$W$^-$ channel. Thus, the whole spectrum shape is 
determined by the W$^+$W$^-$ channel. The branching ratio of the 
direct channel to the  W$^+$W$^-$ channel $f$ is however a single parameter, so the direct $e^+ e^-$ channel from the 
subhalo which has the desirable spectral shape of a peak is also suppressed. The final spectrum fits most of the data, 
but does not have a peak, and is not significantly better than the broad spectrum fit given in Sec. \ref{methods}. 

Is it possible for the direct $e^+ e^-$ channel to dominate over the W$^+$W$^-$ channel? Unfortunately, for 
realistic parameters for CR propagation and dark matter halo, the Galactic halo diffuse dark matter annihilation via 
direct $e^+ e^-$  channel produce fairly hard spectrum, even after the diffusion process, as shown by the cyan colored
curve in Fig. \ref{fig:J_fit_WW}. The curve is broader than that of the nearby subhalo contribution, but still too narrow to 
produce the broad excess.

\begin{figure}[htbp]
\centering{
\includegraphics[scale=0.45]{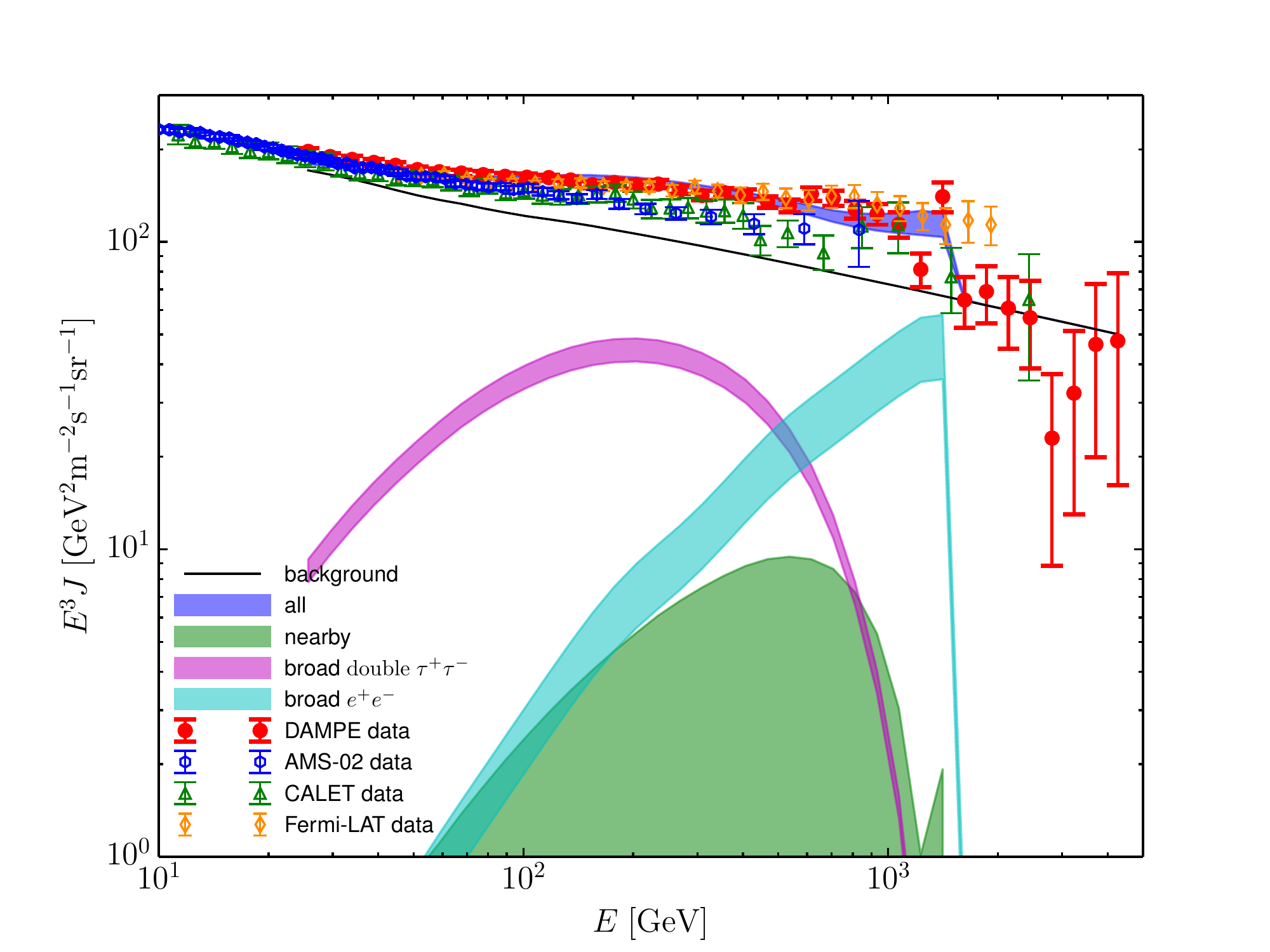}
\caption{Same as Fig. \ref{fig:J_fit_WW}, but for model B.}
\label{fig:J_fit_4tau}
}
\end{figure}

\begin{figure}[htbp]
\centering{
\includegraphics[scale=0.45]{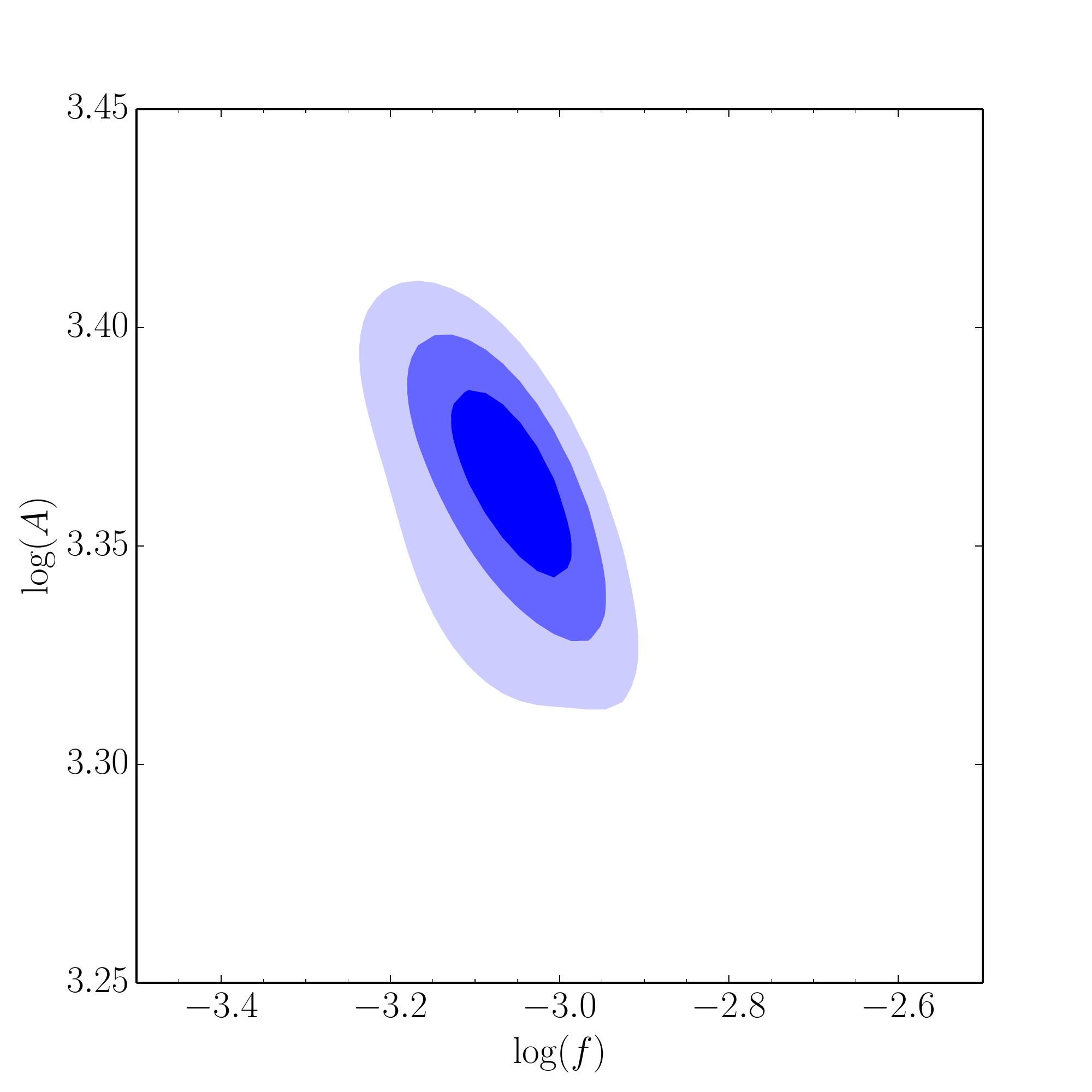}
\caption{Same as Fig. \ref{fig:chi2_2D_b_log_frac2_log_norm1_WW}, but for model B.}
\label{fig:chi2_2D_b_log_frac2_log_norm1_4tau}
}
\end{figure}

{\bf Model B.} In Figs. \ref{fig:J_fit_4tau} and \ref{fig:chi2_2D_b_log_frac2_log_norm1_4tau} we show the corresponding results for model B. 
In this case, the fitting curve passes closer to the peak point, but the spectral shape is more like a cliff than a peak, 
the flux on the left side of the peak is over-predicted. In fact, in this model the main contribution to 
the 1.4 TeV spectrum is not from the nearby subhalo, but instead 
from the direct $e^+e^-$ channel component in the broad spectrum, i.e. from the Galactic halo diffuse dark matter 
annihilation or decay.  Another problem is that the 
flux at energy around several tens GeVs  are under predicted. Though the deviation is not very obvious on this plot, the 
measurement error in that energy range is much smaller. 
In this model,  we obtain $5.6\times10^{-4} \lsim f \lsim 1.2\times10^{-3}$ at $3\sigma$.

In all these global fit models, the contribution from a possible nearby subhalo is limited to $\lsim$10 - 20\%. 
This small contribution is not or just barely apparent in the total flux, and fails to produce the peak-like feature. 

\subsection{Eye Ball Fit Models}  
In the above we have seen that even with nearby subhalo, the global fit models do not produce a 
sharp peak in the spectrum. Here we consider how such a peak might be generated if we relax the requirement on the fit. 

The different components have different spectral shapes. The background electrons have a simple power law shape
and its normalization is determined by using the positron spectrum measured in the AMS-02 experiment. The direct 
$e^+ e^-$,  W$^+$W$^-$ and double $\tau^+ \tau^-$ channels from the Galactic halo diffuse dark matter annihilation also have essentially fixed 
spectral shape, though it depends on the Galactic halo model and the CR propagation parameters. The normalization 
of these depends on the dark matter annihilation cross-section or decay rate. In fact, the shapes are slightly different
for the case of annihilation and the case of decay, as the radial profile of production in the Galactic halo are different, 
though we find that the difference is not large.  On the other hand, for the nearby components the shape depend on the 
distance of the subhalo. This change is most obvious for the direct $e^+ e^-$ channel. By adjusting the branching ratio, one could change the relative contributions of the 
direct, W$^+$W$^-$ and double $\tau^+ \tau^-$ channels, and by adjusting the subhalo distance and mass, one could change the 
relative contribution from the subhalo and the Galactic halo. 
The various different components are summed up to produce the total spectrum. 

We note that the measurement error on the spectrum is smaller at lower energy scales. When performing a global fit 
with natural weight according to Eq.~(\ref{eq:chi2}), the low energy part of the spectrum would largely determine
the fit, that is the reason why the contribution of the direct $e^+ e^-$ component is strongly suppressed
in the above, because the shape of broad (i.e. Galactic diffuse dark matter) direct $e^+ e^-$ component does not fit the lower energy part of the 
spectrum. This is also why the W$^+$W$^-$ component is strongly favored, because its shape agrees extremely well with
the observed excess. Here, to study whether a peak feature can be generated, we will give up the minimal variance fit
based on the $\chi^2$, but use an ``eye ball fit'' to produce the spectrum. Note that even though the spectrum may appear 
to the eye to fit the data well, the actual $\chi^2$ value would be much larger than the global fits we obtained above.

In performing the ``eye ball fit", we shall ignore the data above 2 TeV, where a dip appeared in the $E^3\times {\rm Flux}$ 
plot. In this part the error is too large to draw definite conclusions. 

\begin{figure}[htbp]
\centering{
\includegraphics[scale=0.45]{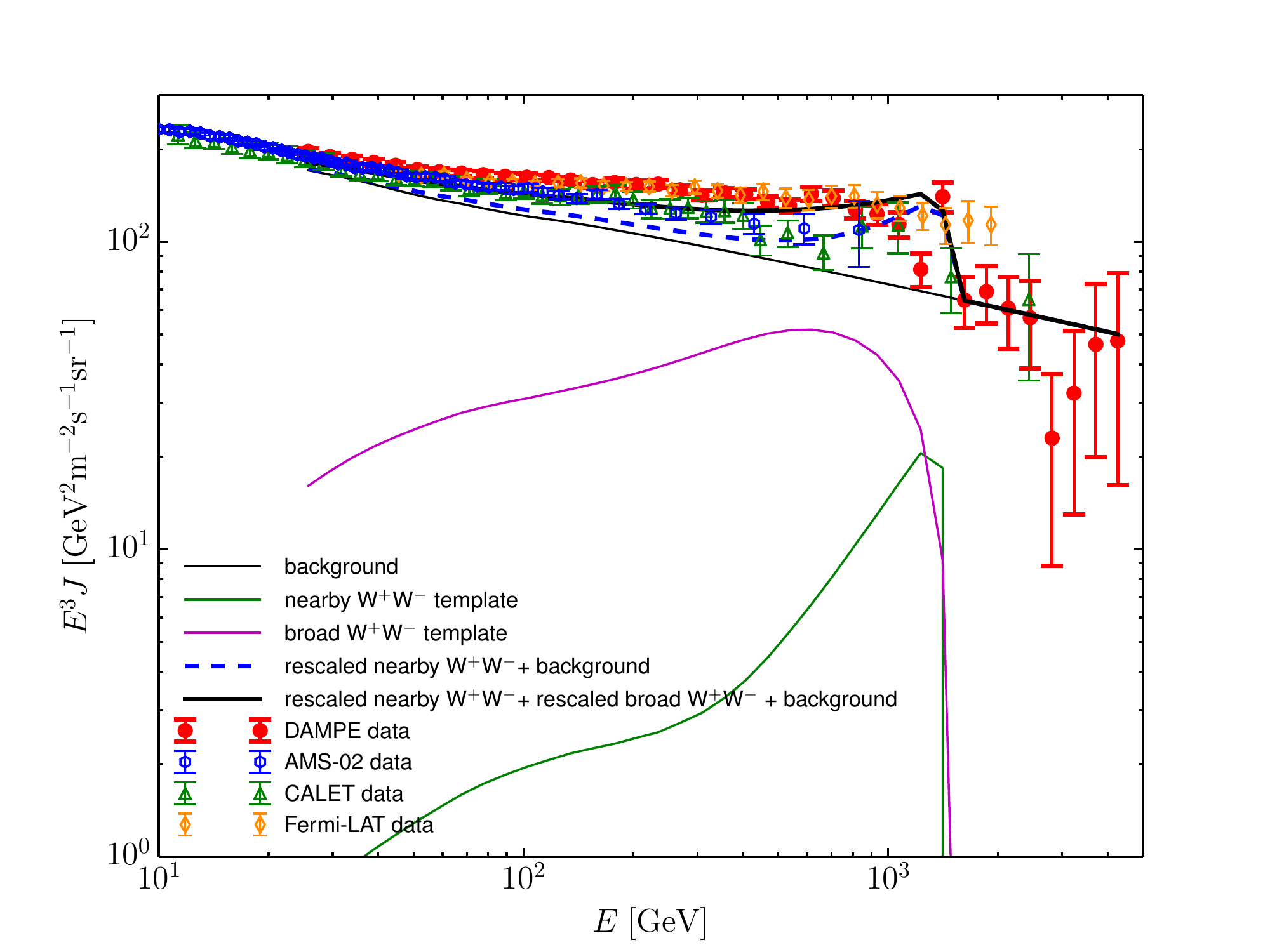}
\caption{An eye ball fit to the DAMPE spectrum with principal contribution from 
the broad and nearby W$^+$W$^-$ components.}
\label{fig:WW}
}
\end{figure}

{\bf Model A.} Since the broad (diffuse dark matter in Galactic halo)  W$^+$W$^-$ component has a shape very well suited for fitting the broad spectrum
excess, we try to produce a spectrum with the broad and nearby (subhalo) W$^+$W$^-$ components as principal contributions 
in addition to the background.  We have tried various combination of the two, and show one example of such 
``eye ball fit"  in Fig. \ref{fig:WW}. We see the broad W$^+$W$^-$ has a relatively flat shape up to 
0.7 TeV, above which it drops rapidly. By contrast, the nearby W$^+$W$^-$ source produce a one sided peak at 
$\sim 1.4$ TeV, it drops to zero above 1.4 TeV, while declines rapidly at smaller 1.4 TeV. If we adjust the relative 
contribution of the two, we could create a slight hill in the spectrum at the desired position. However, we see that
the hill has a gentle slope on the left side, though on the right side it could be very steep. Also, on the low energy 
part the fit is not good: it is obviously below the data point. To get a better fit on the low energy part, one has to 
increase the amplitude of the broad W$^+$W$^-$ component. This can be achieved by increasing the annihilate 
cross-section or decay rate, but then the nearby source component would also increase, increasing the peak 
height while keep the same shape. If one wants to maintain the height by reducing the nearby source component 
(this can be done by adjusting the subhalo mass), the relative amplitude of the two components will be changed, and the 
slope below 1.4 TeV would be even more gentle, making the peak disappear into a one sided slope, which is essentially the 
result of global best fit of model A above. 

\begin{figure}[htbp]
\centering{
\includegraphics[scale=0.45]{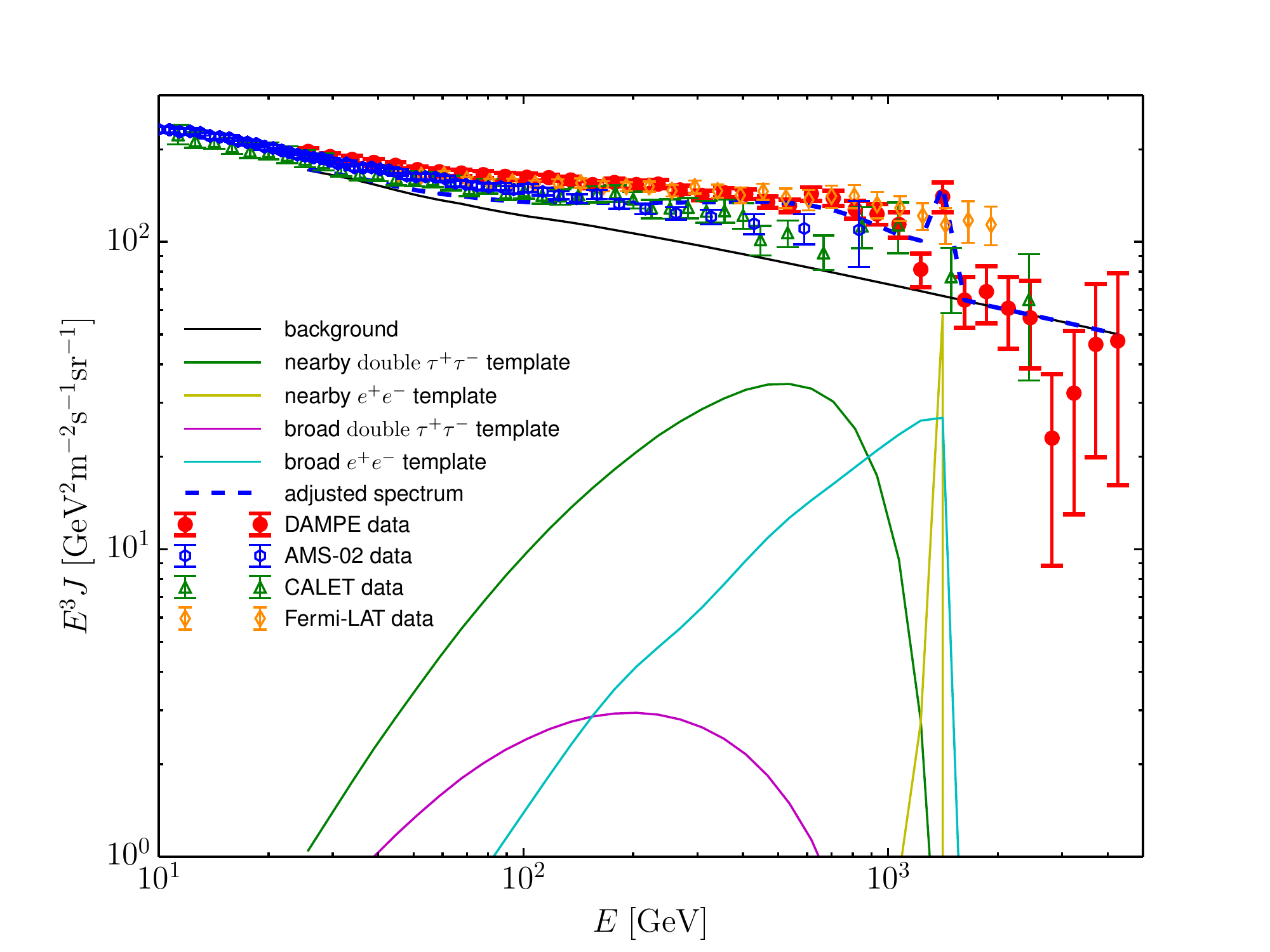}
\caption{An eye ball fit to the DAMPE spectrum with principal contribution from 
the broad and nearby double $\tau^+\tau^-$ and direct $e^+e^-$ components.}
\label{fig:ee4tau}
}
\end{figure}

{\bf Model B.}
Next we consider the fit with the  direct $e^+ e^-$ channels. The advantage of this channel is that the 
nearby source of this channel may generate very sharp peak, though its broad component has the wrong shape to 
fit the broad excess. If we relax the fit to the broad excess though, we could produce a peaked spectrum in this case. 
The double $\tau^+\tau^-$ components which have flatter shapes may also be employed to help improve the fit to the 
low energy broad excess part. An ``eye ball fit'' is shown in Fig. \ref{fig:ee4tau}. Here we see  the total spectrum 
(marked as ``adjusted spectrum") can well reproduce the peak feature at 1.4 TeV. This peak feature is dominated by
contribution from the direct $e^+ e^-$ channel of nearby source. Interestingly, the direct $e^+ e^-$ from the diffuse dark matter in Galactic halo, 
the double $\tau^+\tau^-$ from the diffuse dark matter together make up the contribution to the broad excess at a few hundred GeVs. Nevertheless, 
the total spectrum at below  the few hundred GeVs are still significantly below that of observations. If one wants to improve
the fit to that part from these components, however, one would have to increase the double $\tau^+\tau^-$  contribution, which 
would again make the peak change into a slope or cliff, as the case of global fit model B.

\subsection{Rescaled Background Model}\label{sec:rescaled_bg}

In our analysis of the DAMPE data, we distinguish the ``background", which is the part of the spectrum 
contributed by normal astrophysical sources, and the excess, which we attribute to the contribution of dark matter.  
In the above we have used the experimentally derived background from the AMS-02 experiment. 
Here we rescale the experimentally derived background by a factor $1.13\times{\rm exp}(-E/5 \TeV)$, to explore the possibility 
that  whether required dark matter contribution could be reduced and whether  the 4$\tau$ channel could still fit the DAMPE data.

\begin{figure}[htbp]
\centering{
\includegraphics[scale=0.45]{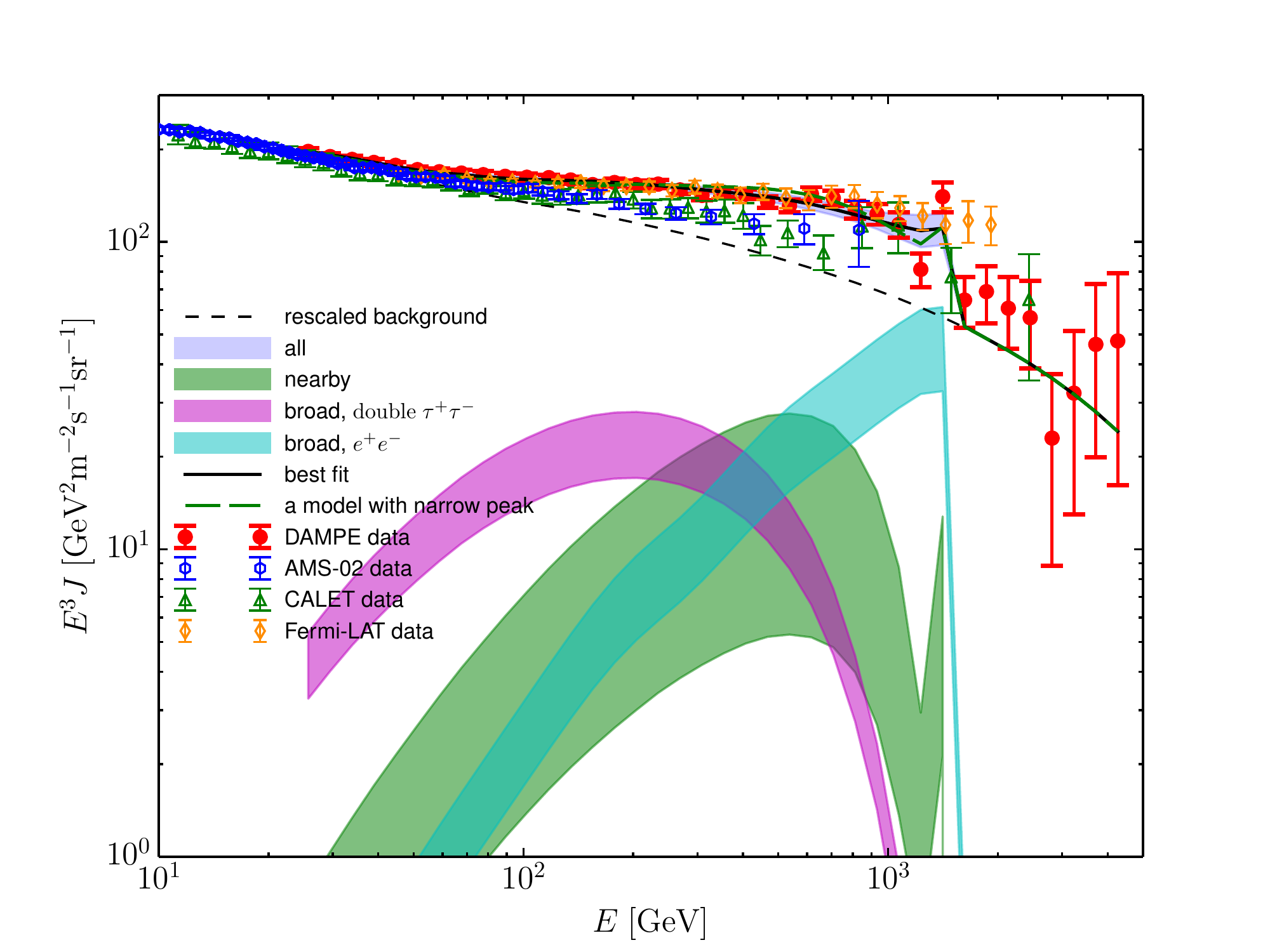}
\caption{A fit to the DAMPE spectrum with adjusted background, for the direct $e^+e^-$ and double $\tau^+\tau^-$ channels.}
\label{fig:rescaledbg}
}
\end{figure}

With this change, the fit to the broad excess can be significantly improved for the model B. 
We show the fit to the model with the rescaled background in Fig. \ref{fig:rescaledbg}. Here, compared with the 
global best fit model of the original background, 
the reduced $\chi^2$ of the best fit model decreased from 2.2 to 0.7.
This is because with the rescaled background, the shape of the 
double $\tau^+ \tau^-$ components have much better match with the broad excess. The best fit model still does not show
the narrow peak, but a model with the narrow peak is only slightly worse fit to the model than the best fit one.
See the green dashed line in Fig. \ref{fig:rescaledbg}.

This example shows that we have to view the various fits with some caution. Systematic uncertainty such as the 
definition of the background  could be very important in our interpretation of the data.

\subsection{A background consistent with the H.E.S.S. 2017 data at $\gtrsim5$ TeV}\label{hessbg}

\begin{figure}[htbp]
\centering{
\includegraphics[scale=0.45]{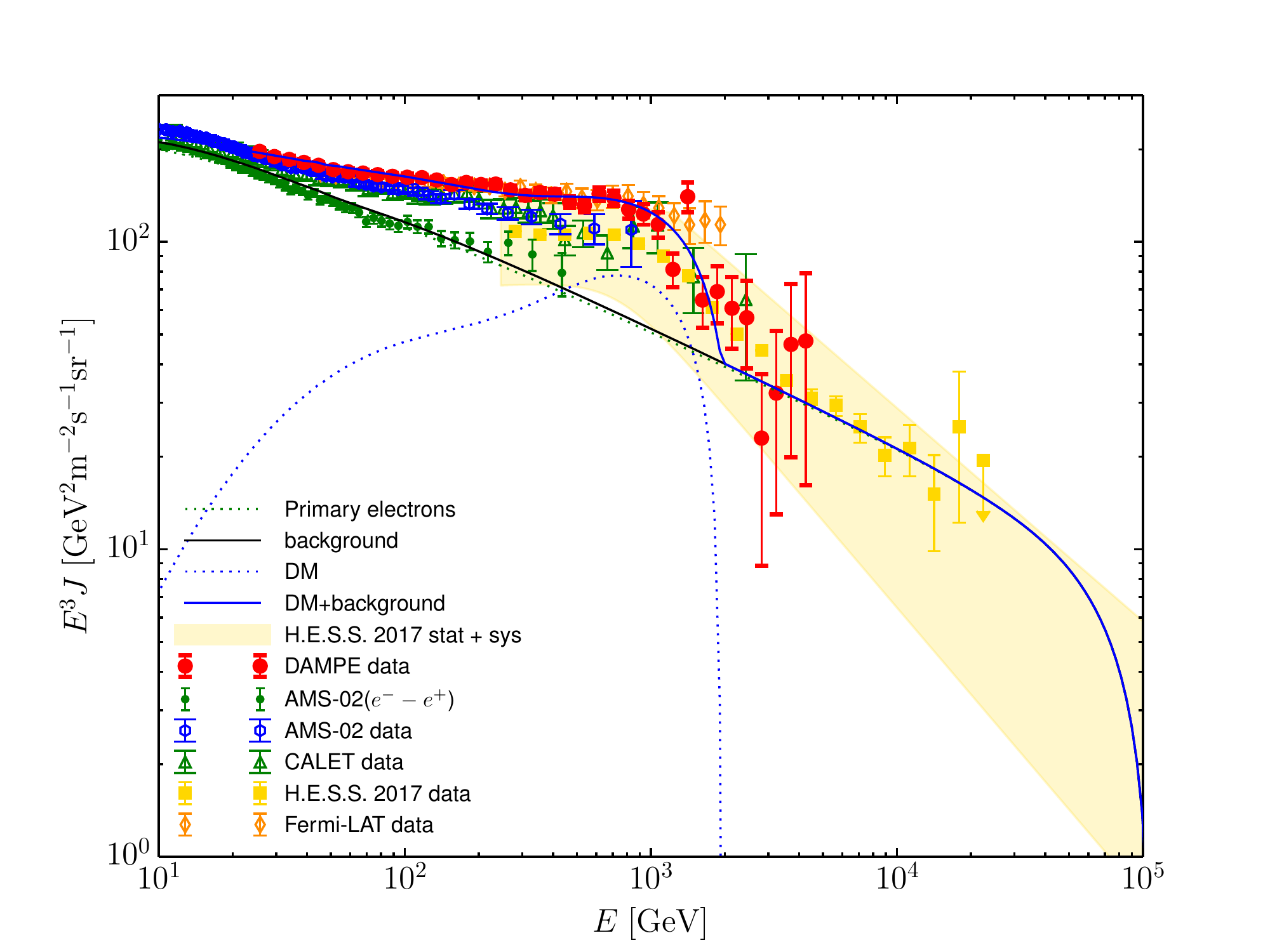}
\caption{A fit to the DAMPE spectrum with H.E.S.S. 2017 data consistent background, for the W$^+$W$^-$  and direct $e^+e^-$ channels.}
\label{fig:hessbg}
}
\end{figure}

In the models described so far, we  have not considered the constraints from the cosmic ray 
observations at energies above $ 5 \TeV$, because these data still have quite large uncertainties. Here 
we consider an astrophysical background with the inclusion of the H.E.S.S. 2017 
data\footnote{\url{https://indico.snu.ac.kr/indico/event/15/session/5/contribution/694/material/slides/0.pdf}}, which results in  
softer spectrum than those used above.  It drops dramatically at $\gtrsim40$ TeV due to the energy loss in propagation through 
the Milky Way.  Similar to Sec. \ref{broad}, we first investigate a model that only includes the diffuse dark matter annihilation from the Galactic halo, to fit the DAMPE broad spectrum excess. We investigate many channels and find model A channels are still the best. The results for model A channels are shown in Fig. \ref{fig:hessbg}. The best-fit dark matter particle mass $m_\chi= 1.9$ TeV and best-fit cross-section $\mean{\sigma v}=8.7\times10^{-23}$ cm$^3$s$^{-1}$, close to the model A in Sec. \ref{broad}, which is 1.7 TeV and $5.8\times10^{-23}$ cm$^3$s$^{-1}$, see the top left panel of Fig. \ref{fig:EplusAnni}. Here we get $\chi^2_{\rm min}/{\rm d.o.f}=1.1$.

\begin{figure}[htbp]
\centering{
\includegraphics[scale=0.45]{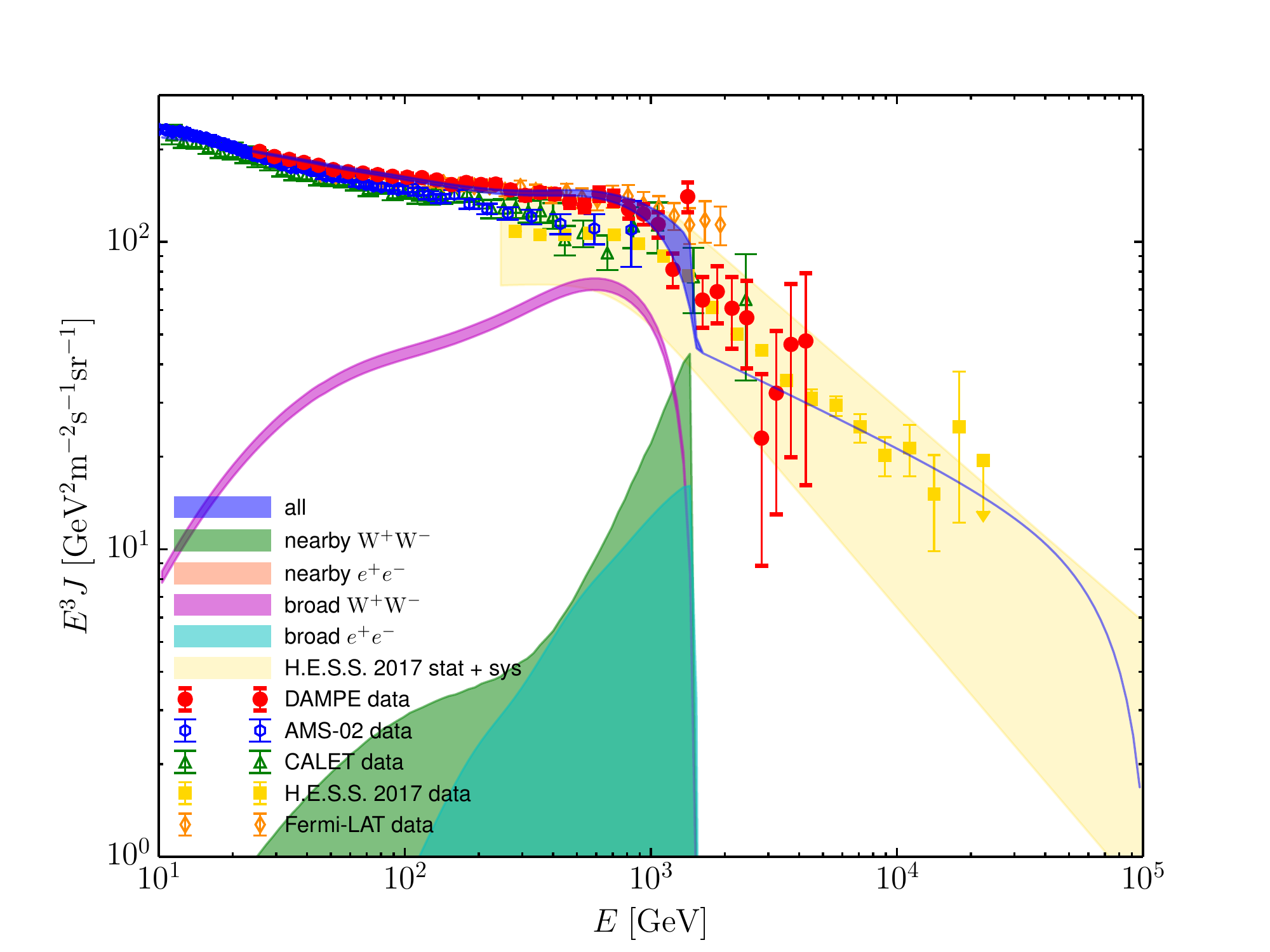}
\caption{Same as Fig. \ref{fig:J_fit_WW}, but a background consistent with H.E.S.S. 2017 data is adopted. The nearby $e^+e^-$ component is not seen because it is too weak that is below the x-axis.}
\label{fig:J_fit_WW_hessbg}.
}
\end{figure}

We then include the nearby source and perform the fitting again as model A in Sec. \ref{peak}. We fix dark matter particle mass to be 1.5 TeV. The results are shown in Fig. \ref{fig:J_fit_WW_hessbg}. Although adopting a different astrophysical background, we get similar conclusions: the annihilation of the diffuse dark matter in Galactic halo via W$^+$W$^-$ channel could explain the broad spectrum excess very well, however the direct $e^+e^-$ channel contributes negligibly; even introducing a nearby source, it is still not easy to generate a sharp peak near the 1.4 TeV. 

\section{The subhalo}\label{subhalo}

We now consider the subhalo which can produce the peak shown above. We let the electrons and positrons of a subhalo plus the background  fit the points near the peak.

\subsection{Decaying dark matter subhalo}

\begin{figure}[htbp]
\centering{
\includegraphics[scale=0.45]{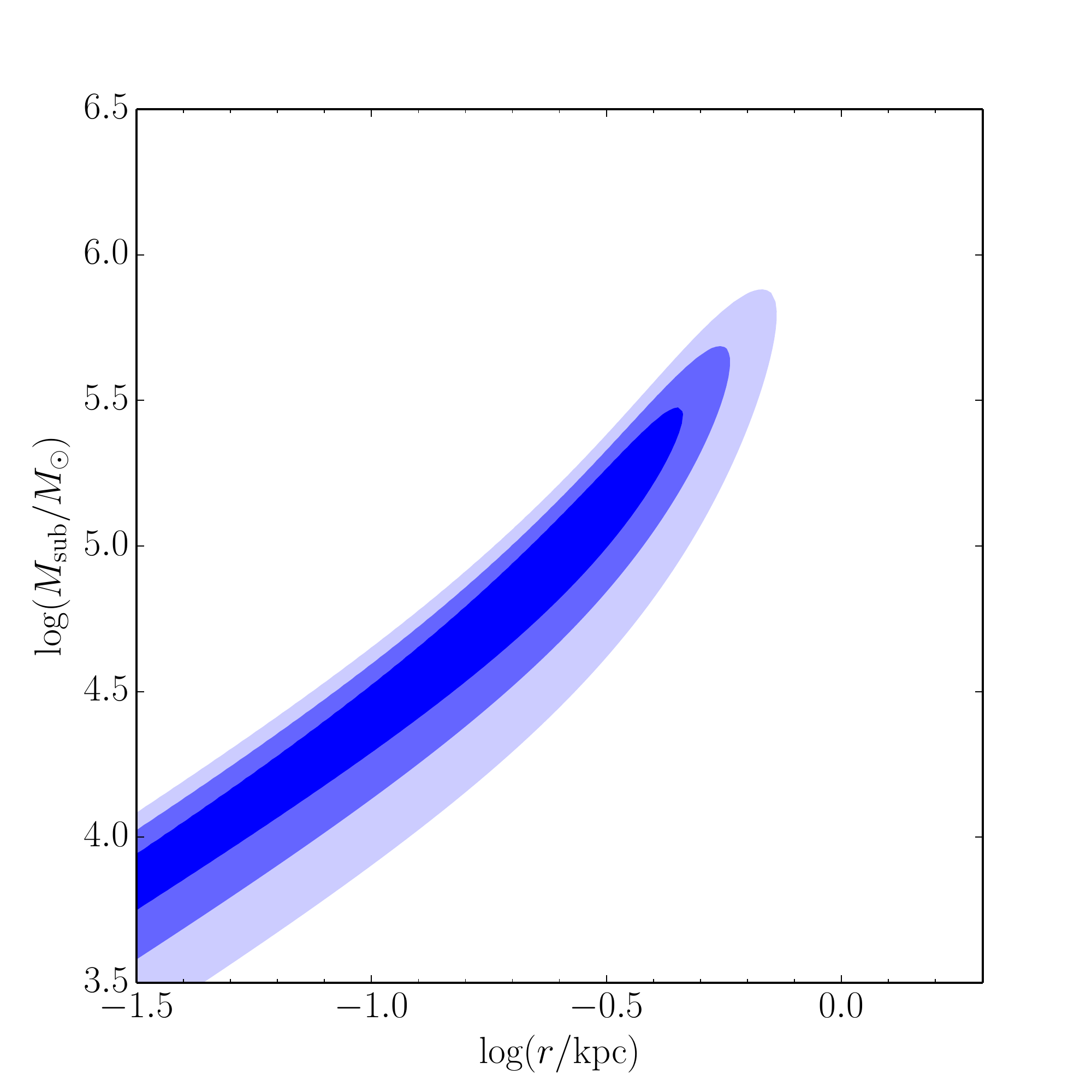}
\caption{The constraints on the log$(r)$ - log$(M_{\rm sub})$ by fitting only the peak for decaying dark matter.}
\label{fig:M_subDecay}
}
\end{figure}

For decaying dark matter with mass $m_\chi=2E_0$, and the lifetime $t_\chi$, 
the injection rate is
\begin{equation}
Q_{\rm decay}=\frac{2M_{\rm sub}}{m_\chi t_\chi}=\frac{M_{\rm sub}}{E_0 t_\chi}=2.6\times10^{39}\frac{M_{\rm sub}}{E_0t_\chi}~{\rm s}^{-1},
\end{equation}
where $M_{\rm sub}$ is subhalo mass in units of solar mass, $E_0$ in units GeV and $t_\chi$ in units 
Universe age ($4.4\times10^{17}$ s).    
If we adopt the life time value $7.3\times10^{25}$ s$^{-1}$, in which the Galactic halo diffuse dark matter contribution could 
produce the broad process,  the result is shown on  Fig.~\ref{fig:M_subDecay}.

We see that for a reasonable distance of 0.1 kpc, the required halo mass is about $10^{4.5} \Msun$.
Using longer dark matter particle lifetime, for example $\sim10^{28}$ s, we would have subhalo mass as high as $10^{6.5}~M_\odot$ at 0.1 kpc.

\subsection{Annihilating dark matter subhalo}

For dark matter annihilation, the density profile of the subhalo is crucial.  
For the subhalo, we assume a power-law density profile (e.g. \cite{Reed2005,Diemand2006,Hooper2017})
\begin{equation}
\rho(r)=\rho_0 r^{-\alpha},\qquad M_{\rm sub}= \frac{4\pi \rho_0}{3-\alpha}r_t^{3-\alpha},
\end{equation}
where $r_t$ is the tidal radius,  and 
\begin{equation}
\rho_0= \frac{(3-\alpha)M_{\rm sub}}{4\pi r_t^{3-\alpha}}
\end{equation}
The inner cutoff radius for cuspy halo is determined by equating the annihilation time scale to dynamical time scale, 
$\tau_d =  \tau_{\rm anni} $, where
\begin{equation}
\tau_d=\frac{r_t}{\sqrt{GM_{\rm sub}/r_t}}, \qquad \tau_{\rm anni}=\frac{ n_\chi(r_c) }{\dot{n}_\chi(r_c)}.
\end{equation}

Finally the injection rate from the annihilating dark matter in the subhalo is given by 
\begin{align}
Q_{\rm anni}&=\int_{r_c}^{r_t}   2\left(\frac{\rho_0c^2}{E_0} r^{-\alpha}\right)^2\mean{\sigma v}  4\pi r^2   dr  \\
&=\begin{cases}
8\pi \left( \frac{\rho_0c^2}{E_0} \right)^2\mean{\sigma v}\frac{\left(r_t^{3-2\alpha}-r_c^{3-2\alpha}\right)}{3-2\alpha}~~~{\rm when}~\alpha\neq1.5\\
8\pi \left( \frac{\rho_0c^2}{E_0} \right)^2\mean{\sigma v}[{\rm ln}(r_t)-{\rm ln}(r_c)]~~~{\rm when}~\alpha=1.5. \nonumber
\end{cases}
\end{align}

\begin{figure}[htbp]
\centering{
\includegraphics[scale=0.45]{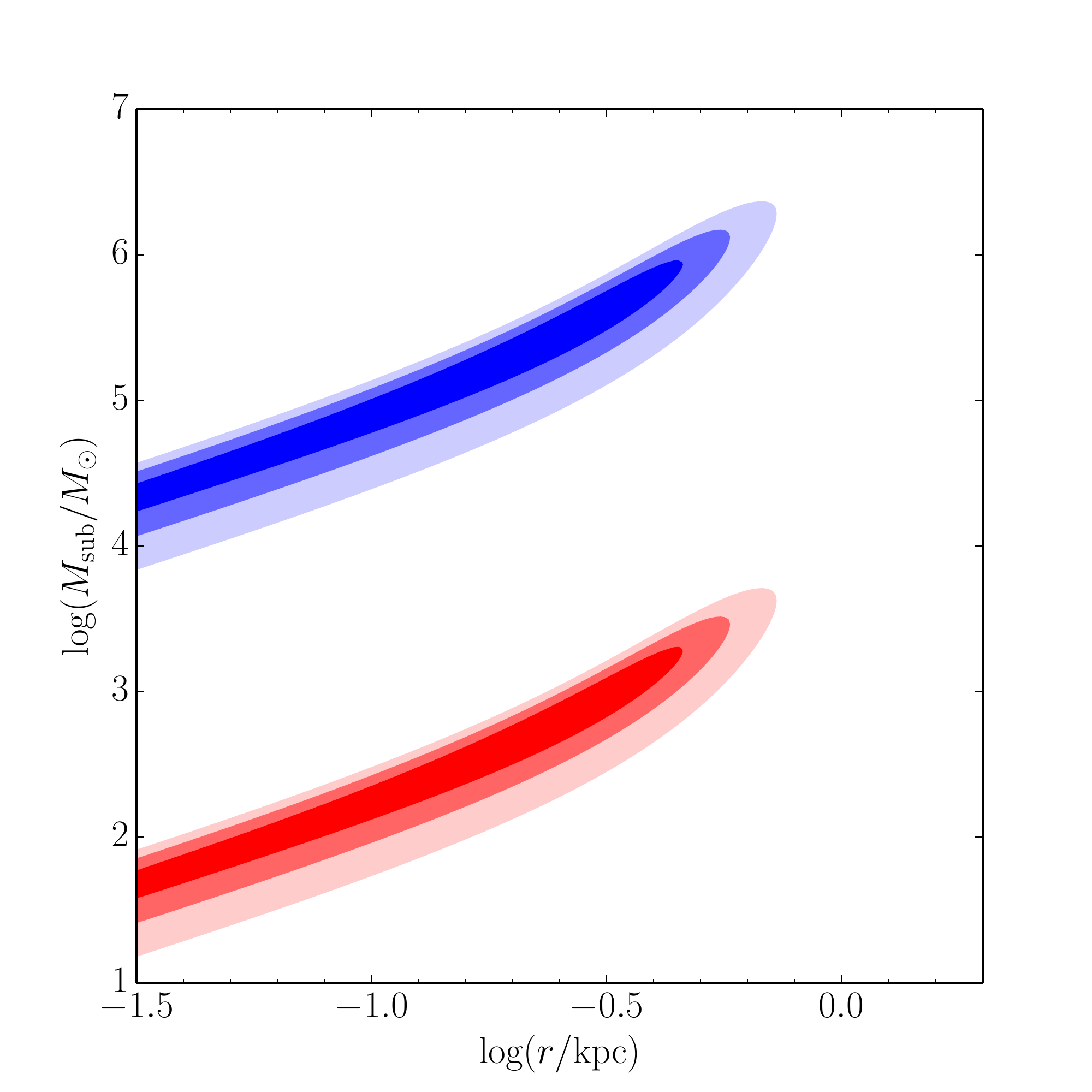}
\caption{The constraints on the log$(r)$ - log$(M_{\rm sub})$ for dark matter annihilation
case, $\mean{\sigma v}=10^{-23}$ cm$^3$~s$^{-1}$ is assumed. The blue contours for $\alpha=1.2$, red for $\alpha=1.7$.}
\label{fig:M_sub}
}
\end{figure}

For annihilation we assume $\mean{\sigma v} = 10^{-23}$ cm$^3$ s$^{-1}$, for which the annihilation of the diffuse dark matter in 
Galactic halo gives the broad excess.  Then, if the peak is due to the subhalo, 
the mass and distance of the subhalo is shown in  Fig. \ref{fig:M_sub}. 
To fit the peak, assuming the subhalo is located at $\sim 0.1$ kpc from the Solar system,
the required subhalo mass is $\sim10^{5} \Msun $ for $\alpha = 1.2$.
 If the halo has a steeper profile, e.g. 
$\alpha=1.7$, then the halo mass could be as small as $\sim 10^{2.5}~M_\odot$. 
A smaller subhalo with very steep density profile could be formed from the large density perturbation peaks ($\delta \rho/\rho\sim0.3$) in the early Universe in some scenarios (e.g. \cite{Yang2016_ultracompact,Yang2013a,Yang2013b}.)
If we instead adopt a thermal cross-section, $\mean{\sigma v}=3\times10^{-26}$ cm$^3$ s$^{-1}$, we then have that
at 0.1 kpc, the required subhalo mass is $\sim10^{7.5}~M_\odot$ for $\alpha=1.2$ and $\sim10^{4.5}~M_\odot$ for $\alpha=1.7$
respectively.

For both the decay and the annihilation models, we obtain the upper limit  $r < 0.53$ kpc ($3\sigma$ confidence level) 
after marginalizing the parameter $Q$. The constraint is mainly due to  achieve the narrow peak with width less than $\sim$0.2 TeV.

\subsection{Inverse Compton Scattering}

If such a subhalo is located near the Solar system, it might also produce strong $\gamma$-ray photons. Even if 
we conservatively assume that photons are not produced in the decay or annihilation, the electrons and positrons 
produced by the subhalo would have ICS with background photons. If a subhalo is a strong electrons/positrons source, it is also a strong ICS source. This would helps to 
identify the subhalo by looking for nearby bright and somewhat extended gamma-ray sources. 

Analogously to Eq. (\ref{eq:ICS}), we obtain the following emissivity for the up scattered photons:
\begin{align}
&\epsilon(\mathcal{E},E_0,Q,r)\approx \frac{1}{4\pi} \int dE\psi \sum _i \frac{3\sigma_T c U_{{\rm rad},i}}{4\gamma^2 \mathcal{E}^2_{0,i}}G(q_i,\Gamma_{e,i}),
\end{align}
where the electron spectrum $\psi$ is from Eq. \ref{psi}.

If the line-of-sight is $\theta$ offset from the center of the subhalo, we have the specific intensity of  photons up scattered  by electrons generated from a source with $Q$ injection rate and distance $r$ from the observer,
\begin{equation}
I_\theta(\mathcal{E},E_0,Q,r,\theta)=\int_0^\infty \epsilon(\mathcal{E},E_0,Q, x'   )dr',
\end{equation}
where  $x'=\sqrt{r^2+r'^2-2rr'{\rm cos}\theta}$.

\begin{figure}[htbp]
\centering{
\includegraphics[scale=0.45]{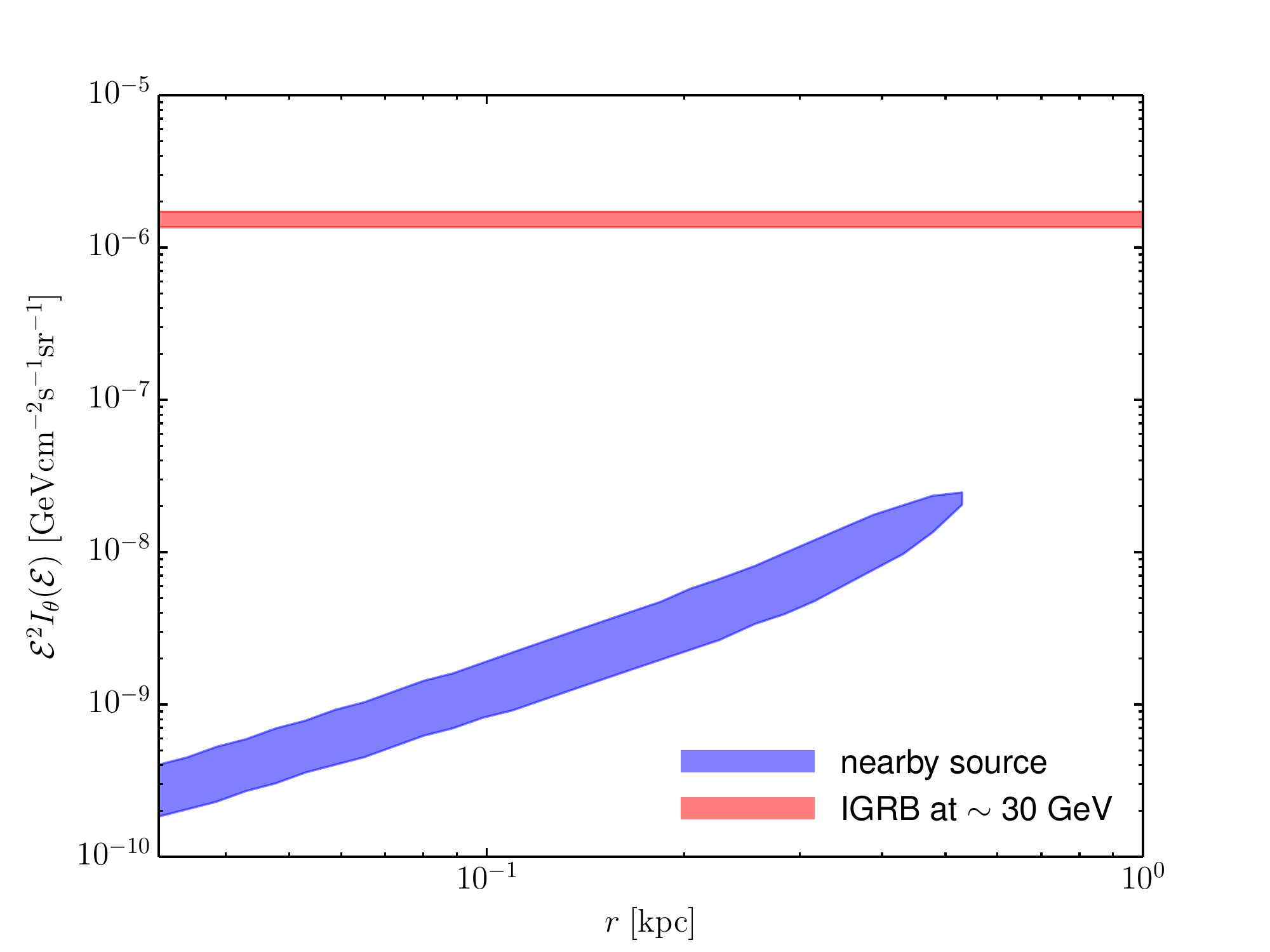}
\caption{The predicted gamma-ray intensity at 30 GeV for a subhalo with electrons/positrons that fits the peak (with $1\sigma$ C.L.), for line-of-sight toward 
the subhalo center. In the panel we also plot the IGRB given by paper  \cite{Ackermann:2014usa}.}
\label{fig:ICS}
}
\end{figure}

We show the ICS  specific intensity at 30 GeV for line-of-sight toward the center of the subhalo in Fig.~\ref{fig:ICS}. It is much smaller than the isotropic gamma radiation background (IGRB) given in \cite{Ackermann:2014usa}. 
The expected flux increases with decreasing distance, this is because to achieve the DAMPE peak, the more nearby source  needs to 
produce fewer number of electrons, hence also generating less ICS radiation. 
It is still challenge to observe such an extended gamma-ray source by existing or even upcoming instruments.

\subsection{Anisotropy}
Finally, we note that if the nearby subhalo is involved, it may generate a large anisotropy in the CR flux. 
This anisotropy, if measured, would be crucial for testing this subhalo scenario, or any similar nearby source scenarios  for the CR origins.
Currently the anisotropy of the DAMPE $\sim$1.4 TeV peak is not analyzed, probably due to the limited particle number: there are only 
93 particles detected in the energy bin between 1.3 - 1.5 TeV. Here we present the theoretical quantitative relation between the dipole 
anisotropy and the distance $r$. When the anisotropy analysis released in future, a fast check of the nearby scenario would be possible.   

\begin{figure}[htbp]
\centering{
\includegraphics[scale=0.45]{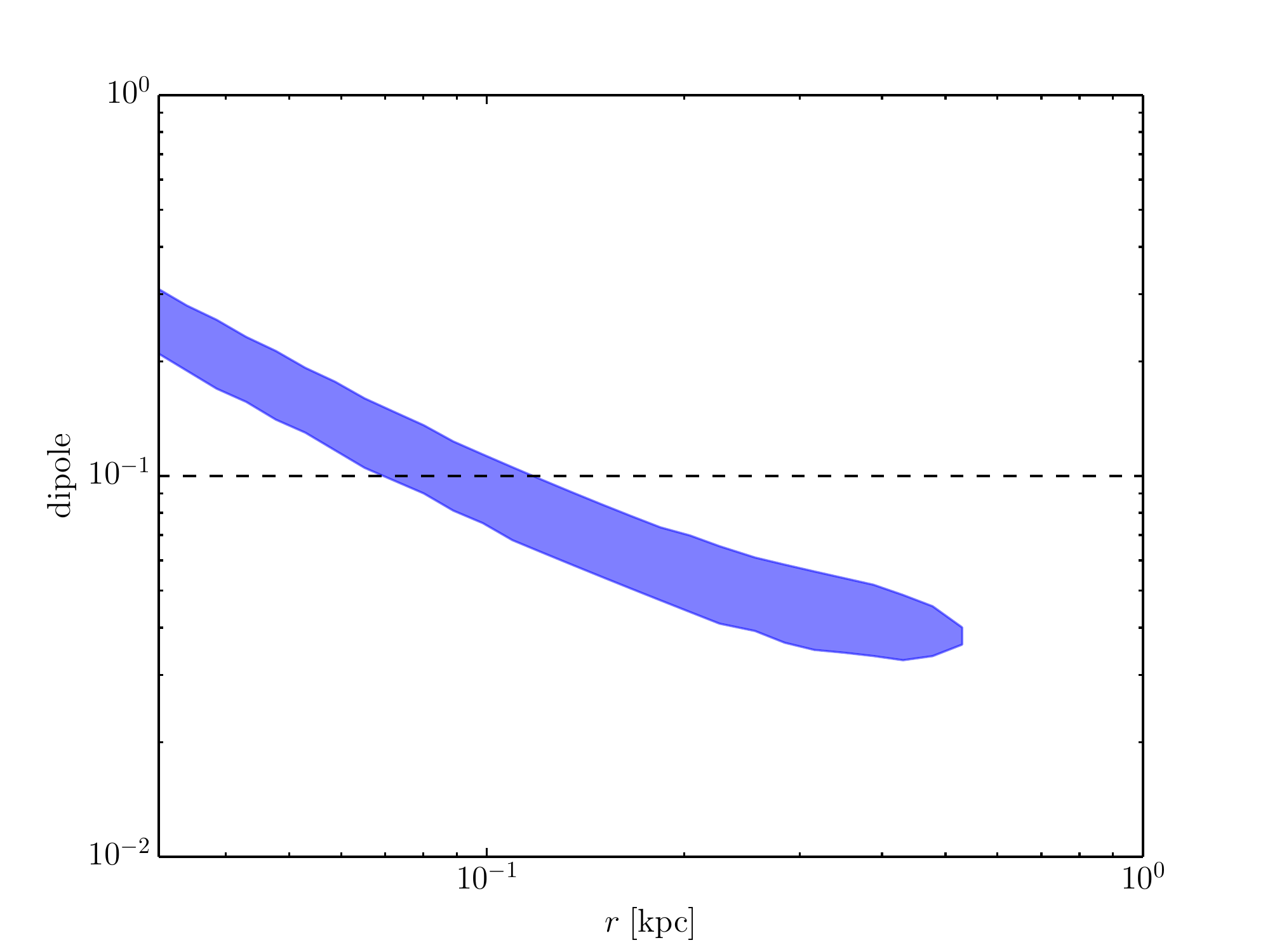}
\caption{The predicted dipole from a nearby source at various distance $r$. We show the range corresponding the $1\sigma$ C.L. fits of the peak. To guide the eye we plot the  dipole$=0.1$ by horzion line.}
\label{fig:dipole}
}
\end{figure}

The amplitude of the dipole anisotropy is \cite{Globus2017}
\begin{equation}
\left(\frac{\delta J}{J}\right)_{\rm max}\sim \frac{3D_{xx}}{c \psi }\left|\frac{d\psi}{dr}\right|.
\label{eq:dipole}
\end{equation}
In Fig. \ref{fig:dipole} we show the predicted averaged dipole of electrons/positrons in the energy bin 1.3 -1.5 TeV, from nearby source that fits the peak. Indeed, if the dipole is measured in future, it would be very helpful for identifying the distance of the nearby source. Obviously, if a dipole $\lsim0.03$ is measured, then it would be in conflict with the requirement by peak width, see Fig. \ref{fig:M_subDecay} and 
\ref{fig:M_sub}, then the nearby source scenario would be ruled out. On the other hand, if a rather high dipole is measured, then it infers that the source is quite close to us. In this case the gravity effects of such a subhalo should be carefully inspected.

\subsection{The probability for the existence of a nearby subhalo}
In this subsection we discuss the probability that our Solar system is occasionally close to a subhalo.
There have been many simulations that investigate the properties of subhalos in Milky Way-size halos and galaxy clusters 
(e.g.  \citep{Aquarius2008, ViaLactea,ViaLacteaI, ViaLacteaII,Gao2011,PhoenixProject,Jiang2016,Bosch2016,Giocoli2008}), 
including the mass function and spatial distribution. Basically, it is found that the mass function is a power-law form. 

For example, in the {\it Aquarius} project \citep{Aquarius2008}, for a host halo with mass $1.839\times10^{12}~M_\odot$ (Milky Way-size), the mass distribution of subhalos within $r_{50}=433.48$ kpc follows the expression
\begin{equation}
\frac{dN_{\rm sub}}{dM_{\rm sub}}=a_0\left(  \frac{M_{\rm sub}}{m_0} \right)^n,
\end{equation}
where $a_0=3.26\times10^{-5}~M_\odot^{-1}$, $m_0=2.52\times10^7~M_\odot$ and $n=-1.9$ \footnote {The simulations give this relation for subhalos above $10^5~M_\odot$, because of the limited resolution, we however extrapolate it down to $10^3~M_\odot$ in this paper.}. Moreover, the spatial distribution of their number density is well described by Einasto profile and is independent of the subhalo mass.  It writes
\begin{align}
&\frac{dn_{\rm sub}}{dM_{\rm sub}}(M_{\rm sub},r)= \nonumber \\
&a_0\left(  \frac{M_{\rm sub}}{m_0} \right)^n \times A_{\rm sub} {\rm exp}\left( -\frac{2}{\alpha}\left[\left(\frac{r}{r_{-2}}\right)^\alpha-1\right]\right),
\label{eq:n_sub}
\end{align}
where $\alpha=0.678$ and $r_{-2}=199$ kpc, and $A_{\rm sub}$ is a normalization factor derived by requiring 
\begin{equation}
\int_0^{r_{50}} dr 4\pi r^2    A_{\rm sub} {\rm exp}\left( -\frac{2}{\alpha}\left[\left(\frac{r}{r_{-2}}\right)^\alpha-1\right]\right)=1.
\end{equation}

\begin{figure}[htbp]
\centering{
\includegraphics[scale=0.45]{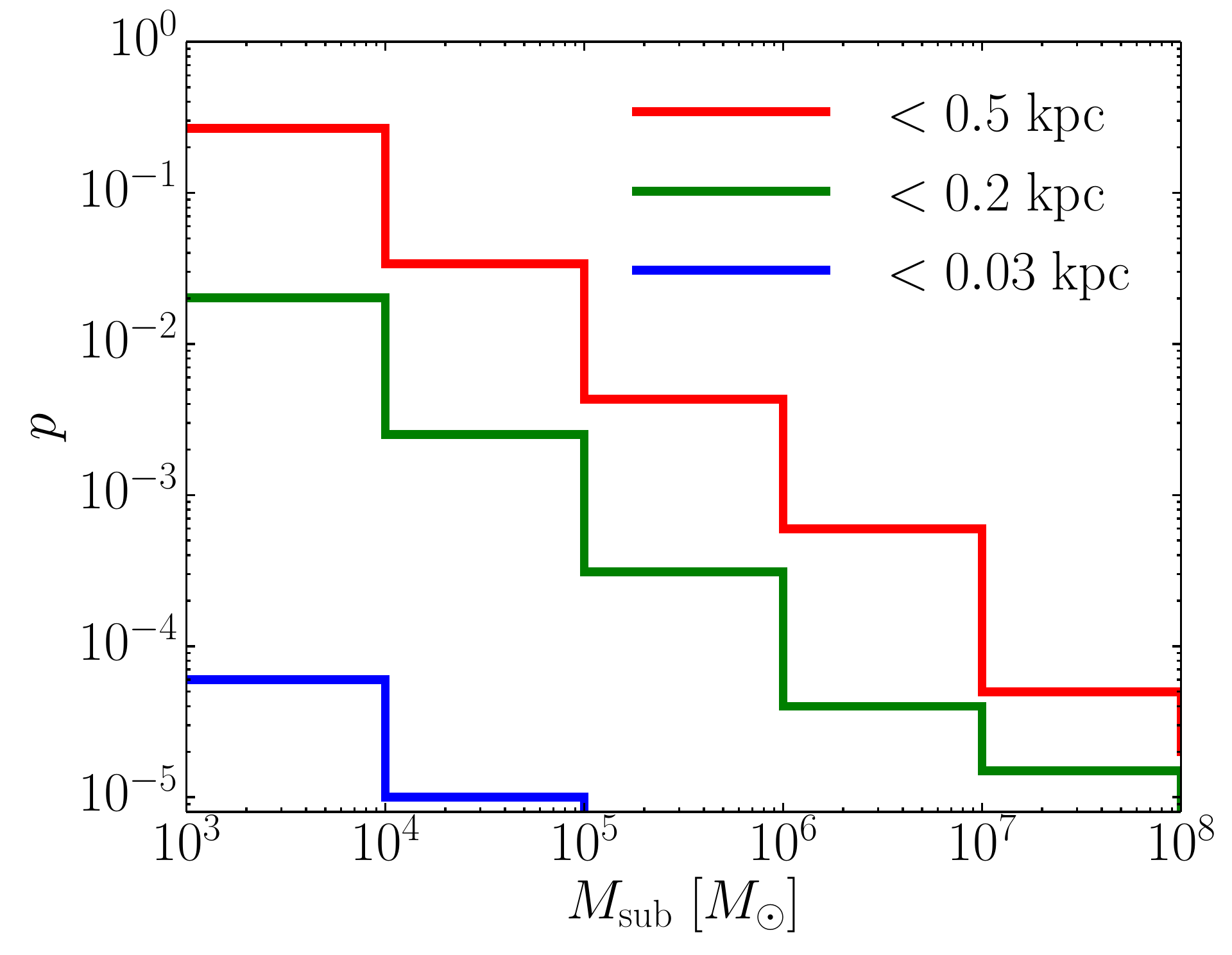}
\caption{The  probability that our Solar system is occasionally close to a subhalo (within distance $r$) with mass between $M_{\rm sub}$ and $10\times M_{\rm sub}$.}
\label{fig:subhalos}
}
\end{figure}

We assume the spatial distribution of subhalos is spherical symmetry, then generate Monte Carlo samples of Milky Way-size halos whose subhalos following the  Eq. (\ref{eq:n_sub}). Then randomly assign the location of our Solar system at the distance 8.5 kpc from the center of the halo.  We then count the 
probability for the existence of a subhalo within different distance $r$. The results are shown in Fig. \ref{fig:subhalos}.

Indeed, the probability that our Solar system is occasionally close to a subhalo is rather small. For example, the probability that a subhalo with mass between $10^5$ and $10^6~M_\odot$ within 0.5 kpc is less than 1\%. It challenges the subhalo scenario for interpreting the CR excess. This is the problem faced by all similar scenarios and actually has already been noticed long ago (e.g. \cite{Brun2009}).
For this reason, when we obtain the constrains on  subhalo mass 
and distance, we do not consider this probability.

In our paper, considering the above effect, together with the assumption of enhanced dark matter annihilation cross-section of the order of  $10^{-23}$ cm$^3$ s$^{-1}$ and steep density profile with slope $\alpha\sim-1.7$, the marginally acceptable parameters (with probability $\sim30\%$) for subhalo mass and distance are: subhalos with mass $\sim10^3-10^4~M_\odot$ at distance 
$\sim0.5$ kpc.

\section{Summary}\label{summary} 

We studied the CR electron and positron spectrum from dark matter annihilation or decay
 in the energy range of the recent DAMPE measurement. 
 We obtained the broad excess in 
 the spectrum from a few tens GeV to about 2 TeV, by subtracting the astrophysical background derived from
AMS-02 $e^--e^+$ data and the CR propagation model.
We found that for both the W$^+$W$^-$ channel and the combination of direct $e^+ e^-$ and double $\tau^+\tau^-$
channels, diffuse dark matter annihilation or decay in the Galactic halo can provide good fits to the broad 
excess in the electron and positron spectrum,  at least at $\gsim 100$ GeV.
This naturally explains the TeV break found in the CR electron and positron  spectrum. However, this 
requires a cross-section of order of $10^{-23}$~cm$^3$ s$^{-1}$, which is larger than the WIMP
cross-section required for obtaining the correct dark matter abundance through thermal decoupling in the early Universe.

In the DAMPE spectrum, there is also a prominent single point peak at $\sim$1.4 TeV. 
Tentatively treating it as real, we investigated how such a peak could be produced, in spite of the energy loss 
while the electrons and positrons diffusing through the interstellar space. 
We considered the possibility that the peak is produced by a nearby dark matter subhalo.
We found that in the global fit models, there is no peak in the spectrum, and inclusion of 
the nearby subhalo does not significantly improve the fit. 
This is because to match the shape of broad spectrum, a suppressed contribution from the direct $e^+e^-$ channel is favored by the 
data. As a result, a nearby subhalo is hard  to generate a narrow peak, as spectrum of electrons and positrons generated by other channels are always broad.

We then relaxed the requirement of fitting the whole spectrum data, and  studied how a spectrum with a $\sim$1.4 TeV peak
could be generated. We found that in the case of W$^+$W$^-$ channel, a peak feature with gentle slope on the low energy
 side and steep slope on the high energy side can be produced. In the case of  double 
 $\tau^+\tau^-$ channel and direct $e^+ e^-$ channel, a sharp peak feature as found in the DAMPE data could be produced. However, in these cases, 
 the fitting to the broad excess from a few tens to a few hundred GeVs are not very good. In these fits, however,
 we have assumed a background derived from the AMS-02 experiment. If we rescaled this background, a better fit 
 can be obtained for the double $\tau^+\tau^-$ and direct $e^+e^-$ channels, even with a peak. 
 Moreover, if we adopted a background spectrum that is consistent with the H.E.S.S. 2017 data when extending to $\gsim 5$ TeV,
 the conclusions will not change except that a bit higher cross-section is required. 
 These show that the precisely measured
 spectral shape from DAMPE could provide stringent constraint on the model, though theoretical uncertainty is 
 a very important factor in making interpretation and obtaining the constraint.

Giving the peak strength, we estimated the mass and distance of the subhalo. We found that  if assuming this is also the source for broad excess, the  required mass for a subhalo 0.1 kpc from our Solar system is $\sim10^{4.5}~M_\odot$ for the decay case, 
and $\sim10^{5}~M_\odot$ ($10^{2.5}~M_\odot$) for the annihilation case with density profile slope $\alpha=1.2$ (1.7). The mass could be even smaller for steeper 
slope. However, after inspecting the radial distribution of subhalos in a dark matter halo with mass close to our Milky Way's host halo, we found that 
the probability that a subhalo with mass in above range happens to locate near the Solar system is very small.

We also considered the inverse Compton scattering by the electrons, and the anisotropy in CR if the subhalo is too close by. 
The inverse Compton scattering radiation is still not observable for current $\gamma$-ray telescopes.

{\bf Postscript}: While we were preparing the draft of this paper, we noted that a paper on the theoretical interpretation of 
the DAMPE result by Q. Yuan et al. \cite{Yuan2017} appeared on the arxiv preprint server, 
which investigated both dark matter and pulsar origin of the DAMPE excess and 1.4 TeV peak, 
though the adopted dark matter models and the methods are somewhat different from ours.

\acknowledgments
We thank Prof. Jin Chang and Dr. Xiaoyuan Huang for helpful discussions.  XLC acknowledges the support of the 
NSFC through grant No. 1633004 and 11373030, the MoST through grant 2016YFE0100300, and the CAS Frontier Science Key Project No. QYZDJ-SSW-SLH017. BY acknowledges the support of NSFC grant 11653003 and the Bairen program from the Chinese Academy of Sciences (CAS). BY and HBJ acknowledge the support of the NSFC-CAS  joint fund for space scientific satellites No. U1738125.

\bibliographystyle{apsrev}
\bibliography{refsall}

\begin{thebibliography}{99}
\expandafter\ifx\csname natexlab\endcsname\relax\def\natexlab#1{#1}\fi
\expandafter\ifx\csname bibnamefont\endcsname\relax
  \def\bibnamefont#1{#1}\fi
\expandafter\ifx\csname bibfnamefont\endcsname\relax
  \def\bibfnamefont#1{#1}\fi
\expandafter\ifx\csname citenamefont\endcsname\relax
  \def\citenamefont#1{#1}\fi
\expandafter\ifx\csname url\endcsname\relax
  \def\url#1{\texttt{#1}}\fi
\expandafter\ifx\csname urlprefix\endcsname\relax\def\urlprefix{URL }\fi
\providecommand{\bibinfo}[2]{#2}
\providecommand{\eprint}[2][]{\url{#2}}

\bibitem[{\citenamefont{{Planck Collaboration}
  et~al.}(2016)\citenamefont{{Planck Collaboration}, {Ade}, {Aghanim},
  {Arnaud}, {Ashdown}, {Aumont}, {Baccigalupi}, {Banday}, {Barreiro},
  {Bartlett} et~al.}}]{Planck2016cos}
\bibinfo{author}{\bibnamefont{{Planck Collaboration}}},
  \bibinfo{author}{\bibfnamefont{P.~A.~R.} \bibnamefont{{Ade}}},
  \bibinfo{author}{\bibfnamefont{N.}~\bibnamefont{{Aghanim}}},
  \bibinfo{author}{\bibfnamefont{M.}~\bibnamefont{{Arnaud}}},
  \bibinfo{author}{\bibfnamefont{M.}~\bibnamefont{{Ashdown}}},
  \bibinfo{author}{\bibfnamefont{J.}~\bibnamefont{{Aumont}}},
  \bibinfo{author}{\bibfnamefont{C.}~\bibnamefont{{Baccigalupi}}},
  \bibinfo{author}{\bibfnamefont{A.~J.} \bibnamefont{{Banday}}},
  \bibinfo{author}{\bibfnamefont{R.~B.} \bibnamefont{{Barreiro}}},
  \bibinfo{author}{\bibfnamefont{J.~G.} \bibnamefont{{Bartlett}}},
  \bibnamefont{et~al.}, \bibinfo{journal}{\aap} \textbf{\bibinfo{volume}{594}},
  \bibinfo{eid}{A13} (\bibinfo{year}{2016}), \eprint{1502.01589}.

\bibitem[{\citenamefont{{Battaglieri} et~al.}(2017)\citenamefont{{Battaglieri},
  {Belloni}, {Chou}, {Cushman}, {Echenard}, {Essig}, {Estrada}, {Feng},
  {Flaugher}, {Fox} et~al.}}]{2017arXiv170704591B}
\bibinfo{author}{\bibfnamefont{M.}~\bibnamefont{{Battaglieri}}},
  \bibinfo{author}{\bibfnamefont{A.}~\bibnamefont{{Belloni}}},
  \bibinfo{author}{\bibfnamefont{A.}~\bibnamefont{{Chou}}},
  \bibinfo{author}{\bibfnamefont{P.}~\bibnamefont{{Cushman}}},
  \bibinfo{author}{\bibfnamefont{B.}~\bibnamefont{{Echenard}}},
  \bibinfo{author}{\bibfnamefont{R.}~\bibnamefont{{Essig}}},
  \bibinfo{author}{\bibfnamefont{J.}~\bibnamefont{{Estrada}}},
  \bibinfo{author}{\bibfnamefont{J.~L.} \bibnamefont{{Feng}}},
  \bibinfo{author}{\bibfnamefont{B.}~\bibnamefont{{Flaugher}}},
  \bibinfo{author}{\bibfnamefont{P.~J.} \bibnamefont{{Fox}}},
  \bibnamefont{et~al.}, \bibinfo{journal}{ArXiv e-prints}
  (\bibinfo{year}{2017}), \eprint{1707.04591}.

\bibitem[{\citenamefont{{Strong} et~al.}(2007)\citenamefont{{Strong},
  {Moskalenko}, and {Ptuskin}}}]{2007ARNPS..57..285S}
\bibinfo{author}{\bibfnamefont{A.~W.} \bibnamefont{{Strong}}},
  \bibinfo{author}{\bibfnamefont{I.~V.} \bibnamefont{{Moskalenko}}},
  \bibnamefont{and} \bibinfo{author}{\bibfnamefont{V.~S.}
  \bibnamefont{{Ptuskin}}}, \bibinfo{journal}{Annual Review of Nuclear and
  Particle Science} \textbf{\bibinfo{volume}{57}}, \bibinfo{pages}{285}
  (\bibinfo{year}{2007}), \eprint{astro-ph/0701517}.

\bibitem[{\citenamefont{{DuVernois} et~al.}(2001)\citenamefont{{DuVernois},
  {Barwick}, {Beatty}, {Bhattacharyya}, {Bower}, {Chaput}, {Coutu}, {de Nolfo},
  {Lowder}, {McKee} et~al.}}]{2001ApJ...559..296D}
\bibinfo{author}{\bibfnamefont{M.~A.} \bibnamefont{{DuVernois}}},
  \bibinfo{author}{\bibfnamefont{S.~W.} \bibnamefont{{Barwick}}},
  \bibinfo{author}{\bibfnamefont{J.~J.} \bibnamefont{{Beatty}}},
  \bibinfo{author}{\bibfnamefont{A.}~\bibnamefont{{Bhattacharyya}}},
  \bibinfo{author}{\bibfnamefont{C.~R.} \bibnamefont{{Bower}}},
  \bibinfo{author}{\bibfnamefont{C.~J.} \bibnamefont{{Chaput}}},
  \bibinfo{author}{\bibfnamefont{S.}~\bibnamefont{{Coutu}}},
  \bibinfo{author}{\bibfnamefont{G.~A.} \bibnamefont{{de Nolfo}}},
  \bibinfo{author}{\bibfnamefont{D.~M.} \bibnamefont{{Lowder}}},
  \bibinfo{author}{\bibfnamefont{S.}~\bibnamefont{{McKee}}},
  \bibnamefont{et~al.}, \bibinfo{journal}{\apj} \textbf{\bibinfo{volume}{559}},
  \bibinfo{pages}{296} (\bibinfo{year}{2001}).

\bibitem[{\citenamefont{{Chang} et~al.}(2008)\citenamefont{{Chang}, {Adams},
  {Ahn}, {Bashindzhagyan}, {Christl}, {Ganel}, {Guzik}, {Isbert}, {Kim},
  {Kuznetsov} et~al.}}]{2008Natur.456..362C}
\bibinfo{author}{\bibfnamefont{J.}~\bibnamefont{{Chang}}},
  \bibinfo{author}{\bibfnamefont{J.~H.} \bibnamefont{{Adams}}},
  \bibinfo{author}{\bibfnamefont{H.~S.} \bibnamefont{{Ahn}}},
  \bibinfo{author}{\bibfnamefont{G.~L.} \bibnamefont{{Bashindzhagyan}}},
  \bibinfo{author}{\bibfnamefont{M.}~\bibnamefont{{Christl}}},
  \bibinfo{author}{\bibfnamefont{O.}~\bibnamefont{{Ganel}}},
  \bibinfo{author}{\bibfnamefont{T.~G.} \bibnamefont{{Guzik}}},
  \bibinfo{author}{\bibfnamefont{J.}~\bibnamefont{{Isbert}}},
  \bibinfo{author}{\bibfnamefont{K.~C.} \bibnamefont{{Kim}}},
  \bibinfo{author}{\bibfnamefont{E.~N.} \bibnamefont{{Kuznetsov}}},
  \bibnamefont{et~al.}, \bibinfo{journal}{\nat} \textbf{\bibinfo{volume}{456}},
  \bibinfo{pages}{362} (\bibinfo{year}{2008}).

\bibitem[{\citenamefont{{Adriani} et~al.}(2011)\citenamefont{{Adriani},
  {Barbarino}, {Bazilevskaya}, {Bellotti}, {Boezio}, {Bogomolov}, {Bongi},
  {Bonvicini}, {Borisov}, {Bottai} et~al.}}]{2011PhRvL.106t1101A}
\bibinfo{author}{\bibfnamefont{O.}~\bibnamefont{{Adriani}}},
  \bibinfo{author}{\bibfnamefont{G.~C.} \bibnamefont{{Barbarino}}},
  \bibinfo{author}{\bibfnamefont{G.~A.} \bibnamefont{{Bazilevskaya}}},
  \bibinfo{author}{\bibfnamefont{R.}~\bibnamefont{{Bellotti}}},
  \bibinfo{author}{\bibfnamefont{M.}~\bibnamefont{{Boezio}}},
  \bibinfo{author}{\bibfnamefont{E.~A.} \bibnamefont{{Bogomolov}}},
  \bibinfo{author}{\bibfnamefont{M.}~\bibnamefont{{Bongi}}},
  \bibinfo{author}{\bibfnamefont{V.}~\bibnamefont{{Bonvicini}}},
  \bibinfo{author}{\bibfnamefont{S.}~\bibnamefont{{Borisov}}},
  \bibinfo{author}{\bibfnamefont{S.}~\bibnamefont{{Bottai}}},
  \bibnamefont{et~al.}, \bibinfo{journal}{Physical Review Letters}
  \textbf{\bibinfo{volume}{106}}, \bibinfo{eid}{201101} (\bibinfo{year}{2011}),
  \eprint{1103.2880}.

\bibitem[{\citenamefont{{Abdo} et~al.}(2009)\citenamefont{{Abdo}, {Ackermann},
  {Ajello}, {Atwood}, {Axelsson}, {Baldini}, {Ballet}, {Barbiellini},
  {Bastieri}, {Battelino} et~al.}}]{2009PhRvL.102r1101A}
\bibinfo{author}{\bibfnamefont{A.~A.} \bibnamefont{{Abdo}}},
  \bibinfo{author}{\bibfnamefont{M.}~\bibnamefont{{Ackermann}}},
  \bibinfo{author}{\bibfnamefont{M.}~\bibnamefont{{Ajello}}},
  \bibinfo{author}{\bibfnamefont{W.~B.} \bibnamefont{{Atwood}}},
  \bibinfo{author}{\bibfnamefont{M.}~\bibnamefont{{Axelsson}}},
  \bibinfo{author}{\bibfnamefont{L.}~\bibnamefont{{Baldini}}},
  \bibinfo{author}{\bibfnamefont{J.}~\bibnamefont{{Ballet}}},
  \bibinfo{author}{\bibfnamefont{G.}~\bibnamefont{{Barbiellini}}},
  \bibinfo{author}{\bibfnamefont{D.}~\bibnamefont{{Bastieri}}},
  \bibinfo{author}{\bibfnamefont{M.}~\bibnamefont{{Battelino}}},
  \bibnamefont{et~al.}, \bibinfo{journal}{Physical Review Letters}
  \textbf{\bibinfo{volume}{102}}, \bibinfo{eid}{181101} (\bibinfo{year}{2009}),
  \eprint{0905.0025}.

\bibitem[{\citenamefont{{Abdollahi} et~al.}(2017)\citenamefont{{Abdollahi},
  {Ackermann}, {Ajello}, {Atwood}, {Baldini}, {Barbiellini}, {Bastieri},
  {Bellazzini}, {Bloom}, {Bonino} et~al.}}]{2017PhRvD..95h2007A}
\bibinfo{author}{\bibfnamefont{S.}~\bibnamefont{{Abdollahi}}},
  \bibinfo{author}{\bibfnamefont{M.}~\bibnamefont{{Ackermann}}},
  \bibinfo{author}{\bibfnamefont{M.}~\bibnamefont{{Ajello}}},
  \bibinfo{author}{\bibfnamefont{W.~B.} \bibnamefont{{Atwood}}},
  \bibinfo{author}{\bibfnamefont{L.}~\bibnamefont{{Baldini}}},
  \bibinfo{author}{\bibfnamefont{G.}~\bibnamefont{{Barbiellini}}},
  \bibinfo{author}{\bibfnamefont{D.}~\bibnamefont{{Bastieri}}},
  \bibinfo{author}{\bibfnamefont{R.}~\bibnamefont{{Bellazzini}}},
  \bibinfo{author}{\bibfnamefont{E.~D.} \bibnamefont{{Bloom}}},
  \bibinfo{author}{\bibfnamefont{R.}~\bibnamefont{{Bonino}}},
  \bibnamefont{et~al.}, \bibinfo{journal}{\prd} \textbf{\bibinfo{volume}{95}},
  \bibinfo{eid}{082007} (\bibinfo{year}{2017}).

\bibitem[{\citenamefont{{Aguilar} et~al.}(2014)\citenamefont{{Aguilar}, {Aisa},
  {Alpat}, {Alvino}, {Ambrosi}, {Andeen}, {Arruda}, {Attig}, {Azzarello},
  {Bachlechner} et~al.}}]{2014PhRvL.113v1102A}
\bibinfo{author}{\bibfnamefont{M.}~\bibnamefont{{Aguilar}}},
  \bibinfo{author}{\bibfnamefont{D.}~\bibnamefont{{Aisa}}},
  \bibinfo{author}{\bibfnamefont{B.}~\bibnamefont{{Alpat}}},
  \bibinfo{author}{\bibfnamefont{A.}~\bibnamefont{{Alvino}}},
  \bibinfo{author}{\bibfnamefont{G.}~\bibnamefont{{Ambrosi}}},
  \bibinfo{author}{\bibfnamefont{K.}~\bibnamefont{{Andeen}}},
  \bibinfo{author}{\bibfnamefont{L.}~\bibnamefont{{Arruda}}},
  \bibinfo{author}{\bibfnamefont{N.}~\bibnamefont{{Attig}}},
  \bibinfo{author}{\bibfnamefont{P.}~\bibnamefont{{Azzarello}}},
  \bibinfo{author}{\bibfnamefont{A.}~\bibnamefont{{Bachlechner}}},
  \bibnamefont{et~al.}, \bibinfo{journal}{Physical Review Letters}
  \textbf{\bibinfo{volume}{113}}, \bibinfo{eid}{221102} (\bibinfo{year}{2014}).

\bibitem[{\citenamefont{{Abramowski} et~al.}(2011)\citenamefont{{Abramowski},
  {Acero}, {Aharonian}, {Akhperjanian}, {Anton}, {Barnacka}, {Barres de
  Almeida}, {Bazer-Bachi}, {Becherini}, {Becker} et~al.}}]{2011PhRvL.106p1301A}
\bibinfo{author}{\bibfnamefont{A.}~\bibnamefont{{Abramowski}}},
  \bibinfo{author}{\bibfnamefont{F.}~\bibnamefont{{Acero}}},
  \bibinfo{author}{\bibfnamefont{F.}~\bibnamefont{{Aharonian}}},
  \bibinfo{author}{\bibfnamefont{A.~G.} \bibnamefont{{Akhperjanian}}},
  \bibinfo{author}{\bibfnamefont{G.}~\bibnamefont{{Anton}}},
  \bibinfo{author}{\bibfnamefont{A.}~\bibnamefont{{Barnacka}}},
  \bibinfo{author}{\bibfnamefont{U.}~\bibnamefont{{Barres de Almeida}}},
  \bibinfo{author}{\bibfnamefont{A.~R.} \bibnamefont{{Bazer-Bachi}}},
  \bibinfo{author}{\bibfnamefont{Y.}~\bibnamefont{{Becherini}}},
  \bibinfo{author}{\bibfnamefont{J.}~\bibnamefont{{Becker}}},
  \bibnamefont{et~al.}, \bibinfo{journal}{Physical Review Letters}
  \textbf{\bibinfo{volume}{106}}, \bibinfo{eid}{161301} (\bibinfo{year}{2011}),
  \eprint{1103.3266}.

\bibitem[{\citenamefont{{Aharonian} et~al.}(2008)\citenamefont{{Aharonian},
  {Akhperjanian}, {Barres de Almeida}, {Bazer-Bachi}, {Becherini}, {Behera},
  {Benbow}, {Bernl{\"o}hr}, {Boisson}, {Bochow} et~al.}}]{2008PhRvL.101z1104A}
\bibinfo{author}{\bibfnamefont{F.}~\bibnamefont{{Aharonian}}},
  \bibinfo{author}{\bibfnamefont{A.~G.} \bibnamefont{{Akhperjanian}}},
  \bibinfo{author}{\bibfnamefont{U.}~\bibnamefont{{Barres de Almeida}}},
  \bibinfo{author}{\bibfnamefont{A.~R.} \bibnamefont{{Bazer-Bachi}}},
  \bibinfo{author}{\bibfnamefont{Y.}~\bibnamefont{{Becherini}}},
  \bibinfo{author}{\bibfnamefont{B.}~\bibnamefont{{Behera}}},
  \bibinfo{author}{\bibfnamefont{W.}~\bibnamefont{{Benbow}}},
  \bibinfo{author}{\bibfnamefont{K.}~\bibnamefont{{Bernl{\"o}hr}}},
  \bibinfo{author}{\bibfnamefont{C.}~\bibnamefont{{Boisson}}},
  \bibinfo{author}{\bibfnamefont{A.}~\bibnamefont{{Bochow}}},
  \bibnamefont{et~al.}, \bibinfo{journal}{Physical Review Letters}
  \textbf{\bibinfo{volume}{101}}, \bibinfo{eid}{261104} (\bibinfo{year}{2008}),
  \eprint{0811.3894}.

\bibitem[{\citenamefont{{Aharonian} et~al.}(2009)\citenamefont{{Aharonian},
  {Akhperjanian}, {Anton}, {Barres de Almeida}, {Bazer-Bachi}, {Becherini},
  {Behera}, {Bernl{\"o}hr}, {Bochow}, {Boisson} et~al.}}]{2009A&A...508..561A}
\bibinfo{author}{\bibfnamefont{F.}~\bibnamefont{{Aharonian}}},
  \bibinfo{author}{\bibfnamefont{A.~G.} \bibnamefont{{Akhperjanian}}},
  \bibinfo{author}{\bibfnamefont{G.}~\bibnamefont{{Anton}}},
  \bibinfo{author}{\bibfnamefont{U.}~\bibnamefont{{Barres de Almeida}}},
  \bibinfo{author}{\bibfnamefont{A.~R.} \bibnamefont{{Bazer-Bachi}}},
  \bibinfo{author}{\bibfnamefont{Y.}~\bibnamefont{{Becherini}}},
  \bibinfo{author}{\bibfnamefont{B.}~\bibnamefont{{Behera}}},
  \bibinfo{author}{\bibfnamefont{K.}~\bibnamefont{{Bernl{\"o}hr}}},
  \bibinfo{author}{\bibfnamefont{A.}~\bibnamefont{{Bochow}}},
  \bibinfo{author}{\bibfnamefont{C.}~\bibnamefont{{Boisson}}},
  \bibnamefont{et~al.}, \bibinfo{journal}{\aap} \textbf{\bibinfo{volume}{508}},
  \bibinfo{pages}{561} (\bibinfo{year}{2009}), \eprint{0905.0105}.

\bibitem[{\citenamefont{{Cirelli} et~al.}(2009)\citenamefont{{Cirelli},
  {Kadastik}, {Raidal}, and {Strumia}}}]{2009NuPhB.813....1C}
\bibinfo{author}{\bibfnamefont{M.}~\bibnamefont{{Cirelli}}},
  \bibinfo{author}{\bibfnamefont{M.}~\bibnamefont{{Kadastik}}},
  \bibinfo{author}{\bibfnamefont{M.}~\bibnamefont{{Raidal}}}, \bibnamefont{and}
  \bibinfo{author}{\bibfnamefont{A.}~\bibnamefont{{Strumia}}},
  \bibinfo{journal}{Nuclear Physics B} \textbf{\bibinfo{volume}{813}},
  \bibinfo{pages}{1} (\bibinfo{year}{2009}), \eprint{0809.2409}.

\bibitem[{\citenamefont{{Dev} et~al.}(2014)\citenamefont{{Dev}, {Ghosh},
  {Okada}, and {Saha}}}]{Dev2014}
\bibinfo{author}{\bibfnamefont{P.~S.~B.} \bibnamefont{{Dev}}},
  \bibinfo{author}{\bibfnamefont{D.~K.} \bibnamefont{{Ghosh}}},
  \bibinfo{author}{\bibfnamefont{N.}~\bibnamefont{{Okada}}}, \bibnamefont{and}
  \bibinfo{author}{\bibfnamefont{I.}~\bibnamefont{{Saha}}},
  \bibinfo{journal}{\prd} \textbf{\bibinfo{volume}{89}}, \bibinfo{eid}{095001}
  (\bibinfo{year}{2014}), \eprint{1307.6204}.

\bibitem[{\citenamefont{{Profumo} et~al.}(2017)\citenamefont{{Profumo},
  {Queiroz}, {Silk}, and {Siqueira}}}]{2017arXiv171103133P}
\bibinfo{author}{\bibfnamefont{S.}~\bibnamefont{{Profumo}}},
  \bibinfo{author}{\bibfnamefont{F.~S.} \bibnamefont{{Queiroz}}},
  \bibinfo{author}{\bibfnamefont{J.}~\bibnamefont{{Silk}}}, \bibnamefont{and}
  \bibinfo{author}{\bibfnamefont{C.}~\bibnamefont{{Siqueira}}},
  \bibinfo{journal}{ArXiv e-prints}  (\bibinfo{year}{2017}),
  \eprint{1711.03133}.

\bibitem[{\citenamefont{{Bergstr{\"o}m}
  et~al.}(2009)\citenamefont{{Bergstr{\"o}m}, {Edsj{\"o}}, and
  {Zaharijas}}}]{2009PhRvL.103c1103B}
\bibinfo{author}{\bibfnamefont{L.}~\bibnamefont{{Bergstr{\"o}m}}},
  \bibinfo{author}{\bibfnamefont{J.}~\bibnamefont{{Edsj{\"o}}}},
  \bibnamefont{and}
  \bibinfo{author}{\bibfnamefont{G.}~\bibnamefont{{Zaharijas}}},
  \bibinfo{journal}{Physical Review Letters} \textbf{\bibinfo{volume}{103}},
  \bibinfo{eid}{031103} (\bibinfo{year}{2009}), \eprint{0905.0333}.

\bibitem[{\citenamefont{{Feng} et~al.}(2010)\citenamefont{{Feng}, {Kaplinghat},
  and {Yu}}}]{2010PhRvD..82h3525F}
\bibinfo{author}{\bibfnamefont{J.~L.} \bibnamefont{{Feng}}},
  \bibinfo{author}{\bibfnamefont{M.}~\bibnamefont{{Kaplinghat}}},
  \bibnamefont{and} \bibinfo{author}{\bibfnamefont{H.-B.} \bibnamefont{{Yu}}},
  \bibinfo{journal}{\prd} \textbf{\bibinfo{volume}{82}}, \bibinfo{eid}{083525}
  (\bibinfo{year}{2010}), \eprint{1005.4678}.

\bibitem[{\citenamefont{{Zaharijas} et~al.}(2010)\citenamefont{{Zaharijas},
  {Cuoco}, {Yang}, {Conrad}, and {for the Fermi-LAT
  collaboration}}}]{2010arXiv1012.0588Z}
\bibinfo{author}{\bibfnamefont{G.}~\bibnamefont{{Zaharijas}}},
  \bibinfo{author}{\bibfnamefont{A.}~\bibnamefont{{Cuoco}}},
  \bibinfo{author}{\bibfnamefont{Z.}~\bibnamefont{{Yang}}},
  \bibinfo{author}{\bibfnamefont{J.}~\bibnamefont{{Conrad}}}, \bibnamefont{and}
  \bibinfo{author}{\bibnamefont{{for the Fermi-LAT collaboration}}},
  \bibinfo{journal}{ArXiv e-prints}  (\bibinfo{year}{2010}),
  \eprint{1012.0588}.

\bibitem[{\citenamefont{{Zhao} et~al.}(2016)\citenamefont{{Zhao}, {Bi}, {Jia},
  {Yin}, and {Zhu}}}]{2016PhRvD..93h3513Z}
\bibinfo{author}{\bibfnamefont{Y.}~\bibnamefont{{Zhao}}},
  \bibinfo{author}{\bibfnamefont{X.-J.} \bibnamefont{{Bi}}},
  \bibinfo{author}{\bibfnamefont{H.-Y.} \bibnamefont{{Jia}}},
  \bibinfo{author}{\bibfnamefont{P.-F.} \bibnamefont{{Yin}}}, \bibnamefont{and}
  \bibinfo{author}{\bibfnamefont{F.-R.} \bibnamefont{{Zhu}}},
  \bibinfo{journal}{\prd} \textbf{\bibinfo{volume}{93}}, \bibinfo{eid}{083513}
  (\bibinfo{year}{2016}), \eprint{1601.02181}.

\bibitem[{\citenamefont{{Liu} et~al.}(2017)\citenamefont{{Liu}, {Bi}, {Lin},
  {Wang}, and {Yin}}}]{2017PhRvD..96b3006L}
\bibinfo{author}{\bibfnamefont{W.}~\bibnamefont{{Liu}}},
  \bibinfo{author}{\bibfnamefont{X.-J.} \bibnamefont{{Bi}}},
  \bibinfo{author}{\bibfnamefont{S.-J.} \bibnamefont{{Lin}}},
  \bibinfo{author}{\bibfnamefont{B.-B.} \bibnamefont{{Wang}}},
  \bibnamefont{and} \bibinfo{author}{\bibfnamefont{P.-F.} \bibnamefont{{Yin}}},
  \bibinfo{journal}{\prd} \textbf{\bibinfo{volume}{96}}, \bibinfo{eid}{023006}
  (\bibinfo{year}{2017}), \eprint{1611.09118}.

\bibitem[{\citenamefont{{Lu} et~al.}(2017)\citenamefont{{Lu}, {Wu}, {Zhang},
  and {Zhou}}}]{2017arXiv171100749L}
\bibinfo{author}{\bibfnamefont{B.-Q.} \bibnamefont{{Lu}}},
  \bibinfo{author}{\bibfnamefont{Y.-L.} \bibnamefont{{Wu}}},
  \bibinfo{author}{\bibfnamefont{W.-H.} \bibnamefont{{Zhang}}},
  \bibnamefont{and} \bibinfo{author}{\bibfnamefont{Y.-F.}
  \bibnamefont{{Zhou}}}, \bibinfo{journal}{ArXiv e-prints}
  (\bibinfo{year}{2017}), \eprint{1711.00749}.

\bibitem[{\citenamefont{{Zhao} et~al.}(2017)\citenamefont{{Zhao}, {Bi}, {Yin},
  and {Zhang}}}]{2017arXiv171104696Z}
\bibinfo{author}{\bibfnamefont{Y.}~\bibnamefont{{Zhao}}},
  \bibinfo{author}{\bibfnamefont{X.-J.} \bibnamefont{{Bi}}},
  \bibinfo{author}{\bibfnamefont{P.-F.} \bibnamefont{{Yin}}}, \bibnamefont{and}
  \bibinfo{author}{\bibfnamefont{X.}~\bibnamefont{{Zhang}}},
  \bibinfo{journal}{ArXiv e-prints}  (\bibinfo{year}{2017}),
  \eprint{1711.04696}.

\bibitem[{\citenamefont{{Lin} et~al.}(2015)\citenamefont{{Lin}, {Yuan}, and
  {Bi}}}]{2015PhRvD..91f3508L}
\bibinfo{author}{\bibfnamefont{S.-J.} \bibnamefont{{Lin}}},
  \bibinfo{author}{\bibfnamefont{Q.}~\bibnamefont{{Yuan}}}, \bibnamefont{and}
  \bibinfo{author}{\bibfnamefont{X.-J.} \bibnamefont{{Bi}}},
  \bibinfo{journal}{\prd} \textbf{\bibinfo{volume}{91}}, \bibinfo{eid}{063508}
  (\bibinfo{year}{2015}), \eprint{1409.6248}.

\bibitem[{\citenamefont{{Malyshev} et~al.}(2009)\citenamefont{{Malyshev},
  {Cholis}, and {Gelfand}}}]{2009PhRvD..80f3005M}
\bibinfo{author}{\bibfnamefont{D.}~\bibnamefont{{Malyshev}}},
  \bibinfo{author}{\bibfnamefont{I.}~\bibnamefont{{Cholis}}}, \bibnamefont{and}
  \bibinfo{author}{\bibfnamefont{J.}~\bibnamefont{{Gelfand}}},
  \bibinfo{journal}{\prd} \textbf{\bibinfo{volume}{80}}, \bibinfo{eid}{063005}
  (\bibinfo{year}{2009}), \eprint{0903.1310}.

\bibitem[{\citenamefont{{Hu} et~al.}(2009)\citenamefont{{Hu}, {Yuan}, {Wang},
  {Fan}, {Zhang}, and {Bi}}}]{2009ApJ...700L.170H}
\bibinfo{author}{\bibfnamefont{H.-B.} \bibnamefont{{Hu}}},
  \bibinfo{author}{\bibfnamefont{Q.}~\bibnamefont{{Yuan}}},
  \bibinfo{author}{\bibfnamefont{B.}~\bibnamefont{{Wang}}},
  \bibinfo{author}{\bibfnamefont{C.}~\bibnamefont{{Fan}}},
  \bibinfo{author}{\bibfnamefont{J.-L.} \bibnamefont{{Zhang}}},
  \bibnamefont{and} \bibinfo{author}{\bibfnamefont{X.-J.} \bibnamefont{{Bi}}},
  \bibinfo{journal}{\apjl} \textbf{\bibinfo{volume}{700}},
  \bibinfo{pages}{L170} (\bibinfo{year}{2009}), \eprint{0901.1520}.

\bibitem[{\citenamefont{{Cholis} and {Hooper}}(2013)}]{2013PhRvD..88b3013C}
\bibinfo{author}{\bibfnamefont{I.}~\bibnamefont{{Cholis}}} \bibnamefont{and}
  \bibinfo{author}{\bibfnamefont{D.}~\bibnamefont{{Hooper}}},
  \bibinfo{journal}{\prd} \textbf{\bibinfo{volume}{88}}, \bibinfo{eid}{023013}
  (\bibinfo{year}{2013}), \eprint{1304.1840}.

\bibitem[{\citenamefont{{Blandford} et~al.}(2014)\citenamefont{{Blandford},
  {Simeon}, and {Yuan}}}]{2014NuPhS.256....9B}
\bibinfo{author}{\bibfnamefont{R.}~\bibnamefont{{Blandford}}},
  \bibinfo{author}{\bibfnamefont{P.}~\bibnamefont{{Simeon}}}, \bibnamefont{and}
  \bibinfo{author}{\bibfnamefont{Y.}~\bibnamefont{{Yuan}}},
  \bibinfo{journal}{Nuclear Physics B Proceedings Supplements}
  \textbf{\bibinfo{volume}{256}}, \bibinfo{pages}{9} (\bibinfo{year}{2014}),
  \eprint{1409.2589}.

\bibitem[{\citenamefont{{Fujita} et~al.}(2009)\citenamefont{{Fujita}, {Kohri},
  {Yamazaki}, and {Ioka}}}]{2009PhRvD..80f3003F}
\bibinfo{author}{\bibfnamefont{Y.}~\bibnamefont{{Fujita}}},
  \bibinfo{author}{\bibfnamefont{K.}~\bibnamefont{{Kohri}}},
  \bibinfo{author}{\bibfnamefont{R.}~\bibnamefont{{Yamazaki}}},
  \bibnamefont{and} \bibinfo{author}{\bibfnamefont{K.}~\bibnamefont{{Ioka}}},
  \bibinfo{journal}{\prd} \textbf{\bibinfo{volume}{80}}, \bibinfo{eid}{063003}
  (\bibinfo{year}{2009}), \eprint{0903.5298}.

\bibitem[{\citenamefont{{Kohri} et~al.}(2016)\citenamefont{{Kohri}, {Ioka},
  {Fujita}, and {Yamazaki}}}]{2016PTEP.2016b1E01K}
\bibinfo{author}{\bibfnamefont{K.}~\bibnamefont{{Kohri}}},
  \bibinfo{author}{\bibfnamefont{K.}~\bibnamefont{{Ioka}}},
  \bibinfo{author}{\bibfnamefont{Y.}~\bibnamefont{{Fujita}}}, \bibnamefont{and}
  \bibinfo{author}{\bibfnamefont{R.}~\bibnamefont{{Yamazaki}}},
  \bibinfo{journal}{Progress of Theoretical and Experimental Physics}
  \textbf{\bibinfo{volume}{2016}}, \bibinfo{eid}{021E01}
  (\bibinfo{year}{2016}), \eprint{1505.01236}.

\bibitem[{\citenamefont{{Blasi} and {Serpico}}(2009)}]{Blasi2009a}
\bibinfo{author}{\bibfnamefont{P.}~\bibnamefont{{Blasi}}} \bibnamefont{and}
  \bibinfo{author}{\bibfnamefont{P.~D.} \bibnamefont{{Serpico}}},
  \bibinfo{journal}{Physical Review Letters} \textbf{\bibinfo{volume}{103}},
  \bibinfo{eid}{081103} (\bibinfo{year}{2009}), \eprint{0904.0871}.

\bibitem[{\citenamefont{{Blasi}}(2009)}]{Blasi2009b}
\bibinfo{author}{\bibfnamefont{P.}~\bibnamefont{{Blasi}}},
  \bibinfo{journal}{Physical Review Letters} \textbf{\bibinfo{volume}{103}},
  \bibinfo{eid}{051104} (\bibinfo{year}{2009}), \eprint{0903.2794}.

\bibitem[{\citenamefont{{Serpico}}(2012)}]{Serpico2012}
\bibinfo{author}{\bibfnamefont{P.~D.} \bibnamefont{{Serpico}}},
  \bibinfo{journal}{Astroparticle Physics} \textbf{\bibinfo{volume}{39}},
  \bibinfo{pages}{2} (\bibinfo{year}{2012}), \eprint{1108.4827}.

\bibitem[{\citenamefont{{Mertsch} and {Sarkar}}(2009)}]{Mertsch2009}
\bibinfo{author}{\bibfnamefont{P.}~\bibnamefont{{Mertsch}}} \bibnamefont{and}
  \bibinfo{author}{\bibfnamefont{S.}~\bibnamefont{{Sarkar}}},
  \bibinfo{journal}{Physical Review Letters} \textbf{\bibinfo{volume}{103}},
  \bibinfo{eid}{081104} (\bibinfo{year}{2009}), \eprint{0905.3152}.

\bibitem[{\citenamefont{{Mertsch} and {Sarkar}}(2011)}]{Mertsch2011}
\bibinfo{author}{\bibfnamefont{P.}~\bibnamefont{{Mertsch}}} \bibnamefont{and}
  \bibinfo{author}{\bibfnamefont{S.}~\bibnamefont{{Sarkar}}},
  \bibinfo{journal}{Physical Review Letters} \textbf{\bibinfo{volume}{107}},
  \bibinfo{eid}{091101} (\bibinfo{year}{2011}), \eprint{1104.3585}.

\bibitem[{\citenamefont{{Donato} et~al.}(2014)\citenamefont{{Donato}, {Cenko},
  {Covino}, {Troja}, {Pursimo}, {Cheung}, {Fox}, {Kutyrev}, {Campana},
  {Fugazza} et~al.}}]{Donato2014}
\bibinfo{author}{\bibfnamefont{D.}~\bibnamefont{{Donato}}},
  \bibinfo{author}{\bibfnamefont{S.~B.} \bibnamefont{{Cenko}}},
  \bibinfo{author}{\bibfnamefont{S.}~\bibnamefont{{Covino}}},
  \bibinfo{author}{\bibfnamefont{E.}~\bibnamefont{{Troja}}},
  \bibinfo{author}{\bibfnamefont{T.}~\bibnamefont{{Pursimo}}},
  \bibinfo{author}{\bibfnamefont{C.~C.} \bibnamefont{{Cheung}}},
  \bibinfo{author}{\bibfnamefont{O.}~\bibnamefont{{Fox}}},
  \bibinfo{author}{\bibfnamefont{A.}~\bibnamefont{{Kutyrev}}},
  \bibinfo{author}{\bibfnamefont{S.}~\bibnamefont{{Campana}}},
  \bibinfo{author}{\bibfnamefont{D.}~\bibnamefont{{Fugazza}}},
  \bibnamefont{et~al.}, \bibinfo{journal}{\apj} \textbf{\bibinfo{volume}{781}},
  \bibinfo{eid}{59} (\bibinfo{year}{2014}), \eprint{1311.6162}.

\bibitem[{\citenamefont{{Mertsch} and {Sarkar}}(2014)}]{Mertsch2014}
\bibinfo{author}{\bibfnamefont{P.}~\bibnamefont{{Mertsch}}} \bibnamefont{and}
  \bibinfo{author}{\bibfnamefont{S.}~\bibnamefont{{Sarkar}}},
  \bibinfo{journal}{\prd} \textbf{\bibinfo{volume}{90}}, \bibinfo{eid}{061301}
  (\bibinfo{year}{2014}), \eprint{1402.0855}.

\bibitem[{\citenamefont{{Tomassetti} and {Donato}}(2012)}]{Tomassetti2012}
\bibinfo{author}{\bibfnamefont{N.}~\bibnamefont{{Tomassetti}}}
  \bibnamefont{and} \bibinfo{author}{\bibfnamefont{F.}~\bibnamefont{{Donato}}},
  \bibinfo{journal}{\aap} \textbf{\bibinfo{volume}{544}}, \bibinfo{eid}{A16}
  (\bibinfo{year}{2012}), \eprint{1203.6094}.

\bibitem[{\citenamefont{{Tomassetti} and {Donato}}(2015)}]{Tomassetti2015}
\bibinfo{author}{\bibfnamefont{N.}~\bibnamefont{{Tomassetti}}}
  \bibnamefont{and} \bibinfo{author}{\bibfnamefont{F.}~\bibnamefont{{Donato}}},
  \bibinfo{journal}{\apjl} \textbf{\bibinfo{volume}{803}}, \bibinfo{eid}{L15}
  (\bibinfo{year}{2015}), \eprint{1502.06150}.

\bibitem[{\citenamefont{{Chang} et~al.}(2017)\citenamefont{{Chang}, {Ambrosi},
  {An}, {Asfandiyarov}, {Azzarello}, {Bernardini}, {Bertucci}, {Cai},
  {Caragiulo}, {Chen} et~al.}}]{2017APh....95....6C}
\bibinfo{author}{\bibfnamefont{J.}~\bibnamefont{{Chang}}},
  \bibinfo{author}{\bibfnamefont{G.}~\bibnamefont{{Ambrosi}}},
  \bibinfo{author}{\bibfnamefont{Q.}~\bibnamefont{{An}}},
  \bibinfo{author}{\bibfnamefont{R.}~\bibnamefont{{Asfandiyarov}}},
  \bibinfo{author}{\bibfnamefont{P.}~\bibnamefont{{Azzarello}}},
  \bibinfo{author}{\bibfnamefont{P.}~\bibnamefont{{Bernardini}}},
  \bibinfo{author}{\bibfnamefont{B.}~\bibnamefont{{Bertucci}}},
  \bibinfo{author}{\bibfnamefont{M.~S.} \bibnamefont{{Cai}}},
  \bibinfo{author}{\bibfnamefont{M.}~\bibnamefont{{Caragiulo}}},
  \bibinfo{author}{\bibfnamefont{D.~Y.} \bibnamefont{{Chen}}},
  \bibnamefont{et~al.}, \bibinfo{journal}{Astroparticle Physics}
  \textbf{\bibinfo{volume}{95}}, \bibinfo{pages}{6} (\bibinfo{year}{2017}),
  \eprint{1706.08453}.

\bibitem[{\citenamefont{{DAMPE collaboration}}(2017)}]{DAMPEpaper}
\bibinfo{author}{\bibnamefont{{DAMPE collaboration}}},
  \bibinfo{journal}{Nature} \textbf{\bibinfo{volume}{551}}
  (\bibinfo{year}{2017}), \eprint{arxiv:1711.10981}.

\bibitem[{\citenamefont{Collaboration et~al.}(2017)\citenamefont{Collaboration,
  :, Abdollahi, Ackermann, Ajello, Atwood, Baldini, Barbiellini, Bastieri,
  Bellazzini et~al.}}]{Abdollahi:2017nat}
\bibinfo{author}{\bibfnamefont{F.-L.} \bibnamefont{Collaboration}},
  \bibinfo{author}{\bibnamefont{:}},
  \bibinfo{author}{\bibfnamefont{S.}~\bibnamefont{Abdollahi}},
  \bibinfo{author}{\bibfnamefont{M.}~\bibnamefont{Ackermann}},
  \bibinfo{author}{\bibfnamefont{M.}~\bibnamefont{Ajello}},
  \bibinfo{author}{\bibfnamefont{W.~B.} \bibnamefont{Atwood}},
  \bibinfo{author}{\bibfnamefont{L.}~\bibnamefont{Baldini}},
  \bibinfo{author}{\bibfnamefont{G.}~\bibnamefont{Barbiellini}},
  \bibinfo{author}{\bibfnamefont{D.}~\bibnamefont{Bastieri}},
  \bibinfo{author}{\bibfnamefont{R.}~\bibnamefont{Bellazzini}},
  \bibnamefont{et~al.}, \bibinfo{journal}{Physical Review D}
  \textbf{\bibinfo{volume}{95}}, \bibinfo{pages}{082007}
  (\bibinfo{year}{2017}), ISSN \bibinfo{issn}{2470-0010}, \eprint{1704.07195}.

\bibitem[{\citenamefont{{Adriani} et~al.}(2017)\citenamefont{{Adriani},
  {Akaike}, {Asano}, {Asaoka}, {Bagliesi}, {Bigongiari}, {Binns}, {Bonechi},
  {Bongi}, {Brogi} et~al.}}]{CALET2017}
\bibinfo{author}{\bibfnamefont{O.}~\bibnamefont{{Adriani}}},
  \bibinfo{author}{\bibfnamefont{Y.}~\bibnamefont{{Akaike}}},
  \bibinfo{author}{\bibfnamefont{K.}~\bibnamefont{{Asano}}},
  \bibinfo{author}{\bibfnamefont{Y.}~\bibnamefont{{Asaoka}}},
  \bibinfo{author}{\bibfnamefont{M.~G.} \bibnamefont{{Bagliesi}}},
  \bibinfo{author}{\bibfnamefont{G.}~\bibnamefont{{Bigongiari}}},
  \bibinfo{author}{\bibfnamefont{W.~R.} \bibnamefont{{Binns}}},
  \bibinfo{author}{\bibfnamefont{S.}~\bibnamefont{{Bonechi}}},
  \bibinfo{author}{\bibfnamefont{M.}~\bibnamefont{{Bongi}}},
  \bibinfo{author}{\bibfnamefont{P.}~\bibnamefont{{Brogi}}},
  \bibnamefont{et~al.}, \bibinfo{journal}{Physical Review Letters}
  \textbf{\bibinfo{volume}{119}}, \bibinfo{eid}{181101} (\bibinfo{year}{2017}),
  \eprint{1712.01711}.

\bibitem[{\citenamefont{Berezinskii et~al.}(1990)\citenamefont{Berezinskii,
  Bulanov, Dogiel, and Ginzburg}}]{Ginzburg:1990sk}
\bibinfo{author}{\bibfnamefont{V.~S.} \bibnamefont{Berezinskii}},
  \bibinfo{author}{\bibfnamefont{S.~V.} \bibnamefont{Bulanov}},
  \bibinfo{author}{\bibfnamefont{V.~A.} \bibnamefont{Dogiel}},
  \bibnamefont{and} \bibinfo{author}{\bibfnamefont{V.~L.}
  \bibnamefont{Ginzburg}}, \emph{\bibinfo{title}{{Astrophysics of cosmic
  rays}}} (\bibinfo{publisher}{North-Holland}, \bibinfo{address}{Amsterdam},
  \bibinfo{year}{1990}), ISBN \bibinfo{isbn}{0444886419}.

\bibitem[{\citenamefont{Strong and Moskalenko}(1998)}]{Strong:1998pw}
\bibinfo{author}{\bibfnamefont{A.~W.} \bibnamefont{Strong}} \bibnamefont{and}
  \bibinfo{author}{\bibfnamefont{I.~V.} \bibnamefont{Moskalenko}},
  \bibinfo{journal}{The Astrophysical Journal} \textbf{\bibinfo{volume}{509}},
  \bibinfo{pages}{212} (\bibinfo{year}{1998}), ISSN \bibinfo{issn}{0004-637X},
  \eprint{9807150}.

\bibitem[{\citenamefont{Blandford and Eichler}(1987)}]{Blandford:1987pw}
\bibinfo{author}{\bibfnamefont{R.}~\bibnamefont{Blandford}} \bibnamefont{and}
  \bibinfo{author}{\bibfnamefont{D.}~\bibnamefont{Eichler}},
  \bibinfo{journal}{Physics Reports} \textbf{\bibinfo{volume}{154}},
  \bibinfo{pages}{1} (\bibinfo{year}{1987}), ISSN \bibinfo{issn}{03701573}.

\bibitem[{\citenamefont{{Vladimirov} et~al.}(2011)\citenamefont{{Vladimirov},
  {Digel}, {J{\'o}hannesson}, {Michelson}, {Moskalenko}, {Nolan}, {Orlando},
  {Porter}, and {Strong}}}]{Vladimirov2011_GALPROP}
\bibinfo{author}{\bibfnamefont{A.~E.} \bibnamefont{{Vladimirov}}},
  \bibinfo{author}{\bibfnamefont{S.~W.} \bibnamefont{{Digel}}},
  \bibinfo{author}{\bibfnamefont{G.}~\bibnamefont{{J{\'o}hannesson}}},
  \bibinfo{author}{\bibfnamefont{P.~F.} \bibnamefont{{Michelson}}},
  \bibinfo{author}{\bibfnamefont{I.~V.} \bibnamefont{{Moskalenko}}},
  \bibinfo{author}{\bibfnamefont{P.~L.} \bibnamefont{{Nolan}}},
  \bibinfo{author}{\bibfnamefont{E.}~\bibnamefont{{Orlando}}},
  \bibinfo{author}{\bibfnamefont{T.~A.} \bibnamefont{{Porter}}},
  \bibnamefont{and} \bibinfo{author}{\bibfnamefont{A.~W.}
  \bibnamefont{{Strong}}}, \bibinfo{journal}{Computer Physics Communications}
  \textbf{\bibinfo{volume}{182}}, \bibinfo{pages}{1156} (\bibinfo{year}{2011}),
  \eprint{1008.3642}.

\bibitem[{\citenamefont{Moskalenko et~al.}(2001)\citenamefont{Moskalenko,
  Strong, Ormes, and Potgieter}}]{Moskalenko:2001ya}
\bibinfo{author}{\bibfnamefont{I.~V.} \bibnamefont{Moskalenko}},
  \bibinfo{author}{\bibfnamefont{A.~W.} \bibnamefont{Strong}},
  \bibinfo{author}{\bibfnamefont{J.~F.} \bibnamefont{Ormes}}, \bibnamefont{and}
  \bibinfo{author}{\bibfnamefont{M.~S.} \bibnamefont{Potgieter}},
  \bibinfo{journal}{The Astrophysical Journal} \textbf{\bibinfo{volume}{565}},
  \bibinfo{pages}{280} (\bibinfo{year}{2001}), ISSN \bibinfo{issn}{0004-637X},
  \eprint{0106567}.

\bibitem[{\citenamefont{Strong and Mattox}(1996)}]{1996A&A...308L..21S}
\bibinfo{author}{\bibfnamefont{A.}~\bibnamefont{Strong}} \bibnamefont{and}
  \bibinfo{author}{\bibfnamefont{J.}~\bibnamefont{Mattox}},
  \bibinfo{journal}{Astron.Astrophys.} \textbf{\bibinfo{volume}{308}},
  \bibinfo{pages}{L21} (\bibinfo{year}{1996}).

\bibitem[{\citenamefont{Aguilar et~al.}(2015)\citenamefont{Aguilar, Aisa,
  Alpat, Alvino, Ambrosi, Andeen, Arruda, Attig, Azzarello, Bachlechner
  et~al.}}]{Aguilar:2015ooa}
\bibinfo{author}{\bibfnamefont{M.}~\bibnamefont{Aguilar}},
  \bibinfo{author}{\bibfnamefont{D.}~\bibnamefont{Aisa}},
  \bibinfo{author}{\bibfnamefont{B.}~\bibnamefont{Alpat}},
  \bibinfo{author}{\bibfnamefont{A.}~\bibnamefont{Alvino}},
  \bibinfo{author}{\bibfnamefont{G.}~\bibnamefont{Ambrosi}},
  \bibinfo{author}{\bibfnamefont{K.}~\bibnamefont{Andeen}},
  \bibinfo{author}{\bibfnamefont{L.}~\bibnamefont{Arruda}},
  \bibinfo{author}{\bibfnamefont{N.}~\bibnamefont{Attig}},
  \bibinfo{author}{\bibfnamefont{P.}~\bibnamefont{Azzarello}},
  \bibinfo{author}{\bibfnamefont{A.}~\bibnamefont{Bachlechner}},
  \bibnamefont{et~al.}, \bibinfo{journal}{Physical Review Letters}
  \textbf{\bibinfo{volume}{114}}, \bibinfo{pages}{171103}
  (\bibinfo{year}{2015}), ISSN \bibinfo{issn}{0031-9007}.

\bibitem[{\citenamefont{Aguilar et~al.}(2016)\citenamefont{Aguilar, {Ali
  Cavasonza}, Ambrosi, Arruda, Attig, Aupetit, Azzarello, Bachlechner, Barao,
  Barrau et~al.}}]{Aguilar:2016vqr}
\bibinfo{author}{\bibfnamefont{M.}~\bibnamefont{Aguilar}},
  \bibinfo{author}{\bibfnamefont{L.}~\bibnamefont{{Ali Cavasonza}}},
  \bibinfo{author}{\bibfnamefont{G.}~\bibnamefont{Ambrosi}},
  \bibinfo{author}{\bibfnamefont{L.}~\bibnamefont{Arruda}},
  \bibinfo{author}{\bibfnamefont{N.}~\bibnamefont{Attig}},
  \bibinfo{author}{\bibfnamefont{S.}~\bibnamefont{Aupetit}},
  \bibinfo{author}{\bibfnamefont{P.}~\bibnamefont{Azzarello}},
  \bibinfo{author}{\bibfnamefont{A.}~\bibnamefont{Bachlechner}},
  \bibinfo{author}{\bibfnamefont{F.}~\bibnamefont{Barao}},
  \bibinfo{author}{\bibfnamefont{A.}~\bibnamefont{Barrau}},
  \bibnamefont{et~al.}, \bibinfo{journal}{Physical Review Letters}
  \textbf{\bibinfo{volume}{117}}, \bibinfo{pages}{231102}
  (\bibinfo{year}{2016}), ISSN \bibinfo{issn}{0031-9007}.

\bibitem[{\citenamefont{Jin et~al.}(2015)\citenamefont{Jin, Wu, and
  Zhou}}]{Jin:2014ica}
\bibinfo{author}{\bibfnamefont{H.-B.} \bibnamefont{Jin}},
  \bibinfo{author}{\bibfnamefont{Y.-L.} \bibnamefont{Wu}}, \bibnamefont{and}
  \bibinfo{author}{\bibfnamefont{Y.-F.} \bibnamefont{Zhou}},
  \bibinfo{journal}{Journal of Cosmology and Astroparticle Physics}
  \textbf{\bibinfo{volume}{2015}}, \bibinfo{pages}{049} (\bibinfo{year}{2015}),
  ISSN \bibinfo{issn}{1475-7516}, \eprint{1410.0171}.

\bibitem[{\citenamefont{Navarro et~al.}(1996)\citenamefont{Navarro, Frenk, and
  White}}]{Navarro:1996gj}
\bibinfo{author}{\bibfnamefont{J.~F.} \bibnamefont{Navarro}},
  \bibinfo{author}{\bibfnamefont{C.~S.} \bibnamefont{Frenk}}, \bibnamefont{and}
  \bibinfo{author}{\bibfnamefont{S.~D.~M.} \bibnamefont{White}},
  \bibinfo{journal}{The Astrophysical Journal} \textbf{\bibinfo{volume}{490}},
  \bibinfo{pages}{493} (\bibinfo{year}{1996}), ISSN \bibinfo{issn}{0004-637X},
  \eprint{9611107}.

\bibitem[{\citenamefont{Bergstrom et~al.}(1997)\citenamefont{Bergstrom, Ullio,
  and Buckley}}]{Bergstrom:1997fj}
\bibinfo{author}{\bibfnamefont{L.}~\bibnamefont{Bergstrom}},
  \bibinfo{author}{\bibfnamefont{P.}~\bibnamefont{Ullio}}, \bibnamefont{and}
  \bibinfo{author}{\bibfnamefont{J.}~\bibnamefont{Buckley}},
  \bibinfo{journal}{Astroparticle Physics} \textbf{\bibinfo{volume}{9}},
  \bibinfo{pages}{44} (\bibinfo{year}{1997}), ISSN \bibinfo{issn}{09276505},
  \eprint{9712318}.

\bibitem[{\citenamefont{Moore et~al.}(1999)\citenamefont{Moore, Ghigna,
  Governato, Lake, Quinn, Stadel, and Tozzi}}]{Moore:1999nt}
\bibinfo{author}{\bibfnamefont{B.}~\bibnamefont{Moore}},
  \bibinfo{author}{\bibfnamefont{S.}~\bibnamefont{Ghigna}},
  \bibinfo{author}{\bibfnamefont{F.}~\bibnamefont{Governato}},
  \bibinfo{author}{\bibfnamefont{G.}~\bibnamefont{Lake}},
  \bibinfo{author}{\bibfnamefont{T.}~\bibnamefont{Quinn}},
  \bibinfo{author}{\bibfnamefont{J.}~\bibnamefont{Stadel}}, \bibnamefont{and}
  \bibinfo{author}{\bibfnamefont{P.}~\bibnamefont{Tozzi}},
  \bibinfo{journal}{The Astrophysical Journal} \textbf{\bibinfo{volume}{524}},
  \bibinfo{pages}{4} (\bibinfo{year}{1999}), ISSN \bibinfo{issn}{0004637X},
  \eprint{9907411}.

\bibitem[{\citenamefont{Einasto}(2009)}]{Einasto:2009zd}
\bibinfo{author}{\bibfnamefont{J.}~\bibnamefont{Einasto}}, in
  \emph{\bibinfo{booktitle}{Astronomy and Astrophysics 2010}}
  (\bibinfo{year}{2009}), p.~\bibinfo{pages}{31}, \eprint{0901.0632}.

\bibitem[{\citenamefont{Sj{\"{o}}strand
  et~al.}(2008)\citenamefont{Sj{\"{o}}strand, Mrenna, and
  Skands}}]{Sjostrand:2007gs}
\bibinfo{author}{\bibfnamefont{T.}~\bibnamefont{Sj{\"{o}}strand}},
  \bibinfo{author}{\bibfnamefont{S.}~\bibnamefont{Mrenna}}, \bibnamefont{and}
  \bibinfo{author}{\bibfnamefont{P.}~\bibnamefont{Skands}},
  \bibinfo{journal}{Computer Physics Communications}
  \textbf{\bibinfo{volume}{178}}, \bibinfo{pages}{852} (\bibinfo{year}{2008}),
  ISSN \bibinfo{issn}{00104655}, \eprint{0710.3820}.

\bibitem[{\citenamefont{Jin et~al.}(2017)\citenamefont{Jin, Wu, and
  Zhou}}]{Jin:2017iwg}
\bibinfo{author}{\bibfnamefont{H.-B.} \bibnamefont{Jin}},
  \bibinfo{author}{\bibfnamefont{Y.-L.} \bibnamefont{Wu}}, \bibnamefont{and}
  \bibinfo{author}{\bibfnamefont{Y.-F.} \bibnamefont{Zhou}}
  (\bibinfo{year}{2017}), \eprint{1701.02213}.

\bibitem[{\citenamefont{Aguilar et~al.}(2014)\citenamefont{Aguilar, Aisa,
  Alvino, Ambrosi, Andeen, Arruda, Attig, Azzarello, Bachlechner, Barao
  et~al.}}]{Aguilar:2014mma}
\bibinfo{author}{\bibfnamefont{M.}~\bibnamefont{Aguilar}},
  \bibinfo{author}{\bibfnamefont{D.}~\bibnamefont{Aisa}},
  \bibinfo{author}{\bibfnamefont{A.}~\bibnamefont{Alvino}},
  \bibinfo{author}{\bibfnamefont{G.}~\bibnamefont{Ambrosi}},
  \bibinfo{author}{\bibfnamefont{K.}~\bibnamefont{Andeen}},
  \bibinfo{author}{\bibfnamefont{L.}~\bibnamefont{Arruda}},
  \bibinfo{author}{\bibfnamefont{N.}~\bibnamefont{Attig}},
  \bibinfo{author}{\bibfnamefont{P.}~\bibnamefont{Azzarello}},
  \bibinfo{author}{\bibfnamefont{A.}~\bibnamefont{Bachlechner}},
  \bibinfo{author}{\bibfnamefont{F.}~\bibnamefont{Barao}},
  \bibnamefont{et~al.}, \bibinfo{journal}{Physical Review Letters}
  \textbf{\bibinfo{volume}{113}}, \bibinfo{pages}{121102}
  (\bibinfo{year}{2014}), ISSN \bibinfo{issn}{0031-9007}.

\bibitem[{\citenamefont{Accardo et~al.}(2014)\citenamefont{Accardo, Aguilar,
  Aisa, Alpat, Alvino, Ambrosi, Andeen, Arruda, Attig, Azzarello
  et~al.}}]{Accardo:2014lma}
\bibinfo{author}{\bibfnamefont{L.}~\bibnamefont{Accardo}},
  \bibinfo{author}{\bibfnamefont{M.}~\bibnamefont{Aguilar}},
  \bibinfo{author}{\bibfnamefont{D.}~\bibnamefont{Aisa}},
  \bibinfo{author}{\bibfnamefont{B.}~\bibnamefont{Alpat}},
  \bibinfo{author}{\bibfnamefont{A.}~\bibnamefont{Alvino}},
  \bibinfo{author}{\bibfnamefont{G.}~\bibnamefont{Ambrosi}},
  \bibinfo{author}{\bibfnamefont{K.}~\bibnamefont{Andeen}},
  \bibinfo{author}{\bibfnamefont{L.}~\bibnamefont{Arruda}},
  \bibinfo{author}{\bibfnamefont{N.}~\bibnamefont{Attig}},
  \bibinfo{author}{\bibfnamefont{P.}~\bibnamefont{Azzarello}},
  \bibnamefont{et~al.}, \bibinfo{journal}{Physical Review Letters}
  \textbf{\bibinfo{volume}{113}}, \bibinfo{pages}{121101}
  (\bibinfo{year}{2014}), ISSN \bibinfo{issn}{0031-9007}.

\bibitem[{\citenamefont{Jin et~al.}(2013)\citenamefont{Jin, Wu, and
  Zhou}}]{Jin:2013nta}
\bibinfo{author}{\bibfnamefont{H.-B.} \bibnamefont{Jin}},
  \bibinfo{author}{\bibfnamefont{Y.-L.} \bibnamefont{Wu}}, \bibnamefont{and}
  \bibinfo{author}{\bibfnamefont{Y.-F.} \bibnamefont{Zhou}},
  \bibinfo{journal}{Journal of Cosmology and Astroparticle Physics}
  \textbf{\bibinfo{volume}{2013}}, \bibinfo{pages}{026} (\bibinfo{year}{2013}),
  ISSN \bibinfo{issn}{1475-7516}, \eprint{1304.1997}.

\bibitem[{\citenamefont{Liu et~al.}(2013)\citenamefont{Liu, Wu, and
  Zhou}}]{Liu:2013vha}
\bibinfo{author}{\bibfnamefont{Z.-P.} \bibnamefont{Liu}},
  \bibinfo{author}{\bibfnamefont{Y.-L.} \bibnamefont{Wu}}, \bibnamefont{and}
  \bibinfo{author}{\bibfnamefont{Y.-F.} \bibnamefont{Zhou}},
  \bibinfo{journal}{Physical Review D} \textbf{\bibinfo{volume}{88}},
  \bibinfo{pages}{096008} (\bibinfo{year}{2013}), ISSN
  \bibinfo{issn}{1550-7998}, \eprint{1305.5438}.

\bibitem[{\citenamefont{Chen et~al.}(2015{\natexlab{a}})\citenamefont{Chen,
  Liang, Wu, and Zhou}}]{Chen:2015uha}
\bibinfo{author}{\bibfnamefont{J.}~\bibnamefont{Chen}},
  \bibinfo{author}{\bibfnamefont{Z.-L.} \bibnamefont{Liang}},
  \bibinfo{author}{\bibfnamefont{Y.-L.} \bibnamefont{Wu}}, \bibnamefont{and}
  \bibinfo{author}{\bibfnamefont{Y.-F.} \bibnamefont{Zhou}},
  \bibinfo{journal}{Journal of Cosmology and Astroparticle Physics}
  \textbf{\bibinfo{volume}{2015}}, \bibinfo{pages}{021}
  (\bibinfo{year}{2015}{\natexlab{a}}), ISSN \bibinfo{issn}{1475-7516},
  \eprint{1505.04031}.

\bibitem[{\citenamefont{Zhou}(2016)}]{Zhou:2016eul}
\bibinfo{author}{\bibfnamefont{Y.-F.} \bibnamefont{Zhou}},
  \bibinfo{journal}{PoS} \textbf{\bibinfo{volume}{DSU2015}},
  \bibinfo{pages}{22} (\bibinfo{year}{2016}).

\bibitem[{\citenamefont{Chen et~al.}(2015{\natexlab{b}})\citenamefont{Chen,
  Huang, and Jin}}]{Chen:2014nea}
\bibinfo{author}{\bibfnamefont{D.}~\bibnamefont{Chen}},
  \bibinfo{author}{\bibfnamefont{J.}~\bibnamefont{Huang}}, \bibnamefont{and}
  \bibinfo{author}{\bibfnamefont{H.-B.} \bibnamefont{Jin}},
  \bibinfo{journal}{The Astrophysical Journal} \textbf{\bibinfo{volume}{811}},
  \bibinfo{pages}{154} (\bibinfo{year}{2015}{\natexlab{b}}), ISSN
  \bibinfo{issn}{1538-4357}, \eprint{1412.2499}.

\bibitem[{\citenamefont{Salucci et~al.}(2010)\citenamefont{Salucci, Nesti,
  Gentile, and Martins}}]{Salucci:2010qr}
\bibinfo{author}{\bibfnamefont{P.}~\bibnamefont{Salucci}},
  \bibinfo{author}{\bibfnamefont{F.}~\bibnamefont{Nesti}},
  \bibinfo{author}{\bibfnamefont{G.}~\bibnamefont{Gentile}}, \bibnamefont{and}
  \bibinfo{author}{\bibfnamefont{C.~F.} \bibnamefont{Martins}},
  \bibinfo{journal}{Astronomy {\&} Astrophysics}
  \textbf{\bibinfo{volume}{523}}, \bibinfo{pages}{6} (\bibinfo{year}{2010}),
  ISSN \bibinfo{issn}{0004-6361}, \eprint{1003.3101}.

\bibitem[{\citenamefont{Strong et~al.}(2000)\citenamefont{Strong, Moskalenko,
  and Reimer}}]{Strong:1998fr}
\bibinfo{author}{\bibfnamefont{A.~W.} \bibnamefont{Strong}},
  \bibinfo{author}{\bibfnamefont{I.~V.} \bibnamefont{Moskalenko}},
  \bibnamefont{and} \bibinfo{author}{\bibfnamefont{O.}~\bibnamefont{Reimer}},
  \bibinfo{journal}{The Astrophysical Journal} \textbf{\bibinfo{volume}{537}},
  \bibinfo{pages}{763} (\bibinfo{year}{2000}), ISSN \bibinfo{issn}{0004-637X},
  \eprint{9811296}.

\bibitem[{\citenamefont{{Moore} et~al.}(1999)\citenamefont{{Moore}, {Ghigna},
  {Governato}, {Lake}, {Quinn}, {Stadel}, and {Tozzi}}}]{1999ApJ...524L..19M}
\bibinfo{author}{\bibfnamefont{B.}~\bibnamefont{{Moore}}},
  \bibinfo{author}{\bibfnamefont{S.}~\bibnamefont{{Ghigna}}},
  \bibinfo{author}{\bibfnamefont{F.}~\bibnamefont{{Governato}}},
  \bibinfo{author}{\bibfnamefont{G.}~\bibnamefont{{Lake}}},
  \bibinfo{author}{\bibfnamefont{T.}~\bibnamefont{{Quinn}}},
  \bibinfo{author}{\bibfnamefont{J.}~\bibnamefont{{Stadel}}}, \bibnamefont{and}
  \bibinfo{author}{\bibfnamefont{P.}~\bibnamefont{{Tozzi}}},
  \bibinfo{journal}{\apjl} \textbf{\bibinfo{volume}{524}}, \bibinfo{pages}{L19}
  (\bibinfo{year}{1999}), \eprint{astro-ph/9907411}.

\bibitem[{\citenamefont{{Klypin} et~al.}(1999)\citenamefont{{Klypin},
  {Kravtsov}, {Valenzuela}, and {Prada}}}]{1999ApJ...522...82K}
\bibinfo{author}{\bibfnamefont{A.}~\bibnamefont{{Klypin}}},
  \bibinfo{author}{\bibfnamefont{A.~V.} \bibnamefont{{Kravtsov}}},
  \bibinfo{author}{\bibfnamefont{O.}~\bibnamefont{{Valenzuela}}},
  \bibnamefont{and} \bibinfo{author}{\bibfnamefont{F.}~\bibnamefont{{Prada}}},
  \bibinfo{journal}{\apj} \textbf{\bibinfo{volume}{522}}, \bibinfo{pages}{82}
  (\bibinfo{year}{1999}), \eprint{astro-ph/9901240}.

\bibitem[{\citenamefont{{Bullock} and
  {Boylan-Kolchin}}(2017)}]{2017ARA&A..55..343B}
\bibinfo{author}{\bibfnamefont{J.~S.} \bibnamefont{{Bullock}}}
  \bibnamefont{and}
  \bibinfo{author}{\bibfnamefont{M.}~\bibnamefont{{Boylan-Kolchin}}},
  \bibinfo{journal}{\araa} \textbf{\bibinfo{volume}{55}}, \bibinfo{pages}{343}
  (\bibinfo{year}{2017}), \eprint{1707.04256}.

\bibitem[{\citenamefont{{Trotta} et~al.}(2011)\citenamefont{{Trotta},
  {J{\'o}hannesson}, {Moskalenko}, {Porter}, {Ruiz de Austri}, and
  {Strong}}}]{Trotta2011}
\bibinfo{author}{\bibfnamefont{R.}~\bibnamefont{{Trotta}}},
  \bibinfo{author}{\bibfnamefont{G.}~\bibnamefont{{J{\'o}hannesson}}},
  \bibinfo{author}{\bibfnamefont{I.~V.} \bibnamefont{{Moskalenko}}},
  \bibinfo{author}{\bibfnamefont{T.~A.} \bibnamefont{{Porter}}},
  \bibinfo{author}{\bibfnamefont{R.}~\bibnamefont{{Ruiz de Austri}}},
  \bibnamefont{and} \bibinfo{author}{\bibfnamefont{A.~W.}
  \bibnamefont{{Strong}}}, \bibinfo{journal}{\apj}
  \textbf{\bibinfo{volume}{729}}, \bibinfo{eid}{106} (\bibinfo{year}{2011}),
  \eprint{1011.0037}.

\bibitem[{\citenamefont{{Haverkorn}}(2015)}]{Haverkorn2015}
\bibinfo{author}{\bibfnamefont{M.}~\bibnamefont{{Haverkorn}}}, in
  \emph{\bibinfo{booktitle}{Magnetic Fields in Diffuse Media}}, edited by
  \bibinfo{editor}{\bibfnamefont{A.}~\bibnamefont{{Lazarian}}},
  \bibinfo{editor}{\bibfnamefont{E.~M.} \bibnamefont{{de Gouveia Dal Pino}}},
  \bibnamefont{and} \bibinfo{editor}{\bibfnamefont{C.}~\bibnamefont{{Melioli}}}
  (\bibinfo{year}{2015}), vol. \bibinfo{volume}{407} of
  \emph{\bibinfo{series}{Astrophysics and Space Science Library}}, p.
  \bibinfo{pages}{483}, \eprint{1406.0283}.

\bibitem[{\citenamefont{{Lefa} et~al.}(2012)\citenamefont{{Lefa}, {Kelner}, and
  {Aharonian}}}]{Lefa2012}
\bibinfo{author}{\bibfnamefont{E.}~\bibnamefont{{Lefa}}},
  \bibinfo{author}{\bibfnamefont{S.~R.} \bibnamefont{{Kelner}}},
  \bibnamefont{and} \bibinfo{author}{\bibfnamefont{F.~A.}
  \bibnamefont{{Aharonian}}}, \bibinfo{journal}{\apj}
  \textbf{\bibinfo{volume}{753}}, \bibinfo{eid}{176} (\bibinfo{year}{2012}),
  \eprint{1205.2929}.

\bibitem[{\citenamefont{{Porter} and {Strong}}(2005)}]{Porter2005}
\bibinfo{author}{\bibfnamefont{T.~A.} \bibnamefont{{Porter}}} \bibnamefont{and}
  \bibinfo{author}{\bibfnamefont{A.~W.} \bibnamefont{{Strong}}},
  \bibinfo{journal}{International Cosmic Ray Conference}
  \textbf{\bibinfo{volume}{4}}, \bibinfo{pages}{77} (\bibinfo{year}{2005}),
  \eprint{astro-ph/0507119}.

\bibitem[{\citenamefont{{Porter} et~al.}(2006)\citenamefont{{Porter},
  {Moskalenko}, and {Strong}}}]{Porter2006}
\bibinfo{author}{\bibfnamefont{T.~A.} \bibnamefont{{Porter}}},
  \bibinfo{author}{\bibfnamefont{I.~V.} \bibnamefont{{Moskalenko}}},
  \bibnamefont{and} \bibinfo{author}{\bibfnamefont{A.~W.}
  \bibnamefont{{Strong}}}, \bibinfo{journal}{\apjl}
  \textbf{\bibinfo{volume}{648}}, \bibinfo{pages}{L29} (\bibinfo{year}{2006}),
  \eprint{astro-ph/0607344}.

\bibitem[{\citenamefont{Staszak and Collaboration}(2015)}]{Staszak:2015kza}
\bibinfo{author}{\bibfnamefont{D.}~\bibnamefont{Staszak}} \bibnamefont{and}
  \bibinfo{author}{\bibfnamefont{f.~t.~V.} \bibnamefont{Collaboration}}, in
  \emph{\bibinfo{booktitle}{Proceedings, 34th International Cosmic Ray
  Conference (ICRC 2015)}} (\bibinfo{year}{2015}), \eprint{1508.06597}.

\bibitem[{\citenamefont{Aharonian et~al.}(2008)}]{Aharonian:2008aa}
\bibinfo{author}{\bibfnamefont{F.}~\bibnamefont{Aharonian}}
  \bibnamefont{et~al.} (\bibinfo{collaboration}{H.E.S.S.}),
  \bibinfo{journal}{Phys. Rev. Lett.} \textbf{\bibinfo{volume}{101}},
  \bibinfo{pages}{261104} (\bibinfo{year}{2008}), \eprint{0811.3894}.

\bibitem[{\citenamefont{Aharonian et~al.}(2009)}]{Aharonian:2009ah}
\bibinfo{author}{\bibfnamefont{F.}~\bibnamefont{Aharonian}}
  \bibnamefont{et~al.} (\bibinfo{collaboration}{H.E.S.S.}),
  \bibinfo{journal}{Astron. Astrophys.} \textbf{\bibinfo{volume}{508}},
  \bibinfo{pages}{561} (\bibinfo{year}{2009}), \eprint{0905.0105}.

\bibitem[{\citenamefont{Ackermann et~al.}(2010)\citenamefont{Ackermann, Ajello,
  Atwood, Baldini, Ballet, Barbiellini, Bastieri, Baughman, Bechtol, Bellardi
  et~al.}}]{Ackermann:2010ij}
\bibinfo{author}{\bibfnamefont{M.}~\bibnamefont{Ackermann}},
  \bibinfo{author}{\bibfnamefont{M.}~\bibnamefont{Ajello}},
  \bibinfo{author}{\bibfnamefont{W.~B.} \bibnamefont{Atwood}},
  \bibinfo{author}{\bibfnamefont{L.}~\bibnamefont{Baldini}},
  \bibinfo{author}{\bibfnamefont{J.}~\bibnamefont{Ballet}},
  \bibinfo{author}{\bibfnamefont{G.}~\bibnamefont{Barbiellini}},
  \bibinfo{author}{\bibfnamefont{D.}~\bibnamefont{Bastieri}},
  \bibinfo{author}{\bibfnamefont{B.~M.} \bibnamefont{Baughman}},
  \bibinfo{author}{\bibfnamefont{K.}~\bibnamefont{Bechtol}},
  \bibinfo{author}{\bibfnamefont{F.}~\bibnamefont{Bellardi}},
  \bibnamefont{et~al.}, \bibinfo{journal}{Physical Review D}
  \textbf{\bibinfo{volume}{82}}, \bibinfo{pages}{092004}
  (\bibinfo{year}{2010}), ISSN \bibinfo{issn}{1550-7998}, \eprint{1008.3999}.

\bibitem[{\citenamefont{{VERITAS Collaboration}
  et~al.}(2017)\citenamefont{{VERITAS Collaboration}, Archambault, Archer,
  Benbow, Bird, Bourbeau, Brantseg, Buchovecky, Buckley, Bugaev
  et~al.}}]{Archambault:2017wyh}
\bibinfo{author}{\bibnamefont{{VERITAS Collaboration}}},
  \bibinfo{author}{\bibfnamefont{S.}~\bibnamefont{Archambault}},
  \bibinfo{author}{\bibfnamefont{A.}~\bibnamefont{Archer}},
  \bibinfo{author}{\bibfnamefont{W.}~\bibnamefont{Benbow}},
  \bibinfo{author}{\bibfnamefont{R.}~\bibnamefont{Bird}},
  \bibinfo{author}{\bibfnamefont{E.}~\bibnamefont{Bourbeau}},
  \bibinfo{author}{\bibfnamefont{T.}~\bibnamefont{Brantseg}},
  \bibinfo{author}{\bibfnamefont{M.}~\bibnamefont{Buchovecky}},
  \bibinfo{author}{\bibfnamefont{J.~H.} \bibnamefont{Buckley}},
  \bibinfo{author}{\bibfnamefont{V.}~\bibnamefont{Bugaev}},
  \bibnamefont{et~al.}, \bibinfo{journal}{Physical Review D}
  \textbf{\bibinfo{volume}{95}}, \bibinfo{pages}{082001}
  (\bibinfo{year}{2017}), ISSN \bibinfo{issn}{2470-0010}, \eprint{1703.04937}.

\bibitem[{\citenamefont{Collaboration}(2015)}]{Ackermann:2015zua}
\bibinfo{author}{\bibfnamefont{F.-L.} \bibnamefont{Collaboration}},
  \bibinfo{journal}{Physical Review Letters} \textbf{\bibinfo{volume}{115}},
  \bibinfo{pages}{231301} (\bibinfo{year}{2015}), ISSN
  \bibinfo{issn}{0031-9007}, \eprint{1503.02641}.

\bibitem[{\citenamefont{{Reed} et~al.}(2005)\citenamefont{{Reed}, {Governato},
  {Verde}, {Gardner}, {Quinn}, {Stadel}, {Merritt}, and {Lake}}}]{Reed2005}
\bibinfo{author}{\bibfnamefont{D.}~\bibnamefont{{Reed}}},
  \bibinfo{author}{\bibfnamefont{F.}~\bibnamefont{{Governato}}},
  \bibinfo{author}{\bibfnamefont{L.}~\bibnamefont{{Verde}}},
  \bibinfo{author}{\bibfnamefont{J.}~\bibnamefont{{Gardner}}},
  \bibinfo{author}{\bibfnamefont{T.}~\bibnamefont{{Quinn}}},
  \bibinfo{author}{\bibfnamefont{J.}~\bibnamefont{{Stadel}}},
  \bibinfo{author}{\bibfnamefont{D.}~\bibnamefont{{Merritt}}},
  \bibnamefont{and} \bibinfo{author}{\bibfnamefont{G.}~\bibnamefont{{Lake}}},
  \bibinfo{journal}{\mnras} \textbf{\bibinfo{volume}{357}}, \bibinfo{pages}{82}
  (\bibinfo{year}{2005}), \eprint{astro-ph/0312544}.

\bibitem[{\citenamefont{{Diemand} et~al.}(2006)\citenamefont{{Diemand},
  {Kuhlen}, and {Madau}}}]{Diemand2006}
\bibinfo{author}{\bibfnamefont{J.}~\bibnamefont{{Diemand}}},
  \bibinfo{author}{\bibfnamefont{M.}~\bibnamefont{{Kuhlen}}}, \bibnamefont{and}
  \bibinfo{author}{\bibfnamefont{P.}~\bibnamefont{{Madau}}},
  \bibinfo{journal}{\apj} \textbf{\bibinfo{volume}{649}}, \bibinfo{pages}{1}
  (\bibinfo{year}{2006}), \eprint{astro-ph/0603250}.

\bibitem[{\citenamefont{{Hooper} and {Witte}}(2017)}]{Hooper2017}
\bibinfo{author}{\bibfnamefont{D.}~\bibnamefont{{Hooper}}} \bibnamefont{and}
  \bibinfo{author}{\bibfnamefont{S.~J.} \bibnamefont{{Witte}}},
  \bibinfo{journal}{\jcap} \textbf{\bibinfo{volume}{4}}, \bibinfo{eid}{018}
  (\bibinfo{year}{2017}), \eprint{1610.07587}.

\bibitem[{\citenamefont{{Yang}}(2016)}]{Yang2016_ultracompact}
\bibinfo{author}{\bibfnamefont{Y.}~\bibnamefont{{Yang}}},
  \bibinfo{journal}{European Physical Journal Plus}
  \textbf{\bibinfo{volume}{131}}, \bibinfo{eid}{432} (\bibinfo{year}{2016}),
  \eprint{1612.06559}.

\bibitem[{\citenamefont{{Yang} et~al.}(2013{\natexlab{a}})\citenamefont{{Yang},
  {Yang}, {Huang}, {Chen}, {Lu}, and {Zong}}}]{Yang2013a}
\bibinfo{author}{\bibfnamefont{Y.}~\bibnamefont{{Yang}}},
  \bibinfo{author}{\bibfnamefont{G.}~\bibnamefont{{Yang}}},
  \bibinfo{author}{\bibfnamefont{X.}~\bibnamefont{{Huang}}},
  \bibinfo{author}{\bibfnamefont{X.}~\bibnamefont{{Chen}}},
  \bibinfo{author}{\bibfnamefont{T.}~\bibnamefont{{Lu}}}, \bibnamefont{and}
  \bibinfo{author}{\bibfnamefont{H.}~\bibnamefont{{Zong}}},
  \bibinfo{journal}{\prd} \textbf{\bibinfo{volume}{87}}, \bibinfo{eid}{083519}
  (\bibinfo{year}{2013}{\natexlab{a}}), \eprint{1206.3750}.

\bibitem[{\citenamefont{{Yang} et~al.}(2013{\natexlab{b}})\citenamefont{{Yang},
  {Yang}, and {Zong}}}]{Yang2013b}
\bibinfo{author}{\bibfnamefont{Y.}~\bibnamefont{{Yang}}},
  \bibinfo{author}{\bibfnamefont{G.}~\bibnamefont{{Yang}}}, \bibnamefont{and}
  \bibinfo{author}{\bibfnamefont{H.}~\bibnamefont{{Zong}}},
  \bibinfo{journal}{\prd} \textbf{\bibinfo{volume}{87}}, \bibinfo{eid}{103525}
  (\bibinfo{year}{2013}{\natexlab{b}}), \eprint{1305.4213}.

\bibitem[{\citenamefont{Ackermann et~al.}(2015)\citenamefont{Ackermann, Ajello,
  Albert, Atwood, Baldini, Ballet, Barbiellini, Bastieri, Bechtol, Bellazzini
  et~al.}}]{Ackermann:2014usa}
\bibinfo{author}{\bibfnamefont{M.}~\bibnamefont{Ackermann}},
  \bibinfo{author}{\bibfnamefont{M.}~\bibnamefont{Ajello}},
  \bibinfo{author}{\bibfnamefont{A.}~\bibnamefont{Albert}},
  \bibinfo{author}{\bibfnamefont{W.~B.} \bibnamefont{Atwood}},
  \bibinfo{author}{\bibfnamefont{L.}~\bibnamefont{Baldini}},
  \bibinfo{author}{\bibfnamefont{J.}~\bibnamefont{Ballet}},
  \bibinfo{author}{\bibfnamefont{G.}~\bibnamefont{Barbiellini}},
  \bibinfo{author}{\bibfnamefont{D.}~\bibnamefont{Bastieri}},
  \bibinfo{author}{\bibfnamefont{K.}~\bibnamefont{Bechtol}},
  \bibinfo{author}{\bibfnamefont{R.}~\bibnamefont{Bellazzini}},
  \bibnamefont{et~al.}, \bibinfo{journal}{The Astrophysical Journal}
  \textbf{\bibinfo{volume}{799}}, \bibinfo{pages}{86} (\bibinfo{year}{2015}),
  ISSN \bibinfo{issn}{1538-4357}, \eprint{1410.3696}.

\bibitem[{\citenamefont{{Globus} and {Piran}}(2017)}]{Globus2017}
\bibinfo{author}{\bibfnamefont{N.}~\bibnamefont{{Globus}}} \bibnamefont{and}
  \bibinfo{author}{\bibfnamefont{T.}~\bibnamefont{{Piran}}},
  \bibinfo{journal}{\apjl} \textbf{\bibinfo{volume}{850}}, \bibinfo{eid}{L25}
  (\bibinfo{year}{2017}), \eprint{1709.10110}.

\bibitem[{\citenamefont{{Springel} et~al.}(2008)\citenamefont{{Springel},
  {Wang}, {Vogelsberger}, {Ludlow}, {Jenkins}, {Helmi}, {Navarro}, {Frenk}, and
  {White}}}]{Aquarius2008}
\bibinfo{author}{\bibfnamefont{V.}~\bibnamefont{{Springel}}},
  \bibinfo{author}{\bibfnamefont{J.}~\bibnamefont{{Wang}}},
  \bibinfo{author}{\bibfnamefont{M.}~\bibnamefont{{Vogelsberger}}},
  \bibinfo{author}{\bibfnamefont{A.}~\bibnamefont{{Ludlow}}},
  \bibinfo{author}{\bibfnamefont{A.}~\bibnamefont{{Jenkins}}},
  \bibinfo{author}{\bibfnamefont{A.}~\bibnamefont{{Helmi}}},
  \bibinfo{author}{\bibfnamefont{J.~F.} \bibnamefont{{Navarro}}},
  \bibinfo{author}{\bibfnamefont{C.~S.} \bibnamefont{{Frenk}}},
  \bibnamefont{and} \bibinfo{author}{\bibfnamefont{S.~D.~M.}
  \bibnamefont{{White}}}, \bibinfo{journal}{\mnras}
  \textbf{\bibinfo{volume}{391}}, \bibinfo{pages}{1685} (\bibinfo{year}{2008}),
  \eprint{0809.0898}.

\bibitem[{\citenamefont{{Diemand}
  et~al.}(2007{\natexlab{a}})\citenamefont{{Diemand}, {Kuhlen}, and
  {Madau}}}]{ViaLactea}
\bibinfo{author}{\bibfnamefont{J.}~\bibnamefont{{Diemand}}},
  \bibinfo{author}{\bibfnamefont{M.}~\bibnamefont{{Kuhlen}}}, \bibnamefont{and}
  \bibinfo{author}{\bibfnamefont{P.}~\bibnamefont{{Madau}}},
  \bibinfo{journal}{\apj} \textbf{\bibinfo{volume}{657}}, \bibinfo{pages}{262}
  (\bibinfo{year}{2007}{\natexlab{a}}), \eprint{astro-ph/0611370}.

\bibitem[{\citenamefont{{Diemand}
  et~al.}(2007{\natexlab{b}})\citenamefont{{Diemand}, {Kuhlen}, and
  {Madau}}}]{ViaLacteaI}
\bibinfo{author}{\bibfnamefont{J.}~\bibnamefont{{Diemand}}},
  \bibinfo{author}{\bibfnamefont{M.}~\bibnamefont{{Kuhlen}}}, \bibnamefont{and}
  \bibinfo{author}{\bibfnamefont{P.}~\bibnamefont{{Madau}}},
  \bibinfo{journal}{\apj} \textbf{\bibinfo{volume}{667}}, \bibinfo{pages}{859}
  (\bibinfo{year}{2007}{\natexlab{b}}), \eprint{astro-ph/0703337}.

\bibitem[{\citenamefont{{Diemand} et~al.}(2008)\citenamefont{{Diemand},
  {Kuhlen}, {Madau}, {Zemp}, {Moore}, {Potter}, and {Stadel}}}]{ViaLacteaII}
\bibinfo{author}{\bibfnamefont{J.}~\bibnamefont{{Diemand}}},
  \bibinfo{author}{\bibfnamefont{M.}~\bibnamefont{{Kuhlen}}},
  \bibinfo{author}{\bibfnamefont{P.}~\bibnamefont{{Madau}}},
  \bibinfo{author}{\bibfnamefont{M.}~\bibnamefont{{Zemp}}},
  \bibinfo{author}{\bibfnamefont{B.}~\bibnamefont{{Moore}}},
  \bibinfo{author}{\bibfnamefont{D.}~\bibnamefont{{Potter}}}, \bibnamefont{and}
  \bibinfo{author}{\bibfnamefont{J.}~\bibnamefont{{Stadel}}},
  \bibinfo{journal}{\nat} \textbf{\bibinfo{volume}{454}}, \bibinfo{pages}{735}
  (\bibinfo{year}{2008}), \eprint{0805.1244}.

\bibitem[{\citenamefont{{Gao} et~al.}(2011)\citenamefont{{Gao}, {Frenk},
  {Boylan-Kolchin}, {Jenkins}, {Springel}, and {White}}}]{Gao2011}
\bibinfo{author}{\bibfnamefont{L.}~\bibnamefont{{Gao}}},
  \bibinfo{author}{\bibfnamefont{C.~S.} \bibnamefont{{Frenk}}},
  \bibinfo{author}{\bibfnamefont{M.}~\bibnamefont{{Boylan-Kolchin}}},
  \bibinfo{author}{\bibfnamefont{A.}~\bibnamefont{{Jenkins}}},
  \bibinfo{author}{\bibfnamefont{V.}~\bibnamefont{{Springel}}},
  \bibnamefont{and} \bibinfo{author}{\bibfnamefont{S.~D.~M.}
  \bibnamefont{{White}}}, \bibinfo{journal}{\mnras}
  \textbf{\bibinfo{volume}{410}}, \bibinfo{pages}{2309} (\bibinfo{year}{2011}),
  \eprint{1006.2882}.

\bibitem[{\citenamefont{{Gao} et~al.}(2012)\citenamefont{{Gao}, {Navarro},
  {Frenk}, {Jenkins}, {Springel}, and {White}}}]{PhoenixProject}
\bibinfo{author}{\bibfnamefont{L.}~\bibnamefont{{Gao}}},
  \bibinfo{author}{\bibfnamefont{J.~F.} \bibnamefont{{Navarro}}},
  \bibinfo{author}{\bibfnamefont{C.~S.} \bibnamefont{{Frenk}}},
  \bibinfo{author}{\bibfnamefont{A.}~\bibnamefont{{Jenkins}}},
  \bibinfo{author}{\bibfnamefont{V.}~\bibnamefont{{Springel}}},
  \bibnamefont{and} \bibinfo{author}{\bibfnamefont{S.~D.~M.}
  \bibnamefont{{White}}}, \bibinfo{journal}{\mnras}
  \textbf{\bibinfo{volume}{425}}, \bibinfo{pages}{2169} (\bibinfo{year}{2012}),
  \eprint{1201.1940}.

\bibitem[{\citenamefont{{Jiang} and {van den Bosch}}(2016)}]{Jiang2016}
\bibinfo{author}{\bibfnamefont{F.}~\bibnamefont{{Jiang}}} \bibnamefont{and}
  \bibinfo{author}{\bibfnamefont{F.~C.} \bibnamefont{{van den Bosch}}},
  \bibinfo{journal}{\mnras} \textbf{\bibinfo{volume}{458}},
  \bibinfo{pages}{2848} (\bibinfo{year}{2016}).

\bibitem[{\citenamefont{{van den Bosch} and {Jiang}}(2016)}]{Bosch2016}
\bibinfo{author}{\bibfnamefont{F.~C.} \bibnamefont{{van den Bosch}}}
  \bibnamefont{and} \bibinfo{author}{\bibfnamefont{F.}~\bibnamefont{{Jiang}}},
  \bibinfo{journal}{\mnras} \textbf{\bibinfo{volume}{458}},
  \bibinfo{pages}{2870} (\bibinfo{year}{2016}).

\bibitem[{\citenamefont{{Giocoli} et~al.}(2008)\citenamefont{{Giocoli},
  {Tormen}, and {van den Bosch}}}]{Giocoli2008}
\bibinfo{author}{\bibfnamefont{C.}~\bibnamefont{{Giocoli}}},
  \bibinfo{author}{\bibfnamefont{G.}~\bibnamefont{{Tormen}}}, \bibnamefont{and}
  \bibinfo{author}{\bibfnamefont{F.~C.} \bibnamefont{{van den Bosch}}},
  \bibinfo{journal}{\mnras} \textbf{\bibinfo{volume}{386}},
  \bibinfo{pages}{2135} (\bibinfo{year}{2008}), \eprint{0712.1563}.

\bibitem[{\citenamefont{{Brun} et~al.}(2009)\citenamefont{{Brun}, {Delahaye},
  {Diemand}, {Profumo}, and {Salati}}}]{Brun2009}
\bibinfo{author}{\bibfnamefont{P.}~\bibnamefont{{Brun}}},
  \bibinfo{author}{\bibfnamefont{T.}~\bibnamefont{{Delahaye}}},
  \bibinfo{author}{\bibfnamefont{J.}~\bibnamefont{{Diemand}}},
  \bibinfo{author}{\bibfnamefont{S.}~\bibnamefont{{Profumo}}},
  \bibnamefont{and} \bibinfo{author}{\bibfnamefont{P.}~\bibnamefont{{Salati}}},
  \bibinfo{journal}{\prd} \textbf{\bibinfo{volume}{80}}, \bibinfo{eid}{035023}
  (\bibinfo{year}{2009}), \eprint{0904.0812}.

\bibitem[{\citenamefont{{Yuan} et~al.}(2017)\citenamefont{{Yuan}, {Feng},
  {Yin}, {Fan}, {Bi}, {Cui}, {Dong}, {Guo}, {Fang}, {Hu} et~al.}}]{Yuan2017}
\bibinfo{author}{\bibfnamefont{Q.}~\bibnamefont{{Yuan}}},
  \bibinfo{author}{\bibfnamefont{L.}~\bibnamefont{{Feng}}},
  \bibinfo{author}{\bibfnamefont{P.-F.} \bibnamefont{{Yin}}},
  \bibinfo{author}{\bibfnamefont{Y.-Z.} \bibnamefont{{Fan}}},
  \bibinfo{author}{\bibfnamefont{X.-J.} \bibnamefont{{Bi}}},
  \bibinfo{author}{\bibfnamefont{M.-Y.} \bibnamefont{{Cui}}},
  \bibinfo{author}{\bibfnamefont{T.-K.} \bibnamefont{{Dong}}},
  \bibinfo{author}{\bibfnamefont{Y.-Q.} \bibnamefont{{Guo}}},
  \bibinfo{author}{\bibfnamefont{K.}~\bibnamefont{{Fang}}},
  \bibinfo{author}{\bibfnamefont{H.-B.} \bibnamefont{{Hu}}},
  \bibnamefont{et~al.}, \bibinfo{journal}{ArXiv e-prints}
  (\bibinfo{year}{2017}), \eprint{1711.10989}.

\end{thebibliography}

\end{document}